\documentclass[review,12pt]{elsarticle}

\usepackage{lineno}

\usepackage{graphicx}
\usepackage{amsfonts}
\usepackage{amssymb}
\usepackage{float}
\usepackage{latexsym}
\usepackage{amsmath}

\usepackage{subfigure}
\usepackage[euler]{textgreek}
\usepackage{textcomp}
\usepackage{CJK}

\usepackage{algorithm}
\PassOptionsToPackage{noend}{algpseudocode}
\usepackage[end]{algpseudocode}

\usepackage[table]{xcolor}

\graphicspath{ {./figure/}}
\begin{document}

\begin{frontmatter}

\title{JAX-Nano: A GPU-Accelerated Differentiable Phonon Boltzmann Transport Solver for Nanoscale Heat Transfer}

%\author[add1]{Bo Zhang\corref{cor1}}
%\ead{bzhang@niu.edu}
%\cortext[cor1]{Corresponding author. 
%	Address:  
% 	DeKalb, IL 60115, USA
% }
%\address[add1]{Department of Mechanical Engineering, Northern Illinois University, DeKalb, IL 60115, USA}

\author[add1]{Bo Zhang\corref{cor1}}
\ead{bzhang@niu.edu}

\author[add2]{Amy Zhaomei Zhang}
\ead{azhang3@imsa.edu}

\cortext[cor1]{Corresponding author.
    Address:
    DeKalb, IL 60115, USA
}

\address[add1]{Department of Mechanical Engineering,
Northern Illinois University, DeKalb, IL 60115, USA}

\address[add2]{
Illinois Mathematics and Science Academy, Aurora, IL, 60506, USA}

%===============================================================================
%	Abstract
%===============================================================================
\begin{abstract}
As characteristic dimensions of semiconductor devices approach phonon mean free paths, heat transport enters ballistic and quasi-ballistic regimes in which Fourier's law becomes inadequate. The phonon Boltzmann transport equation (BTE) provides a physics-based description of these nonequilibrium effects, but its high-dimensional computational structure poses substantial challenges for numerical simulation and inverse modeling. 
This work presents JAX-Nano, a GPU-accelerated, end-to-end differentiable phonon BTE solver for nanoscale heat transfer. 
The solver implements
steady-state and transient BTE formulations within
a unified finite-volume computational framework,
supporting structured and unstructured meshes,
mode-resolved phonon transport, scattering,
volumetric heat generation, and multiple boundary
conditions. Steady-state transport is solved using
sparse iterative methods, while transient transport
is advanced using explicit and implicit
time-integration schemes. Automatic differentiation
is integrated directly into the discretized BTE
solver, enabling gradient computation with respect
to transport properties, boundary conditions,
heat-generation rates, and geometric parameters
without separately derived adjoint equations.
Forward solutions are validated against analytical
benchmarks and established BTE solvers across
diffusive-to-ballistic transport regimes. The
applicability of JAX-Nano to complex three-dimensional
geometries is demonstrated through thermal transport
simulations of single-fin, double-fin, and arrayed
fin field-effect transistor (FinFET) configurations.
The differentiable capabilities of the solver are
demonstrated through gradient-based single- and
multi-parameter inverse identification, including
the simultaneous recovery of mode-resolved phonon
relaxation times and spatially distributed
volumetric heat-generation rates. Both steady-state
and transient inverse problems are investigated.
Computational benchmarks quantify GPU performance
and scalability with respect to spatial resolution,
angular discretization, and the number of phonon
bands, including comparisons with an established
MPI-based phonon BTE solver. The results demonstrate
that JAX-Nano integrates three-dimensional phonon
transport simulation, automatic differentiation,
and gradient-based inverse parameter identification
within a unified GPU-accelerated computational
framework, providing a foundation for future
sensitivity analysis and optimization of nanoscale
thermal systems.
\end{abstract}


\end{frontmatter}

%\linenumbers

%===============================================================================
%	Introduction
%===============================================================================
\section{Introduction}
Nanoscale heat transport~\cite{Wilson_2014a, Cepellotti_2015a, Chen_2021a} is important in a broad range of applications, including micro- and nanoelectronic devices~\cite{Tien_1994a, Sverdrup_1999a, Chhowalla_2016a, Cao_2023a, Raj_2027a}, thin films~\cite{Chen_1997a, Goodson_1999a, Venkatasubramanian_2001a, Ghosh_2010a}, superlattices~\cite{Simkin_2000a, Koh_2009a, Garg_2011a, Ravichandran_2014a}, nanomaterials~\cite{Hochbaum_2008a, Seol_2010a, Balandin_2011a}, thermoelectrics~\cite{Rowe_2005a, Delaire_2011a}, and short-pulse laser heating~\cite{Wang_2016a, Lindenberg_2017a}. Continued miniaturization and three-dimensional integration of semiconductor devices~\cite{Lundstrom_2006a} have further intensified thermal-management challenges in modern electronic systems. 
As characteristic dimensions approach the mean free paths of heat-carrying phonons, the Knudsen number, $Kn=\Lambda/L$, becomes non-negligible, where $\Lambda$ and $L$ denote the phonon mean free path and characteristic length scale, respectively. Under these conditions, thermal transport can depart substantially from the diffusive regime underlying Fourier's law~\cite{Majumdar_1993a, Chen_2005b, Cahill_2014a, Bao_2018a, Warzoha_2021a, Qian_2021a}. 
Ballistic and quasi-ballistic transport~\cite{Siemens_2010a, Hu_2015a, Zhang_2021a}, phonon-boundary interactions~\cite{Shao_2018a, Ravichandran_2018a, Singh_2011a}, and localized heat generation~\cite{Hu_2024a} can produce strongly nonequilibrium thermal behavior that is not adequately represented by conventional diffusion-based models. Accurate prediction of heat transport at these scales therefore requires computational descriptions that explicitly account for phonon propagation and scattering~\cite{Schleeh_2015a, Murthy_2002a, Hu_2022b}.

The limitations of Fourier's law at micro- and nanoscales have motivated the development and application of alternative approaches for modeling heat transport, including molecular dynamics (MD)~\cite{Volz_1999a, Schelling_2002a}, nonequilibrium Green's function (NEGF) methods~\cite{Mingo_2003a, Wang_2014a, Chen_2022a}, and the phonon Boltzmann transport equation (BTE)~\cite{Hu_2024a, Zou_2001a}. 
While atomistic methods such as MD and quantum transport methods such as NEGF provide detailed descriptions of nanoscale thermal transport, their computational cost generally limits their direct application to large device-scale systems. 
Conversely, continuum heat-diffusion models become inadequate when characteristic dimensions are comparable to phonon mean free paths and non-diffusive transport becomes significant~\cite{Bao_2018a}. 
The phonon BTE provides a physics-based framework that bridges these transport regimes and describes heat transport across diffusive, quasi-ballistic, and ballistic regimes. In conjunction with first-principles lattice-dynamics calculations, the BTE can incorporate mode-resolved phonon properties and scattering processes~\cite{Broido_2007a, Tadano_2014a}. However, the BTE is a high-dimensional integrodifferential equation that, in its transient spectral form, involves seven independent variables: time, three spatial coordinates, two angular coordinates, and phonon frequency. In contrast to Fourier's law, the BTE resolves the evolution of the phonon population over both physical and phonon-state spaces and can explicitly account for phonon scattering, boundary interactions, and volumetric heat generation.

Both deterministic and stochastic approaches have been developed for solving the phonon BTE. Deterministic approaches commonly employ the discrete ordinates method (DOM)~\cite{Ali_2014a, Modest_2021a} for angular discretization together with the finite-volume method (FVM)~\cite{Murthy_2002a, Tryggvason_2011a, Mazumder_2015a, Mangani_2016a, Zhang_2020a} for spatial discretization, whereas stochastic approaches are typically based on Monte Carlo methods~\cite{Mazumder_2001a, Mittal_2010a, Pathak_2021a, Shao_2022a}. 
Advances in first-principles phonon calculations and numerical algorithms have further enabled increasingly realistic simulations of nanoscale thermal transport~\cite{Shiomi_2011a, Shiga_2012a, Peng_2022a}. 
Several computational tools, including ShengBTE~\cite{Li_2014a}, almaBTE~\cite{Carrete_2017a}, OpenBTE~\cite{Romano_2021a}, MCBTE~\cite{Pathak_2021a}, P-TRANS~\cite{Shao_2022a}, and GiftBTE~\cite{Hu_2024a}, have further expanded the capability for predicting thermal conductivity and simulating phonon transport in bulk materials and complex geometries. 
Nevertheless, the need to resolve phonon transport over physical space together with directional and modal degrees of freedom makes three-dimensional device-scale BTE simulations computationally demanding.

Most existing phonon-BTE computations are formulated primarily as forward problems: 
material properties, geometry, boundary conditions, and heat-source characteristics 
are prescribed, and the resulting temperature and heat-flux fields are 
calculated~\cite{Hu_2024a, Li_2014a, Carrete_2017a, Romano_2021a, 
Pathak_2021a, Shao_2022a}. Many emerging problems in nanoscale thermal science, 
however, require the reverse connection between thermal observations and the 
parameters responsible for them. Examples include identifying uncertain transport 
or boundary parameters from temperature measurements, determining sensitivities 
of device temperatures to geometric variations, and optimizing nanoscale structures 
for improved thermal management. These tasks require derivatives of quantities of 
interest with respect to potentially many physical and design parameters.

Inverse heat-transfer problems have been extensively studied using parameter estimation,
sequential, sensitivity, gradient, regularization, and adjoint-based
methods~\cite{Beck_1985a, Hsieh_1986a, Jarny_1991a, Alifanov_1994a,
Sawaf_1995a, Kim_2003a, Kim_2003b, Anderson_2006a, Malinowski_2014a}. These approaches have
been applied to the estimation of unknown boundary conditions, heat fluxes,
thermophysical properties, heat-transfer coefficients, and geometric parameters
from thermal observations. Related inverse and sensitivity methods have also been
developed for nanoscale phonon transport. Kim and Daniel~\cite{Kim_2004a}
derived sensitivity, adjoint, and gradient equations for a one-dimensional inverse
heat-conduction problem governed by the phonon BTE, while P\'eraud and
Hadjiconstantinou~\cite{Peraud_2015a} developed an adjoint formulation of the
linearized phonon BTE for efficient deviational Monte Carlo calculations.
Romano and Johnson~\cite{Romano_2022a} employed adjoint-based sensitivity
analysis of the phonon BTE for inverse design and topology optimization of
nondiffusive thermal transport. More recently, Chen et al.~\cite{Chen_2023a}
combined phonon-transport modeling with scientific machine learning to infer
frequency-dependent interfacial transmission and relaxation properties from
ultrafast electron-diffraction measurements.

Despite these developments, efficient sensitivity evaluation remains challenging for 
general multidimensional phonon-BTE simulations. Conventional finite-difference 
sensitivity analysis requires additional BTE solutions for perturbed parameters and 
can therefore become computationally expensive for high-dimensional inverse and design 
problems. Adjoint methods can substantially reduce the cost of gradient evaluation, 
but generally require the derivation and implementation of an adjoint formulation consistent with the underlying discretized solver.

Differentiable scientific computing~\cite{Um_2020a, Holl_2020a, Bezgin_2023a, Xue_2023a, Zhang_2026a} provides an automated approach to gradient evaluation by integrating automatic differentiation (AD) directly with physics-based numerical solvers.
Unlike physics-informed neural networks (PINNs) and related neural PDE approaches, which represent the solution using a trainable neural network constrained by the governing equations~\cite{Raissi_2019a, Lu_2021b, Karniadakis_2021a, Jin_2021a, Cuomo_2022a, Yu_2022a}, a differentiable numerical solver~\cite{Bezgin_2023a, Xue_2023a, Zhang_2026a} retains the conventional numerical discretization of the physical model while allowing derivatives to be propagated through the computational procedure. 
Reverse-mode AD is particularly attractive for inverse problems and optimization because gradients of a scalar objective with respect to many parameters can, in principle, be evaluated without independently perturbing each parameter. Differentiable numerical methods have recently been explored in computational fluid dynamics~\cite{Zhang_2026a, Sirignano_2020a, Kochkov_2021a, List_2022a, Bezgin_2023a, Zhang_2023c}, finite-element analysis~\cite{Xue_2023a}, molecular dynamics~\cite{Schoenholz_2021a}, nanoscale heat transfer~\cite{Zhang_2024a}, density functional theory~\cite{Li_2021a}, and other physics-based computational problems~\cite{Um_2020a, Holl_2020a}, demonstrating the potential of combining conventional numerical simulation with gradient-based inference and optimization. GPU-oriented differentiable programming environments such as JAX~\cite{jax2018github} further provide a unified setting for automatic differentiation, just-in-time compilation, vectorized computation, and accelerator execution.

Extending this paradigm to phonon transport presents several challenges that are distinct from those encountered in conventional continuum PDE solvers. A deterministic phonon BTE calculation couples spatial transport with directional and modal phonon representations, relaxation processes, local-equilibrium reconstruction, and boundary interactions. Both steady-state iterative solutions and transient time integration must express these operations within a computational graph suitable for differentiation. Complex three-dimensional geometries further introduce boundary treatments whose derivatives must remain numerically meaningful. Most importantly, a derivative returned by AD is the exact derivative of the implemented discrete computation, but it is not automatically an accurate approximation of the sensitivity of the underlying physical problem~\cite{Zhang_2026b}. Gradient reliability may depend on spatial and angular resolution, iterative convergence, temporal discretization, and numerical treatment of boundaries. Consequently, forward-solution validation alone is insufficient for establishing the reliability of a differentiable BTE solver; the computed sensitivities themselves must also be independently verified.

In this work, we develop JAX-Nano, a GPU-accelerated, end-to-end
differentiable phonon BTE solver for nanoscale heat transfer. JAX-Nano
incorporates steady-state and transient BTE formulations within a unified
computational architecture and treats phonon transport, relaxation-time
scattering, temperature-dependent local equilibrium, volumetric heat
generation, and phonon boundary conditions in a differentiable numerical
framework. The implementation uses JAX to combine accelerator-oriented
computation with automatic differentiation, allowing derivatives of thermal
quantities and objective functions to be evaluated with respect to physical,
boundary, heat-source, and geometric parameters without separately deriving
an adjoint solver. Particular emphasis is placed on verifying not only the
forward thermal solution but also the numerical accuracy and convergence
behavior of the resulting gradients. Three-dimensional FinFET thermal
transport~\cite{Pop_2006a, Shrivastava_2012a, Kompala_2016a, Rajeev_2018a, Ahn_2018a, Rezgui_2021a, Nasri_2021a, Duan_2024a, Hu_2025a, Rao_2025a, Hu_2026a} is further considered to demonstrate the applicability of the
framework to complex nanoscale device geometries and gradient-based inverse
parameter identification.

The principal contributions of this work are
summarized as follows:
\begin{itemize}

    \item Development of a unified,
    GPU-accelerated, and differentiable
    finite-volume framework for steady-state
    and transient phonon Boltzmann transport,
    supporting structured and unstructured
    spatial discretizations, multiple boundary
    conditions, and mode-resolved phonon
    transport.

    \item Integration of sparse iterative
    solution methods for steady-state transport
    and explicit and implicit time-integration
    schemes for transient transport within
    a JAX-based computational framework,
    enabling efficient three-dimensional
    phonon BTE simulations on GPUs.

    \item Implementation of end-to-end
    automatic differentiation through the
    discretized phonon BTE solver, enabling
    gradient evaluation with respect to
    transport properties, boundary conditions,
    volumetric heat sources, and geometric
    parameters without separately deriving
    adjoint equations.

    \item Systematic verification and validation
    across diffusive-to-ballistic transport
    regimes, including comparisons with
    analytical solutions and established
    numerical methods, as well as
    three-dimensional simulations of
    complex FinFET configurations.

    \item Demonstration of gradient-based
    single- and multi-parameter inverse
    identification, together with quantitative
    evaluation of GPU computational performance
    and scalability against an established
    MPI-based phonon BTE solver.

\end{itemize}

%%===============================================================================
%	Methodology
%===============================================================================
\section{JAX-Nano Computational Framework}
\label{sec:approach}

\begin{figure}
\begin{center}
\includegraphics [width=1.\columnwidth]{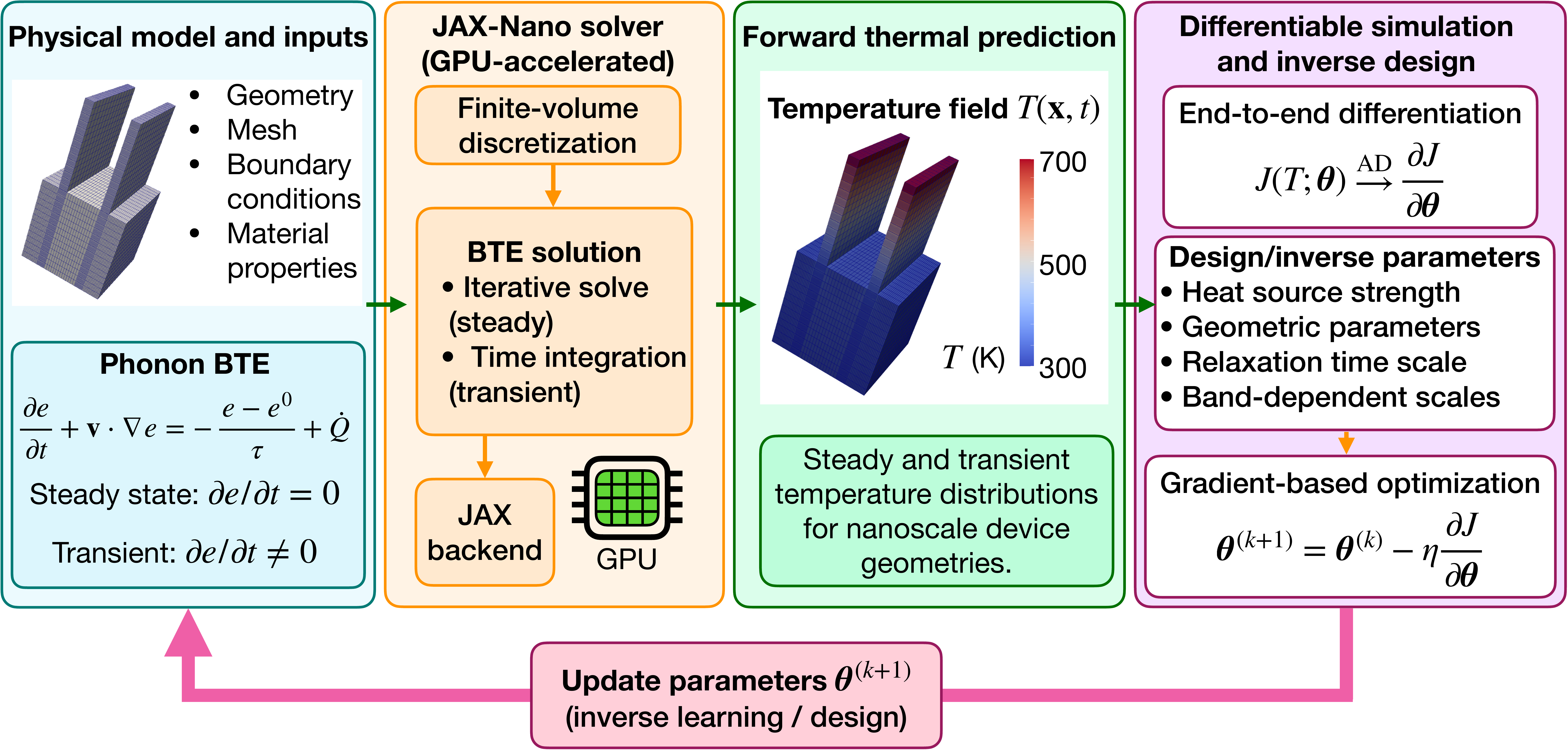}
\end{center}
\caption{Schematic of the JAX-Nano framework for GPU-accelerated differentiable phonon transport simulation and inverse design. The framework integrates physical geometry, boundary conditions, and material properties with a finite-volume phonon Boltzmann transport solver for steady and transient thermal predictions. The JAX-based implementation enables end-to-end automatic differentiation of temperature-based objectives with respect to physical and design parameters, allowing gradient-based inverse learning and optimization through an iterative parameter-update loop.}
\label{fig:schematic}
\end{figure}  

JAX-Nano provides a unified computational framework for steady-state
and transient phonon transport, GPU-accelerated computation, automatic
differentiation, and gradient-based inverse modeling. As illustrated
in Fig.~\ref{fig:schematic}, the framework connects physical inputs
and the finite-volume phonon BTE solver with forward thermal prediction
and end-to-end differentiation. Both steady-state and transient
formulations, including explicit and implicit time integration, are
implemented within the same JAX-based computational architecture.

\subsection{Phonon Boltzmann Transport Equation}
\label{sec:phonon_bte}
Phonon heat transport is described by the Boltzmann transport equation (BTE) under the relaxation-time approximation (RTA),
\begin{equation}
\frac{\partial e}{\partial t}+\mathbf{v}_g\cdot\nabla e
=\frac{e^{\mathrm{eq}}-e}{\tau}+S,
\label{eq:phonon_bte_general}
\end{equation}
where $e=e(\mathbf{x},\omega,p,\mathbf{s},t)$ is the directional spectral phonon energy distribution per unit solid angle and angular frequency, $\mathbf{x}$ is position, $\omega$ is angular frequency, $p$ denotes the phonon polarization branch, $\mathbf{s}$ is the propagation direction, and $\mathbf{v}_g$ is the group velocity. The relaxation time $\tau$ characterizes scattering, while $S$ represents external energy generation. The frequency integration is restricted to the physical phonon spectrum $[\omega_{\min,p},\omega_{\max,p}]$ for each branch.

In the linearized formulation, the isotropic equilibrium distribution is
\begin{equation}
e^{\mathrm{eq}}(\omega,p,T_L)=\frac{C_{\omega p}T_L}{4\pi},
\label{eq:phonon_equilibrium}
\end{equation}
where $C_{\omega p}$ is the spectral volumetric heat capacity per unit angular frequency and $T_L$ is the lattice temperature.

The local energy-equivalent temperature $T$ is defined by matching the total phonon energy to the corresponding linearized equilibrium energy~\cite{Kittel_1953a, Ziman_1960a},
\begin{equation}
\sum_p\int_{\omega_{\min,p}}^{\omega_{\max,p}}
\int_{4\pi}e\,d\Omega\,d\omega
=T\sum_p\int_{\omega_{\min,p}}^{\omega_{\max,p}}
C_{\omega p}\,d\omega.
\label{eq:phonon_temperature_energy_balance}
\end{equation}
Here, $\Omega$ denotes the solid angle in direction space, and
$d\Omega$ is the differential solid-angle element. Integration over
all phonon propagation directions on the unit sphere is represented by
$\int_{4\pi}(\cdot)\,d\Omega$. In spherical angular coordinates,
\begin{equation}
d\Omega=\sin\theta\,d\theta\,d\phi,
\qquad
\int_{4\pi}d\Omega=4\pi,
\label{eq:solid_angle}
\end{equation}
where $\theta\in[0,\pi]$ is the polar angle and
$\phi\in[0,2\pi)$ is the azimuthal angle.
Accordingly,
\begin{equation}
T(\mathbf{x},t)=
\frac{\displaystyle\sum_p\int_{\omega_{\min,p}}^{\omega_{\max,p}}
\int_{4\pi}e(\mathbf{x},\omega,p,\mathbf{s},t)\,d\Omega\,d\omega}
{\displaystyle\sum_p\int_{\omega_{\min,p}}^{\omega_{\max,p}}
C_{\omega p}\,d\omega}.
\label{eq:phonon_general_temperature}
\end{equation}

The lattice temperature $T_L$ appearing in the RTA collision term is determined by requiring that phonon scattering conserve energy,
\begin{equation}
\sum_p\int_{\omega_{\min,p}}^{\omega_{\max,p}}
\int_{4\pi}\frac{e^{\mathrm{eq}}-e}{\tau_{\omega p}}
\,d\Omega\,d\omega=0.
\label{eq:phonon_scattering_energy_conservation}
\end{equation}
Substitution of Eq.~\eqref{eq:phonon_equilibrium} gives
\begin{equation}
T_L(\mathbf{x},t)=
\frac{\displaystyle\sum_p\int_{\omega_{\min,p}}^{\omega_{\max,p}}
\int_{4\pi}\frac{e(\mathbf{x},\omega,p,\mathbf{s},t)}{\tau_{\omega p}}
\,d\Omega\,d\omega}
{\displaystyle\sum_p\int_{\omega_{\min,p}}^{\omega_{\max,p}}
\frac{C_{\omega p}}{\tau_{\omega p}}\,d\omega}.
\label{eq:phonon_lattice_temperature}
\end{equation}
In general, $T$ and $T_L$ differ when relaxation times vary across phonon modes. 

The macroscopic heat-flux vector is obtained as the first velocity moment of the energy distribution,
\begin{equation}
\mathbf{q}(\mathbf{x},t)=
\sum_p\int_{\omega_{\min,p}}^{\omega_{\max,p}}
\int_{4\pi}\mathbf{v}_g e\,d\Omega\,d\omega.
\label{eq:phonon_heat_flux}
\end{equation}

For isotropic volumetric heat generation $Q(\mathbf{x},t)$, a heat-capacity-weighted source can be written as
\begin{equation}
S(\mathbf{x},\omega,p,\mathbf{s},t)=
\frac{C_{\omega p}}{4\pi\displaystyle\sum_{p'}
\int_{\omega_{\min,p'}}^{\omega_{\max,p'}}C_{\omega' p'}\,d\omega'}
Q(\mathbf{x},t),
\label{eq:phonon_source}
\end{equation}
which satisfies
\begin{equation}
\sum_p\int_{\omega_{\min,p}}^{\omega_{\max,p}}
\int_{4\pi}S\,d\Omega\,d\omega=Q(\mathbf{x},t).
\label{eq:phonon_source_conservation}
\end{equation}
The same physical formulation applies to steady and transient transport; the time derivative in Eq.~\eqref{eq:phonon_bte_general} vanishes in the steady limit. 

\subsection{Band Discretization}
\label{sec:band_discretization}
The continuous phonon spectrum is discretized into a finite
number of representative bands using the mean-free-path-based
method implemented in GiftBTE by Hu et
al.~\cite{Hu_2024a,Hu_2024b}. The input phonon properties,
including mode heat capacity, group-velocity magnitude,
and relaxation time, are obtained from lattice-dynamics
calculations. For each phonon mode $m$, the mean free path
is defined as
\begin{equation}
\Lambda_m=v_m\tau_m,
\label{eq:phonon_mode_mfp}
\end{equation}
where $v_m$ and $\tau_m$ denote the group-velocity magnitude
and relaxation time of mode $m$, respectively.

The mean-free-path domain
$[\Lambda_{\min},\Lambda_{\max}]$ is partitioned into
$N_b$ intervals, with each interval defining a
representative phonon band. Let $\mathcal{B}_b$ denote
the set of phonon modes assigned to band $b$.
The effective band heat capacity, group velocity,
and relaxation time are calculated as
\begin{equation}
C_b=\sum_{m\in\mathcal{B}_b}C_m,
\label{eq:band_heat_capacity}
\end{equation}
\begin{equation}
v_b=
\frac{\displaystyle\sum_{m\in\mathcal{B}_b}C_m v_m}
{\displaystyle\sum_{m\in\mathcal{B}_b}C_m},
\label{eq:band_group_velocity}
\end{equation}
and
\begin{equation}
\tau_b=
\frac{\displaystyle\sum_{m\in\mathcal{B}_b}
C_m v_m^2\tau_m}
{\displaystyle v_b
\sum_{m\in\mathcal{B}_b}C_m v_m},
\label{eq:band_relaxation_time}
\end{equation}
where $C_m$ is the heat capacity associated with
mode $m$. The resulting $C_b$, $v_b$, and $\tau_b$
represent the effective transport properties of
band $b$. This band representation assumes isotropic
phonon properties, such that the group-velocity
magnitude and relaxation time are independent of
propagation direction~\cite{Yang_2013a}.

With this discretization, the frequency integrals
in the continuous phonon BTE are approximated by
summations over the representative bands:
\begin{equation}
\sum_p\int_{\omega_{\min,p}}^{\omega_{\max,p}}
\frac{e}{\tau_{\omega p}}\,d\omega
\approx
\sum_{b=1}^{N_b}\frac{e_b}{\tau_b},
\label{eq:band_energy_discretization}
\end{equation}
and
\begin{equation}
\sum_p\int_{\omega_{\min,p}}^{\omega_{\max,p}}
\frac{C_{\omega p}}{\tau_{\omega p}}\,d\omega
\approx
\sum_{b=1}^{N_b}\frac{C_b}{\tau_b}.
\label{eq:band_heat_capacity_discretization}
\end{equation}
Here, $e_b$ denotes the band-resolved directional
energy distribution per unit solid angle, obtained
by aggregating the spectral contributions associated
with band $b$. The resulting band properties are
used in JAX-Nano for the subsequent directional
and spatial discretizations.

\subsection{Directional Discretization}
\label{sec:directional_discretization}
The angular dependence of the phonon energy distribution is discretized
using the Gauss--Legendre quadrature procedure implemented in GiftBTE by
Hu et al.~\cite{Hu_2024a,Hu_2024b}. Integration over the full solid angle
is approximated by a weighted summation over discrete propagation directions:
\begin{equation}
\int_{4\pi}e_b\,d\Omega
=\int_0^\pi\int_0^{2\pi}e_b\sin\theta\,d\phi\,d\theta
\approx\sum_{\alpha=1}^{N_\alpha}w_\alpha e_{b\alpha},
\label{eq:directional_quadrature}
\end{equation}
where $\alpha$ indexes the discrete propagation direction,
$e_{b\alpha}$ is the band-resolved directional phonon energy distribution,
and $w_\alpha$ is its angular quadrature weight.

For each discrete direction, the unit propagation vector is
\begin{equation}
\mathbf{s}_\alpha=\left(
\sin\theta_\alpha\cos\phi_\alpha,\,
\sin\theta_\alpha\sin\phi_\alpha,\,
\cos\theta_\alpha\right),
\label{eq:discrete_direction_vector}
\end{equation}
where $\theta_\alpha$ and $\phi_\alpha$ denote the polar and azimuthal
angles, respectively. The associated quadrature weight is
\begin{equation}
w_\alpha=w_{\theta,\alpha}w_{\phi,\alpha}\sin\theta_\alpha,
\label{eq:directional_weight}
\end{equation}
where $w_{\theta,\alpha}$ and $w_{\phi,\alpha}$ are the
one-dimensional Gauss--Legendre quadrature weights mapped to the corresponding polar and azimuthal angular intervals, respectively.

Quadrature points and weights are first evaluated over
$\theta\in[0,\pi/2]$ and $\phi\in[0,\pi/2]$.
Propagation directions in the remaining angular regions are generated
by symmetry. For $N_\theta$ polar and $N_\phi$ azimuthal quadrature
points in the initial angular region, the total number of discrete
directions is
\begin{equation}
N_\alpha=8N_\theta N_\phi.
\label{eq:number_directions}
\end{equation}
The angular quadrature weights satisfy
\begin{equation}
\sum_{\alpha=1}^{N_\alpha}w_\alpha=4\pi.
\label{eq:angular_weight_normalization}
\end{equation}
The resulting discrete directions and weights are used to evaluate
angular integrals in the phonon transport equation, including the
temperature and heat-flux calculations.

The discrete equilibrium distribution is
\begin{equation}
e_b^{\mathrm{eq}}=
\frac{C_b T_L}{4\pi},
\label{eq:discrete_equilibrium_energy}
\end{equation}
where $C_b$ is the band heat capacity and
$T_L$ is the scattering-equilibrium lattice temperature determined by enforcing energy conservation in the relaxation-time approximation.

The corresponding discrete lattice temperature is
\begin{equation}
T_L=
\frac{
\displaystyle\sum_{b=1}^{N_b}
\sum_{\alpha=1}^{N_\alpha}
w_\alpha e_{b\alpha}/\tau_b
}{
\displaystyle\sum_{b=1}^{N_b}
C_b/\tau_b
},
\label{eq:discrete_lattice_temperature}
\end{equation}
while the energy-equivalent temperature is
\begin{equation}
T=
\frac{
\displaystyle\sum_{b=1}^{N_b}
\sum_{\alpha=1}^{N_\alpha}
w_\alpha e_{b\alpha}
}{
\displaystyle\sum_{b=1}^{N_b}C_b
}.
\label{eq:discrete_energy_temperature}
\end{equation}
When the relaxation time is identical for
all phonon bands, these two temperatures coincide.

The group-velocity vector associated with band \(b\) and direction \(\alpha\) is $\mathbf{v}_{b\alpha}=v_b\mathbf{s}_\alpha$, where $v_b$ is the band-dependent group-velocity magnitude.
The macroscopic heat-flux vector is evaluated as
\begin{equation}
\mathbf{q}=
\sum_{b=1}^{N_b}
\sum_{\alpha=1}^{N_\alpha}
w_\alpha\mathbf{v}_{b\alpha}e_{b\alpha}.
\label{eq:discrete_heat_flux}
\end{equation}
These band- and direction-resolved quantities
provide the basis for the subsequent spatial
finite-volume discretization.

\subsection{Finite-Volume Discretization}
\label{sec:finite_volume}

\begin{figure}
\begin{center}
\includegraphics [width=1.\columnwidth]{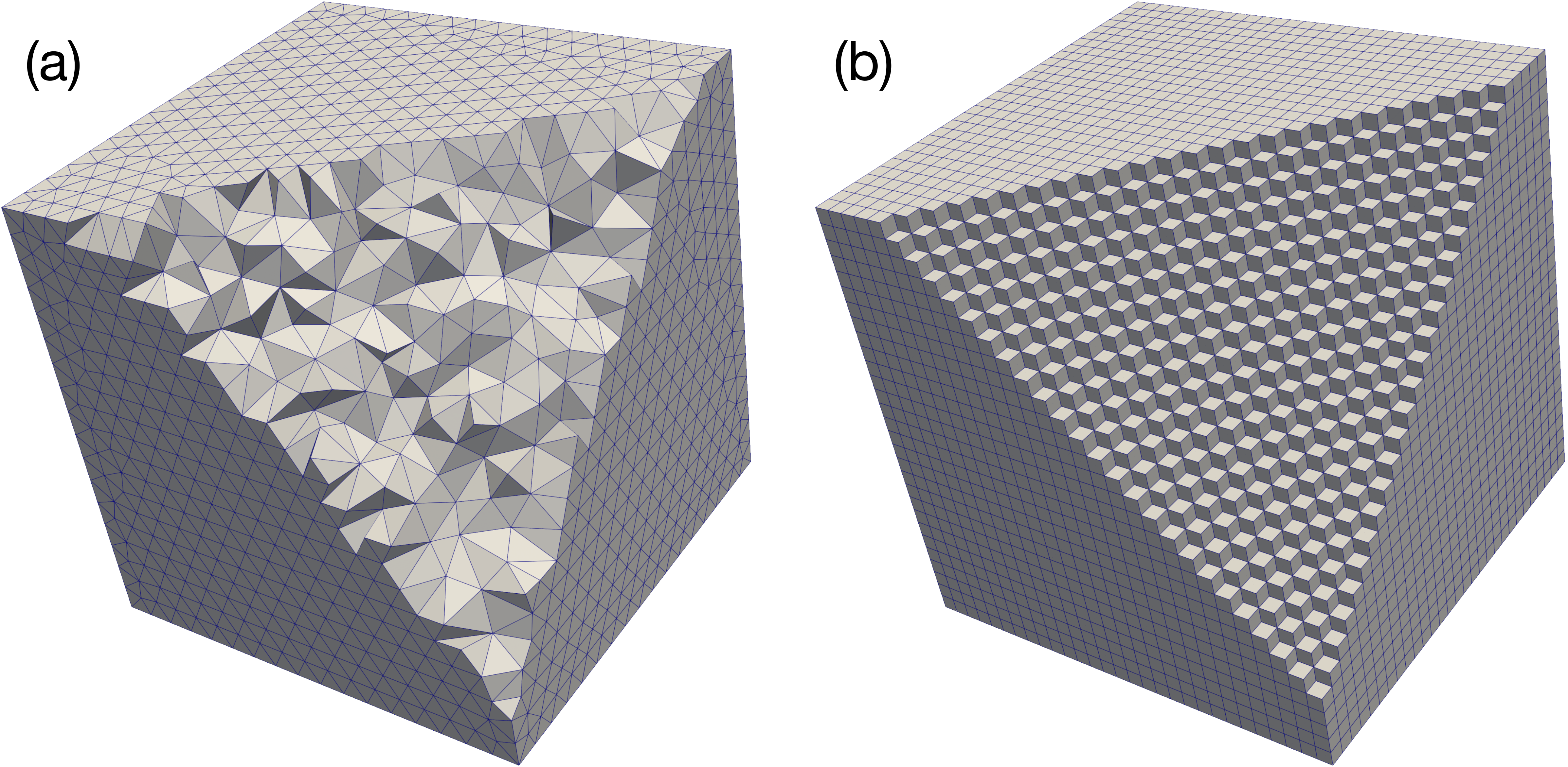}
\end{center}
\caption{Three-dimensional computational meshes employed in JAX-Nano:
(a) unstructured tetrahedral mesh and
(b) structured hexahedral mesh.
Cutaway views reveal the internal cell arrangements.
Both mesh types are supported by the same cell-centered
finite-volume formulation.}
\label{fig:meshes}
\end{figure}  

The computational domain is discretized using a cell-centered finite-volume formulation supporting both structured and unstructured meshes. Figure~\ref{fig:meshes} illustrates representative three-dimensional structured hexahedral and unstructured tetrahedral meshes, with cutaway views exposing their internal cell arrangements. Both mesh types employ the same conservative control-volume balance, while the geometric information and face connectivity are stored in a form suitable for face-based numerical flux evaluation.

Let $\mathcal{D}$ denote the computational domain, partitioned into
$N_c$ nonoverlapping control volumes $\mathcal{D}_i$, such that
$\mathcal{D}=\bigcup_{i=1}^{N_c}\mathcal{D}_i$.
Each cell $i$ has volume $V_i$, centroid $\mathbf{x}_i$, and a set
of faces $\mathcal{F}_i$. For each face $f\in\mathcal{F}_i$,
$A_{if}$, $\mathbf{n}_{if}$, and $\mathbf{x}_{if}$ denote the face
area, outward unit normal, and face centroid, respectively.
The outward-oriented face area vector is defined as
$\mathbf{S}_{if}=A_{if}\mathbf{n}_{if}$.
Interior faces are shared by adjacent control volumes, whereas
boundary faces are treated according to
Section~\ref{sec:boundary_conditions}.

Combining the band and directional discretizations
introduced in Sections~\ref{sec:band_discretization}
and~\ref{sec:directional_discretization},
the continuous phonon BTE in
Eq.~\eqref{eq:phonon_bte_general}
is expressed in terms of representative phonon
bands $b$ and discrete propagation directions
$\alpha$ as
\begin{equation}
\frac{\partial e_{b\alpha}}{\partial t}
+\mathbf{v}_{b\alpha}\cdot\nabla e_{b\alpha}
=
\frac{e_b^{\mathrm{eq}}-e_{b\alpha}}{\tau_b}
+S_b,
\label{eq:fvm_bte}
\end{equation}
where $S_b$ denotes the isotropic heat-source
contribution associated with band $b$.
The band-dependent group-velocity magnitude
$v_b$ is assumed spatially uniform within
each homogeneous material region.

Integrating Eq.~\eqref{eq:fvm_bte} over the control
volume $\mathcal{D}_i$ and applying the divergence theorem
yields the semi-discrete finite-volume balance
\begin{equation}
V_i\frac{d e_{ib\alpha}}{dt}
+\sum_{f\in\mathcal{F}_i}F_{ifb\alpha}
=V_i\left(
\frac{e_{ib}^{\mathrm{eq}}-e_{ib\alpha}}{\tau_b}
+S_{ib}
\right),
\label{eq:fvm_balance}
\end{equation}
where $V_i$ is the volume of control-volume $\mathcal{D}_i$,
$\mathcal{F}_i$ denotes its set of faces, and $e_{ib\alpha}$ is the
cell-averaged directional energy distribution.
The oriented numerical flux through face $f$ is
defined as
\begin{equation}
F_{ifb\alpha}
=v_b\left(\mathbf{s}_{\alpha}\cdot\mathbf{S}_{if}\right)
e_{ifb\alpha},
\label{eq:fvm_flux}
\end{equation}
where $\mathbf{S}_{if}$ is the outward-oriented face
area vector and $e_{ifb\alpha}$ is the reconstructed
face energy distribution. The evaluation of
$e_{ifb\alpha}$ distinguishes the central and
upwind flux discretizations described below.

To account for localized heat generation, JAX-Nano
prescribes a cellwise volumetric heat-generation
rate $Q_i$ according to the location of each cell
centroid within specified heat-generating regions.
Assuming isotropic heat generation, the source is
distributed among phonon bands using spectral
fractions $\eta_b$, giving
\begin{equation}
S_{ib}=\frac{\eta_b Q_i}{4\pi},
\qquad \sum_b\eta_b=1.
\label{eq:fvm_heat_source_distribution}
\end{equation}
Here, $\eta_b$ is the normalized heat-capacity fraction
of phonon band $b$, defined as
\begin{equation}
\eta_b=\frac{C_b}{\sum_{b'}C_{b'}},
\label{eq:fvm_heat_capacity_fraction}
\end{equation}
where $C_b$ denotes the heat capacity of phonon band $b$.
For angular quadrature weights $w_\alpha$ satisfying
$\sum_\alpha w_\alpha=4\pi$, this allocation ensures
that the prescribed volumetric heat-generation rate
is recovered:
\begin{equation}
\sum_b\sum_\alpha w_\alpha S_{ib}=Q_i.
\label{eq:fvm_heat_source}
\end{equation}
For source-free simulations, $Q_i=0$.

\paragraph{Corrected Green--Gauss gradient}
Cell-centered gradients are evaluated using a corrected
Green--Gauss approximation~\cite{Moukalled_2016a} based on face contributions.
For an interior face $f$ shared by neighboring cells
$i$ and $j$, the reconstructed face distribution is
\begin{equation}
e_{ifb\alpha}^{\mathrm{GG}}
=\frac{e_{ib\alpha}+e_{jb\alpha}}{2}
+\frac{\nabla e_{ib\alpha}+\nabla e_{jb\alpha}}{2}
\cdot\left(
\mathbf{x}_{if}
-\frac{\mathbf{x}_i+\mathbf{x}_j}{2}
\right),
\label{eq:fvm_gg_face}
\end{equation}
where $e_{ib\alpha}$ and $e_{jb\alpha}$ are the
cell-averaged directional energy distributions
in cells $i$ and $j$, respectively,
$\mathbf{x}_i$ and $\mathbf{x}_j$ are their
cell centroids, and $\mathbf{x}_{if}$ is the
centroid of the shared face. The quantities
$\nabla e_{ib\alpha}$ and $\nabla e_{jb\alpha}$
denote the available cell-centered gradient
estimates. The second term corrects for the
displacement of the face centroid from the
midpoint between neighboring cell centroids,
which is particularly relevant for skewed
unstructured meshes.
The updated cell-centered gradient is assembled
from interior- and boundary-face contributions:
\begin{equation}
(\nabla e_{ib\alpha})^{\mathrm{new}}
=\frac{1}{V_i}\left(
\sum_{f\in\mathcal{F}_i^{\mathrm{int}}}
e_{ifb\alpha}^{\mathrm{GG}}\mathbf{S}_{if}
+
\sum_{f\in\mathcal{F}_i^{\mathrm{bnd}}}
e_{ifb\alpha}^{\mathrm{BC}}\mathbf{S}_{if}
\right),
\label{eq:fvm_gg_gradient}
\end{equation}
where $\mathcal{F}_i^{\mathrm{int}}$ and
$\mathcal{F}_i^{\mathrm{bnd}}$ denote the sets
of interior and boundary faces of cell $i$,
respectively. The quantity
$e_{ifb\alpha}^{\mathrm{BC}}$ represents the
boundary-face distribution determined by the
prescribed boundary condition, accounting for
outgoing and incoming phonon propagation directions.
For interior cells, only interior-face contributions
are present, whereas boundary-adjacent cells
incorporate contributions from both interior
and boundary faces. Since the reconstructed
face distributions depend on the available
cell-centered gradients, the procedure provides
a geometrically corrected Green--Gauss gradient
update.

\paragraph{Central numerical flux}
For structured meshes, the directional energy distributions
are reconstructed independently from the neighboring cell
centers to the centroid of their shared interior face using
the cell-centered gradients obtained from the corrected Green--Gauss procedure:
\begin{equation}
\begin{aligned}
e_{i\to f,b\alpha}
&=e_{ib\alpha}
+\nabla e_{ib\alpha}\cdot
(\mathbf{x}_{if}-\mathbf{x}_i),\\
e_{j\to f,b\alpha}
&=e_{jb\alpha}
+\nabla e_{jb\alpha}\cdot
(\mathbf{x}_{if}-\mathbf{x}_j).
\end{aligned}
\label{eq:fvm_central_reconstruction}
\end{equation}
Here, $e_{i\to f,b\alpha}$ and
$e_{j\to f,b\alpha}$ denote the directional energy
distributions reconstructed at face $f$ from cells
$i$ and $j$, respectively. The numerical face
distribution is obtained by arithmetic averaging,
and the corresponding flux is
\begin{equation}
e_{ifb\alpha}^{\mathrm{cen}}
=\frac{e_{i\to f,b\alpha}+e_{j\to f,b\alpha}}{2},
\qquad
F_{ifb\alpha}^{\mathrm{cen}}
=(\mathbf{v}_{b\alpha}\cdot\mathbf{S}_{if})
e_{ifb\alpha}^{\mathrm{cen}}.
\label{eq:fvm_central_flux}
\end{equation}
This reconstructed central flux is implemented
for structured meshes without applying a
Barth--Jespersen-type limiter~\cite{Barth_1989a}.

\paragraph{Second-order upwind flux with Barth--Jespersen-type limiting}
For both structured and unstructured meshes, JAX-Nano
employs a second-order upwind flux based on piecewise-linear
reconstruction with Barth--Jespersen-type limiting
\cite{Barth_1989a}. For each cell, phonon band,
and propagation direction, local bounds are determined
from the cell itself and its valid interior neighbors
$\mathcal{N}(i)$:
\begin{equation}
\begin{aligned}
e_{ib\alpha}^{\min}
&=\min_{k\in\{i\}\cup\mathcal{N}(i)}e_{kb\alpha},\\
e_{ib\alpha}^{\max}
&=\max_{k\in\{i\}\cup\mathcal{N}(i)}e_{kb\alpha}.
\end{aligned}
\label{eq:fvm_bj_bounds}
\end{equation}
For an interior face $f$, the unlimited reconstruction
increment is defined as
$\Delta_{ifb\alpha}
=\nabla e_{ib\alpha}\cdot(\mathbf{x}_{if}-\mathbf{x}_i)$.
The facewise Barth--Jespersen-type limiter coefficient
is then evaluated as
\begin{equation}
\phi_{ifb\alpha}=
\begin{cases}
\min\!\left(1,
\dfrac{e_{ib\alpha}^{\max}-e_{ib\alpha}}
{\Delta_{ifb\alpha}}\right),
&\Delta_{ifb\alpha}>0,\\[6pt]
\min\!\left(1,
\dfrac{e_{ib\alpha}^{\min}-e_{ib\alpha}}
{\Delta_{ifb\alpha}}\right),
&\Delta_{ifb\alpha}<0,\\[6pt]
1,&\Delta_{ifb\alpha}=0.
\end{cases}
\label{eq:fvm_bj_face}
\end{equation}
The cellwise limiter is obtained by taking the minimum
over all valid interior faces:
\begin{equation}
\phi_{ib\alpha}
=\min_{f\in\mathcal{F}_i^{\mathrm{int}}}
\phi_{ifb\alpha},
\qquad
0\leq\phi_{ib\alpha}\leq1.
\label{eq:fvm_bj_cell}
\end{equation}
The limited gradients are used for face reconstruction
on both structured and unstructured meshes. The same
limited gradients are also employed to reconstruct
outgoing distributions at boundary faces, although
boundary faces are excluded from the limiter calculation.
For an interior face shared by cells $i$ and $j$,
the reconstructed distributions are
\begin{equation}
\begin{aligned}
e_{i\to f,b\alpha}^{\mathrm{lim}}
&=e_{ib\alpha}
+\phi_{ib\alpha}\nabla e_{ib\alpha}\cdot
(\mathbf{x}_{if}-\mathbf{x}_i),\\
e_{j\to f,b\alpha}^{\mathrm{lim}}
&=e_{jb\alpha}
+\phi_{jb\alpha}\nabla e_{jb\alpha}\cdot
(\mathbf{x}_{if}-\mathbf{x}_j).
\end{aligned}
\label{eq:fvm_limited_reconstruction}
\end{equation}
The upwind face distribution is selected according
to the sign of the normal phonon group velocity:
\begin{equation}
e_{ifb\alpha}^{\mathrm{up}}=
\begin{cases}
e_{i\to f,b\alpha}^{\mathrm{lim}},
&\mathbf{v}_{b\alpha}\cdot\mathbf{S}_{if}\geq0,\\
e_{j\to f,b\alpha}^{\mathrm{lim}},
&\mathbf{v}_{b\alpha}\cdot\mathbf{S}_{if}<0.
\end{cases}
\label{eq:fvm_upwind}
\end{equation}
The corresponding numerical flux is
\begin{equation}
F_{ifb\alpha}^{\mathrm{up}}
=(\mathbf{v}_{b\alpha}\cdot\mathbf{S}_{if})
e_{ifb\alpha}^{\mathrm{up}}.
\label{eq:fvm_upwind_flux}
\end{equation}
When $\phi_{ib\alpha}=0$, the reconstruction from
cell $i$ reduces to its cell-averaged distribution,
whereas $\phi_{ib\alpha}=1$ retains the unlimited
linear reconstruction.

\paragraph{Finite-volume formulation and mesh treatment}
With either numerical flux, the semidiscrete finite-volume
formulation of the phonon BTE is expressed as
\begin{equation}
\frac{d e_{ib\alpha}}{dt}
=-\frac{1}{V_i}\sum_{f\in\mathcal{F}_i}F_{ifb\alpha}
+\frac{e_{ib}^{\mathrm{eq}}-e_{ib\alpha}}{\tau_b}
+S_{ib}.
\label{eq:fvm_semidiscrete}
\end{equation}
The same spatial discretization framework is employed
for steady and transient phonon transport. In steady
state, the time derivative vanishes, whereas transient
simulations retain the temporal derivative and employ
the selected time-integration method. Volumetric heat
generation is incorporated as a cell-local source term
rather than a face flux. Interior-face contributions
are assembled using neighboring-cell connectivity
and consistently oriented face-area vectors. At boundary
faces, the numerical flux combines outgoing reconstructed
distributions with incoming distributions determined
by the corresponding boundary conditions, as detailed
in Section~\ref{sec:boundary_conditions}.
Although the geometric representation and connectivity
data differ between structured hexahedral and
unstructured tetrahedral meshes, the underlying
finite-volume balance remains unchanged. The
second-order upwind flux with Barth--Jespersen-type
limiting is implemented for both structured and
unstructured meshes, while the reconstructed central
flux is implemented for structured meshes only.
Consequently, comparisons with GiftBTE should be
interpreted in terms of the specific mesh and numerical
flux configurations employed in each benchmark.

\subsection{Boundary Conditions}
\label{sec:boundary_conditions}

\begin{figure}
\centering
\includegraphics[width=\columnwidth]{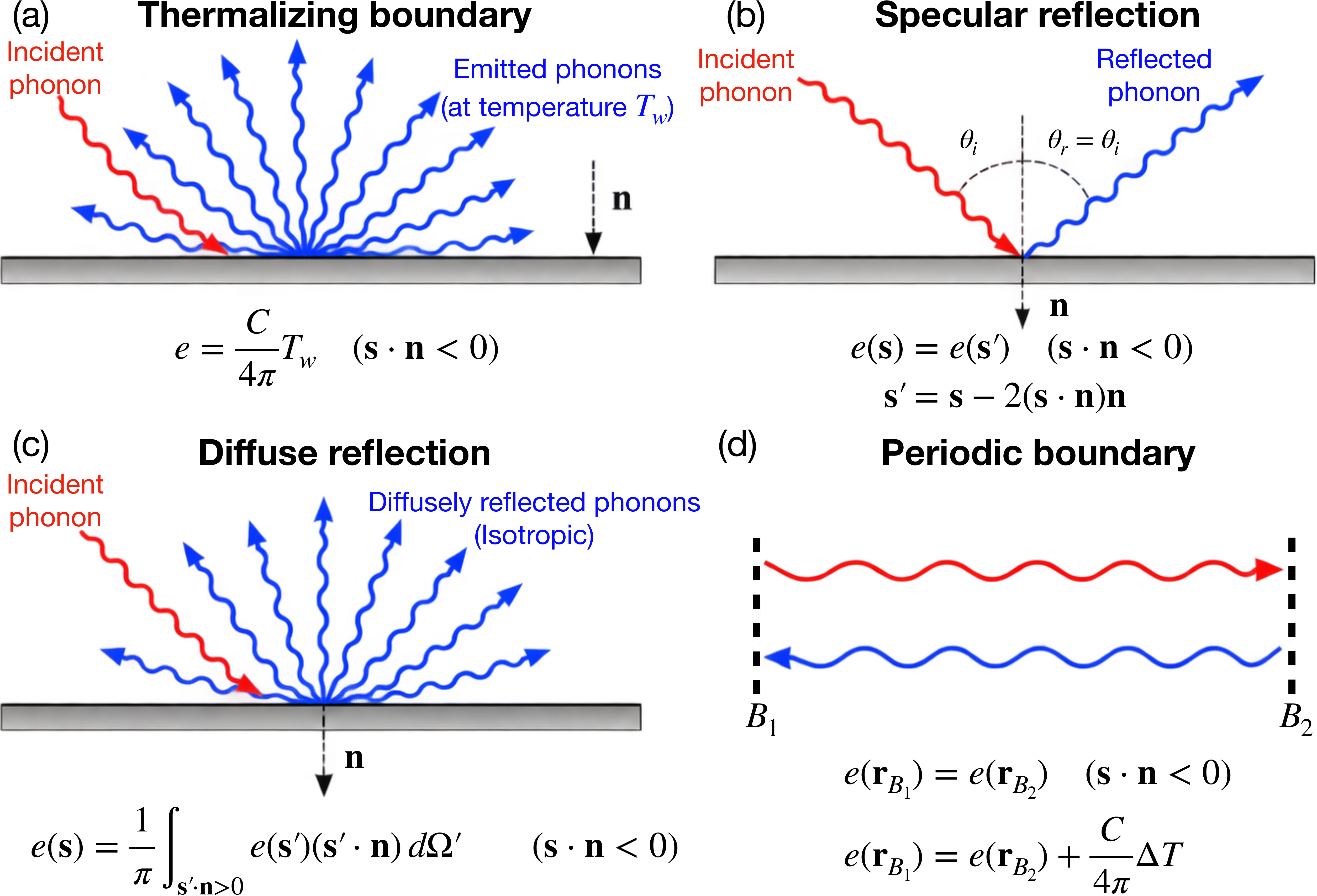}
\caption{
Schematic illustration of the phonon boundary conditions
implemented in JAX-Nano:
(a) thermalizing boundary, where incident phonons are absorbed
and equilibrium phonons are emitted isotropically at the
prescribed wall temperature $T_w$;
(b) specular reflection, where the reflection angle equals
the incidence angle;
(c) diffuse reflection, where incident phonons are
redistributed isotropically over the inward hemisphere,
resulting in zero net normal heat flux across the adiabatic
boundary; and
(d) periodic boundary, where $B_1$ and $B_2$ denote two
corresponding periodic boundaries across which phonon energy
distributions are mapped while preserving their propagation
directions, optionally incorporating a prescribed temperature
difference $\Delta T$.
Red and blue wavy arrows indicate incident and emitted or
reflected phonon propagation directions, respectively.
}
\label{fig:bte_bc}
\end{figure}

Accurate treatment of phonon-boundary interactions is essential
for modeling nanoscale heat transport, particularly when the
characteristic length scale becomes comparable to the phonon
mean free path. JAX-Nano incorporates four types of boundary
conditions: thermalizing, specular reflection, diffuse reflection,
and periodic boundaries, as schematically illustrated in
Fig.~\ref{fig:bte_bc}. These conditions account for phonon
thermalization, elastic boundary scattering, and transport
across periodic interfaces.

Let $\mathbf{n}$ denote the outward unit normal vector of
the computational domain, and let $\mathbf{s}$ represent the
unit phonon propagation direction. Accordingly,
$\mathbf{s}\cdot\mathbf{n}>0$ identifies phonons leaving the
computational domain, whereas
$\mathbf{s}\cdot\mathbf{n}<0$ identifies phonons entering
the domain from a boundary. The boundary conditions are
specified for the incoming phonon energy distribution
$e_b(\mathbf{x},\mathbf{s},t)$, where $b$ denotes the
phonon band.

\paragraph{Thermalizing boundary}
At a thermalizing boundary, phonons incident on the wall
are absorbed, and the wall emits phonons according to
its local equilibrium distribution at a prescribed
temperature $T_w$. Under the isotropic phonon approximation,
the incoming directional energy distribution is given by
\begin{equation}
e_b(\mathbf{x}_w,\mathbf{s},t)
=
\frac{C_b}{4\pi}T_w,
\qquad
\mathbf{s}\cdot\mathbf{n}<0,
\label{eq:bc_thermalizing}
\end{equation}
where $C_b$ is the volumetric heat capacity associated
with phonon band $b$, and $\mathbf{x}_w$ denotes a point
on the boundary. The emitted energy distribution is
independent of the propagation direction and therefore
is isotropic over the inward hemisphere. Unlike an
adiabatic reflecting boundary, the thermalizing boundary
can exchange energy with the computational domain,
allowing a nonzero normal heat flux.

\paragraph{Specular reflection}
At a specularly reflecting boundary, phonons undergo
mirror-like reflection, preserving their energy and
the magnitude of their propagation velocity. The
reflected direction is determined by
\begin{equation}
\mathbf{s}
=
\mathbf{s}'
-
2(\mathbf{s}'\cdot\mathbf{n})\mathbf{n},
\label{eq:bc_specular_direction}
\end{equation}
where $\mathbf{s}'$ and $\mathbf{s}$ denote the incident
and reflected directions, respectively. Consequently,
the incidence and reflection angles satisfy
$\theta_i=\theta_r$. The corresponding boundary condition
is expressed as
\begin{equation}
e_b(\mathbf{x}_w,\mathbf{s},t)
=
e_b(\mathbf{x}_w,\mathbf{s}',t),
\qquad
\mathbf{s}\cdot\mathbf{n}<0.
\label{eq:bc_specular}
\end{equation}
Specular reflection preserves the directional information
of the incident phonons and produces zero net normal
energy flux across an impermeable boundary.

\paragraph{Diffuse reflection}
At a diffusely reflecting boundary, incident phonons
are scattered into the inward hemisphere with an
isotropic directional energy distribution, independent
of their incident propagation directions. For an
adiabatic wall, the reflected distribution is determined
by enforcing conservation of energy, such that the
incident and reflected normal energy fluxes balance.
For isotropic phonon group velocity, the diffuse
boundary condition can be written as
\begin{equation}
e_b(\mathbf{x}_w,\mathbf{s},t)
=
\frac{1}{\pi}
\int_{\mathbf{s}'\cdot\mathbf{n}>0}
e_b(\mathbf{x}_w,\mathbf{s}',t)
(\mathbf{s}'\cdot\mathbf{n})
\,d\Omega',
\qquad
\mathbf{s}\cdot\mathbf{n}<0,
\label{eq:bc_diffuse}
\end{equation}
where $d\Omega'$ is the differential solid angle
associated with the incident direction. The reflected
distribution is independent of $\mathbf{s}$, while its
magnitude is determined by the incident energy flux.
Thus, diffuse reflection randomizes the propagation
directions while maintaining zero net normal heat
flux across the adiabatic boundary. Although both
thermalizing and diffuse boundaries produce isotropic
incoming energy distributions, the former is governed
by the prescribed wall temperature, whereas the latter
is determined by the incident phonon energy flux.

\paragraph{Periodic boundary}
Periodic boundary conditions connect corresponding
locations on opposite boundaries of the computational
domain, allowing phonons to propagate continuously
through a periodically repeated geometry. For two
corresponding boundary locations
$\mathbf{x}_{B_1}$ and $\mathbf{x}_{B_2}$, the
periodic condition is expressed as
\begin{equation}
e_b(\mathbf{x}_{B_1},\mathbf{s},t)
=
e_b(\mathbf{x}_{B_2},\mathbf{s},t).
\label{eq:bc_periodic}
\end{equation}
The phonon propagation direction is preserved across
the periodic interface. For a linearized energy
formulation with a prescribed temperature difference
$\Delta T$ between corresponding boundaries, a
shifted periodic condition may also be imposed:
\begin{equation}
e_b(\mathbf{x}_{B_1},\mathbf{s},t)
=
e_b(\mathbf{x}_{B_2},\mathbf{s},t)
+
\frac{C_b}{4\pi}\Delta T,
\label{eq:bc_periodic_shifted}
\end{equation}
where the sign of $\Delta T$ follows the chosen
temperature-difference convention. Such conditions
enable the modeling of periodically repeated structures
and transport driven by an imposed macroscopic
temperature gradient.

\subsection{Steady-State Solution Method}
\label{sec:steady_solver}
For steady-state phonon transport, the temporal derivative in Eq.~\eqref{eq:fvm_semidiscrete} vanishes. The resulting finite-volume equations are solved using a delta-form defect-correction approach, following the general iterative framework of GiftBTE~\cite{Hu_2024a}. At each iteration, the residual is evaluated using reconstructed face distributions, while a first-order upwind transport-relaxation operator is employed to compute corrections to the directional phonon energy densities. This separation enables higher-order spatial accuracy through flux reconstruction while retaining a sparse first-order correction operator.

For each phonon band $b$ and propagation direction $\alpha$, the correction equation is expressed as
\begin{equation}
\mathbf{A}^{b\alpha}\Delta\mathbf{e}_{b\alpha}^{(k+1)}=\mathbf{b}_{b\alpha}^{(k)},
\label{eq:steady_linear_system}
\end{equation}
where $\Delta e_{ib\alpha}^{(k+1)}=e_{ib\alpha}^{(k+1)}-e_{ib\alpha}^{(k)}$ denotes the energy-density correction in cell $i$.

Unlike the unscaled transport-relaxation formulation, JAX-Nano employs a correction operator scaled by the band-dependent relaxation time $\tau_b$. The diagonal matrix coefficient for cell $i$ is
\begin{equation}
A_{ii}^{b\alpha}=1+\frac{\tau_b}{V_i}\sum_{f\in\mathcal{F}_i}(\mathbf{v}_{b\alpha}\cdot\mathbf{S}_{if})\Theta(\mathbf{v}_{b\alpha}\cdot\mathbf{S}_{if}),
\label{eq:steady_matrix_diagonal}
\end{equation}
while the off-diagonal coefficient associated with an interior face shared by cells $i$ and $j$ is
\begin{equation}
A_{ij}^{b\alpha}=\frac{\tau_b}{V_i}(\mathbf{v}_{b\alpha}\cdot\mathbf{S}_{if})\Theta(-\mathbf{v}_{b\alpha}\cdot\mathbf{S}_{if}).
\label{eq:steady_matrix_offdiagonal}
\end{equation}
Here, $\Theta$ denotes the Heaviside function, and $\mathcal F_i$ includes both interior and boundary faces of cell $i$. 
The diagonal coefficient accounts for local relaxation and outgoing transport, whereas the off-diagonal coefficient represents incoming transport from neighboring cells. Contributions from multiple faces connecting the same cell pair are accumulated. The upwind selection is determined by the sign of the normal phonon group velocity.

The right-hand-side vector is constructed from the residual of the reconstructed finite-volume equations,
\begin{equation}
b_{ib\alpha}^{(k)}=-\frac{\tau_b}{V_i}\sum_{f\in\mathcal{F}_i}F_{ifb\alpha}^{(k)}-e_{ib\alpha}^{(k)}+e_{ib}^{\mathrm{eq},(k)}+\tau_b S_{ib},
\label{eq:steady_rhs}
\end{equation}
where $F_{ifb\alpha}^{(k)}$ denotes the numerical face flux and $S_{ib}$ represents the volumetric source term for band $b$. The numerical fluxes are evaluated using the reconstructed face distributions described in Section~\ref{sec:finite_volume}. In contrast to the least-squares reconstruction adopted in GiftBTE, JAX-Nano employs corrected Green--Gauss gradients. The first-order upwind approximation is restricted to the correction operator, while the reconstructed fluxes determine the residual of the higher-order spatial discretization. Boundary-face fluxes are included in the residual, whereas incoming boundary corrections are treated explicitly in the linearization.

The correction systems associated with different phonon bands and propagation directions are assembled into a block-diagonal sparse matrix,
\begin{equation}
\mathbf{A}=\operatorname{diag}\left(\mathbf{A}^{11},\ldots,\mathbf{A}^{b\alpha},\ldots\right).
\label{eq:steady_block_matrix}
\end{equation}
The resulting global correction system is solved using the biconjugate gradient stabilized (BiCGSTAB) method with Jacobi preconditioning,
\begin{equation}
\mathbf{M}^{-1}=\operatorname{diag}(\mathbf{A})^{-1}.
\label{eq:steady_preconditioner}
\end{equation}
Unlike the sparse direct LU factorization employed in GiftBTE, JAX-Nano uses an iterative linear solver based on sparse matrix--vector products. These products are implemented in JAX through indexed coefficient multiplication and scatter-add accumulation, without constructing dense matrices. The sparse correction matrix and Jacobi preconditioner are assembled once and reused throughout the outer steady-state iterations.

Following each linear solve, the directional energy densities are updated according to
\begin{equation}
e_{ib\alpha}^{(k+1)}=e_{ib\alpha}^{(k)}+\Delta e_{ib\alpha}^{(k+1)}.
\label{eq:steady_energy_update}
\end{equation}
The equilibrium energy densities are then updated by enforcing energy conservation of the phonon scattering operator. Subsequently, the macroscopic temperature and heat flux are recovered from the updated directional energy distributions.

Convergence is assessed using the normalized root-mean-square changes in temperature and heat-flux magnitude between successive iterations,
\begin{equation}
\varepsilon_T^{(k+1)}=\frac{\sqrt{\frac{1}{N_c}\sum_{i=1}^{N_c}\left(T_i^{(k+1)}-T_i^{(k)}\right)^2}}{\max_i |T_i^{(k+1)}|},
\label{eq:steady_error_temperature}
\end{equation}
and
\begin{equation}
\varepsilon_q^{(k+1)}=\frac{\sqrt{\frac{1}{N_c}\sum_{i=1}^{N_c}\left(\|\mathbf{q}_i^{(k+1)}\|-\|\mathbf{q}_i^{(k)}\|\right)^2}}{\max_i\|\mathbf{q}_i^{(k+1)}\|},
\label{eq:steady_error_heatflux}
\end{equation}
where $N_c$ denotes the total number of computational cells. The iterative solution is considered converged when both errors satisfy their prescribed tolerances,
\begin{equation}
\varepsilon_T^{(k+1)}\leq\varepsilon_T^{\mathrm{tol}},\qquad\varepsilon_q^{(k+1)}\leq\varepsilon_q^{\mathrm{tol}}.
\label{eq:steady_convergence_criteria}
\end{equation}

To enable JAX-native execution, the outer steady-state iteration is implemented using \texttt{jax.lax.scan}, which maintains the evolving phonon energy distributions, macroscopic fields, and convergence state over a fixed maximum number of iterations. Within each scan step, \texttt{jax.lax.cond} conditionally executes the defect-correction update or preserves the current solution once both convergence criteria are satisfied. Although the scan retains a fixed iteration structure, conditional execution bypasses subsequent gradient reconstruction, flux evaluation, and sparse linear solves after convergence. The complete computational procedure is summarized in Algorithm~\ref{alg:steady_solver}, where $K_{\max}$ denotes the maximum
number of outer iterations, $c^k$ is a Boolean convergence
flag, $\mathcal{C}^k$ represents the iteration state
containing the directional energy densities, equilibrium
distributions, temperature, heat flux, and convergence
flag, and $\mathcal{H}$ denotes the recorded convergence
history.

\begin{algorithm}
\caption{Steady-state phonon BTE solver in JAX-Nano.}
\label{alg:steady_solver}
\begin{algorithmic}[1]
\Require Phonon properties, mesh, angular quadrature, boundary conditions,
maximum iteration count $K_{\max}$, and tolerances
$\varepsilon_T^{\mathrm{tol}}$, $\varepsilon_q^{\mathrm{tol}}$
\State Initialize $\mathbf e^0$, $\mathbf e_{\mathrm{eq}}^0$, $\mathbf T^0$, $\mathbf q^0$,
and $c^0=\mathrm{false}$
\State Assemble sparse upwind correction matrix $\mathbf A$ and Jacobi
preconditioner $\mathbf M^{-1}$
\State Define $\mathcal C^0=(\mathbf e^0,\mathbf e_{\mathrm{eq}}^0,
\mathbf T^0,\mathbf q^0,c^0)$
\Function{ActiveStep}{$\mathcal C^k$}
  \State $\nabla\mathbf e^k\gets$ corrected Green--Gauss reconstruction
  \State $\mathbf F^k\gets$ reconstructed interior and boundary face-flux sums
  \State $\mathbf r^k\gets
  -(\boldsymbol\tau/\mathbf V)\odot\mathbf F^k
  -(\mathbf e^k-\mathbf e_{\mathrm{eq}}^k) +\boldsymbol{\tau}\odot\mathbf S^k$
  \State $\Delta\mathbf e^{k+1}\gets
  \operatorname{BiCGSTAB}(\mathbf A,\mathbf r^k,\mathbf M^{-1})$
  \State $\mathbf e^{k+1}\gets\mathbf e^k+\Delta\mathbf e^{k+1}$
  \State Update $\mathbf e_{\mathrm{eq}}^{k+1}$, $\mathbf T^{k+1}$,
  and $\mathbf q^{k+1}$
  \State Evaluate $\varepsilon_T^{k+1}$ and $\varepsilon_q^{k+1}$
  \State $c^{k+1}\gets
  (\varepsilon_T^{k+1}\leq\varepsilon_T^{\mathrm{tol}})\land
  (\varepsilon_q^{k+1}\leq\varepsilon_q^{\mathrm{tol}})$
  \State \Return $\mathcal C^{k+1}$ and
  $(\varepsilon_T^{k+1},\varepsilon_q^{k+1})$
\EndFunction
\Function{InactiveStep}{$\mathcal C^k$}
  \State \Return unchanged $\mathcal C^k$ and inactive diagnostics
\EndFunction
\Function{ScanStep}{$\mathcal C^k,k$}
  \State \Return $\texttt{lax.cond}(c^k,\textsc{InactiveStep},
  \textsc{ActiveStep},\mathcal C^k)$
\EndFunction
\State $(\mathcal C^{K_{\max}},\mathcal H)\gets
\texttt{lax.scan}(\textsc{ScanStep},\mathcal C^0,K_{\max})$
\State \Return $\mathbf e$, $\mathbf T$, $\mathbf q$, and convergence history $\mathcal H$
\end{algorithmic}
\end{algorithm}

\subsection{Transient Solution Methods}
To simulate time-dependent phonon transport, JAX-Nano incorporates both explicit and implicit time-integration schemes within the finite-volume framework. The explicit formulation employs a first-order forward-Euler method, in which the directional phonon energy distributions are advanced directly using the transport and relaxation terms evaluated at the current time level. In contrast, the implicit formulation adopts a first-order backward-Euler method, with the resulting nonlinear algebraic system solved through defect-correction iterations and a sparse iterative linear solver. Both approaches employ the same spatial discretization and boundary-condition treatments, providing complementary strategies for transient phonon transport simulations with different stability and computational requirements. The numerical formulations of these methods are presented in the following subsections.

\subsubsection{Explicit Time Integration}
\label{sec:transient_explicit}
The semi-discrete transient phonon BTE derived in the preceding section is integrated in time using the first-order forward-Euler scheme. The temporal derivative of the directional energy density is approximated as
\begin{equation}
\label{eq:transient_explicit_derivative}
\left.
\frac{\mathrm{d}e_{ib\alpha}}{\mathrm{d}t}
\right|_{t=t^n}
\approx
\frac{e_{ib\alpha}^{n+1}-e_{ib\alpha}^{n}}
{\Delta t},
\end{equation}
where $\Delta t$ denotes the time-step size and the superscripts $n$ and $n+1$ represent consecutive time levels. Substitution into the semi-discrete finite-volume equation yields
\begin{equation}
\label{eq:transient_explicit_update}
\begin{aligned}
e_{ib\alpha}^{n+1}
={}&
e_{ib\alpha}^{n}
-
\frac{\Delta t}{V_i}
\sum_{f\in\mathcal{F}_i}
F_{ifb\alpha}^{n}
&-
\frac{\Delta t}{\tau_b}
\left(
e_{ib\alpha}^{n}
-
e_{ib}^{\mathrm{eq},n}
\right)
+
\Delta t\,S_{ib}^{n}.
\end{aligned}
\end{equation}

At each time step, the directional energy distributions are reconstructed at the cell faces using the spatial discretization and boundary-condition treatments described previously. The resulting numerical fluxes, together with the relaxation and prescribed source terms, are evaluated at the current time level and used in Eq.~\eqref{eq:transient_explicit_update} to advance the directional energy densities. The equilibrium energy distributions and macroscopic quantities, including temperature and heat flux, are subsequently updated. This procedure is repeated until the prescribed final simulation time is reached, without requiring nonlinear iterations or linear system solves.

The explicit time step is selected using a Courant--Friedrichs--Lewy (CFL) criterion~\cite{Tannehill_1997a},
\begin{equation}
\label{eq:transient_explicit_stability}
\Delta t
=
C_{\mathrm{CFL}}
\frac{\Delta x_{\min}}{v_{\max}},
\end{equation}
where $C_{\mathrm{CFL}}$ is the prescribed CFL number, $\Delta x_{\min}$ is the minimum characteristic mesh spacing, and $v_{\max}$ is the maximum phonon group velocity among all bands. The CFL number is selected according to the spatial resolution and stability requirements of each simulation. Because phonon relaxation is also treated explicitly, the time-step size must additionally provide adequate resolution of the relevant relaxation timescales.

The explicit transient solution procedure is summarized in Algorithm~\ref{alg:explicit_solver}. The time-stepping loop can be implemented using \texttt{jax.lax.scan}, enabling JAX compilation and automatic differentiation through the temporal evolution. For simulations seeking a steady-state solution, \texttt{jax.lax.cond} can additionally be employed to bypass subsequent updates once prescribed convergence criteria are satisfied.

Although the explicit method avoids linear system solves, its stability restrictions may necessitate a large number of time steps for fine spatial discretizations or short phonon relaxation times. To alleviate these limitations, a fully implicit time-integration method is introduced in the following subsection.

\begin{algorithm}
\caption{Explicit transient phonon BTE solver in JAX-Nano.}
\label{alg:explicit_solver}
\begin{algorithmic}[1]
\Require Phonon properties, mesh, angular quadrature, boundary conditions,
final time $t_{\mathrm{end}}$, and prescribed CFL number $C_{\mathrm{CFL}}$
\State Initialize $\mathbf e^0$, $\mathbf e_{\mathrm{eq}}^0$,
$\mathbf T^0$, $\mathbf q^0$, and $t^0=0$
\State Set $\Delta t=C_{\mathrm{CFL}}\Delta x_{\min}/v_{\max}$
\State Set $N_t=\lceil t_{\mathrm{end}}/\Delta t\rceil$
\State Define $\mathcal C^0=(\mathbf e^0,\mathbf e_{\mathrm{eq}}^0,
\mathbf T^0,\mathbf q^0,t^0)$
\Function{TimeStep}{$\mathcal C^n$}
  \State $\Delta t_n\gets\min(\Delta t,t_{\mathrm{end}}-t^n)$
  \State $\nabla\mathbf e^n\gets$ corrected Green--Gauss reconstruction
  \State $\mathbf F^n\gets$ reconstructed interior and boundary face-flux sums
  \State $\mathbf R^n\gets -\mathbf F^n/\mathbf V
  -(\mathbf e^n-\mathbf e_{\mathrm{eq}}^n)/\boldsymbol\tau+\mathbf S^n$
  \State $\mathbf e^{n+1}\gets\mathbf e^n+\Delta t_n\mathbf R^n$
  \State Update $\mathbf e_{\mathrm{eq}}^{n+1}$, $\mathbf T^{n+1}$,
  and $\mathbf q^{n+1}$
  \State $t^{n+1}\gets t^n+\Delta t_n$
  \State \Return $\mathcal C^{n+1}$
\EndFunction
\Function{ScanStep}{$\mathcal C^n,n$}
  \State $\mathcal C^{n+1}\gets\textsc{TimeStep}(\mathcal C^n)$
  \State \Return $\mathcal C^{n+1}$ and selected diagnostics
\EndFunction
\State $(\mathcal C^{N_t},\mathcal H)\gets
\texttt{lax.scan}(\textsc{ScanStep},\mathcal C^0,N_t)$
\State \Return $\mathbf e$, $\mathbf T$, $\mathbf q$, and transient history $\mathcal H$
\end{algorithmic}
\end{algorithm}

\subsubsection{Implicit Time Integration}
\label{sec:transient_implicit}
To alleviate the restrictive time-step requirements of explicit
integration, JAX-Nano also employs a fully implicit backward-Euler
scheme for transient phonon transport. In this formulation, the
transport, relaxation, and source terms are evaluated at the new
time level $t^{n+1}$, yielding the discretized phonon BTE
\begin{equation}
\label{eq:transient_implicit_update}
\begin{aligned}
\frac{e_{ib\alpha}^{n+1}-e_{ib\alpha}^{n}}{\Delta t}
={}&-\frac{1}{V_i}
\sum_{f\in\mathcal F_i}F_{ifb\alpha}^{n+1}
&-\frac{e_{ib\alpha}^{n+1}
-e_{ib}^{\mathrm{eq},n+1}}{\tau_b}
+S_{ib}^{n+1}.
\end{aligned}
\end{equation}
Because the reconstructed numerical fluxes and local equilibrium
distributions depend on the unknown solution at time level $n+1$,
Eq.~\eqref{eq:transient_implicit_update} defines a nonlinear system.

Following the steady-state solution strategy, this system is solved by nonlinear defect-correction iterations. At each physical time step, the initial iterate is the solution at the preceding time level, $e_{ib\alpha}^{(0)}=e_{ib\alpha}^{n}$. The relaxation-time-scaled residual at nonlinear iteration $k$ is
\begin{equation}
\label{eq:transient_implicit_residual}
\begin{aligned}
r_{ib\alpha}^{(k)}
={}&-\frac{\tau_b}{\Delta t}\left(e_{ib\alpha}^{(k)}-e_{ib\alpha}^{n}\right)
-\frac{\tau_b}{V_i}\sum_{f\in\mathcal F_i}F_{ifb\alpha}^{(k)}
&-\left(e_{ib\alpha}^{(k)}-e_{ib}^{\mathrm{eq},(k)}\right)
+\tau_b S_{ib}^{(k)}.
\end{aligned}
\end{equation}
Here, the superscript $(k)$ denotes a nonlinear iteration within the current physical time step. The residual vanishes when the backward-Euler equations are satisfied. The source term is included for simulations with internal heating and is set to zero otherwise.

The directional energy distribution is updated by solving
\begin{equation}
\label{eq:transient_implicit_correction}
\mathbf A_{\mathrm{tr}}^{b\alpha}\Delta\mathbf e_{b\alpha}^{(k+1)}
=\mathbf r_{b\alpha}^{(k)},
\qquad
\mathbf e_{b\alpha}^{(k+1)}
=\mathbf e_{b\alpha}^{(k)}+\Delta\mathbf e_{b\alpha}^{(k+1)},
\end{equation}
where $\mathbf A_{\mathrm{tr}}^{b\alpha}$ is the sparse correction matrix for phonon band $b$ and propagation direction $\alpha$. As in the steady-state formulation, the correction matrix uses a first-order upwind approximation, whereas the residual retains the reconstructed higher-order face fluxes. Backward-Euler integration adds a diagonal contribution,
\begin{equation}
\label{eq:transient_implicit_matrix}
\mathbf A_{\mathrm{tr}}^{b\alpha}
=\mathbf A_{\mathrm{steady}}^{b\alpha}
+\frac{\tau_b}{\Delta t}\mathbf I,
\end{equation}
where $\mathbf A_{\mathrm{steady}}^{b\alpha}$ is the corresponding steady-state upwind correction matrix and $\mathbf I$ is the identity matrix. For fixed mesh geometry, phonon properties, and time-step size, this matrix can be reused across nonlinear iterations and physical time steps.

The sparse correction systems are solved using the biconjugate gradient stabilized (BiCGSTAB) method with Jacobi preconditioning. JAX-compatible sparse matrix--vector operations enable integration of these solves into the GPU-accelerated framework. At each nonlinear iteration, the reconstructed gradient and face fluxes are evaluated, and the local equilibrium energy distribution is recomputed from the current iterate. Convergence is assessed using the relative correction norm,
\begin{equation}
\label{eq:transient_implicit_convergence}
\varepsilon_{\mathrm{NL}}^{(k+1)}
=\frac{\left\|\Delta\mathbf e^{(k+1)}\right\|_2}
{\left\|\mathbf e^{(k+1)}\right\|_2+10^{-30}}.
\end{equation}
The nonlinear corrections continue until this error falls below the prescribed tolerance or the maximum number of iterations is reached. In the nested-scan implementation, \texttt{jax.lax.scan} organizes the physical time steps and the bounded nonlinear iterations, while \texttt{jax.lax.cond} skips further correction solves once convergence is detected. After the final correction at each time step, the equilibrium distribution and reconstructed gradient are recomputed for the accepted state, and its temperature and heat flux are evaluated.

Algorithm~\ref{alg:implicit_solver} summarizes the implementation
of the implicit transient solver. Here, $\mathcal C$ denotes the
state carried between iterations, while $N_t$ and $K_{\max}$
specify the number of physical time steps and the maximum number
of nonlinear corrections per time step, respectively.
The nonlinear convergence tolerance is denoted by
$\varepsilon_{\mathrm{NL}}^{\mathrm{tol}}$, and $\mathbf M^{-1}$
represents the Jacobi preconditioner. The transient history
$\mathcal H$ stores the selected time-dependent outputs.

\begin{algorithm}
\caption{Implicit transient phonon BTE solver in JAX-Nano.}
\label{alg:implicit_solver}
\begin{algorithmic}[1]
\Require Phonon properties, mesh, angular quadrature, boundary conditions, initial state, $\Delta t$, $N_t$, $K_{\max}$, and $\varepsilon_{\mathrm{NL}}^{\mathrm{tol}}$
\State Initialize $\mathbf e^0$, $\mathbf e_{\mathrm{eq}}^0$, $\mathbf T^0$, and $\mathbf q^0$
\State Assemble sparse steady-state upwind correction matrix $\mathbf A_{\mathrm{steady}}$
\State $\mathbf A_{\mathrm{tr}}\gets\mathbf A_{\mathrm{steady}}+\operatorname{diag}(\boldsymbol\tau/\Delta t)$; construct Jacobi preconditioner $\mathbf M^{-1}$
\Function{CorrectionStep}{$\mathcal C^{(k)},\mathbf e^n$}
  \State $\nabla\mathbf e^{(k)}\gets$ corrected Green--Gauss reconstruction
  \State $\mathbf F^{(k)}\gets$ reconstructed interior and boundary face-flux sums
  \State Update equilibrium energy $\mathbf e_{\mathrm{eq}}^{(k)}$
  \State $\mathbf r^{(k)}\gets -(\boldsymbol\tau/\Delta t)\odot(\mathbf e^{(k)}-\mathbf e^n)-(\boldsymbol\tau/\mathbf V)\odot\mathbf F^{(k)}-(\mathbf e^{(k)}-\mathbf e_{\mathrm{eq}}^{(k)})+\boldsymbol\tau\odot\mathbf S^{(k)}$
  \State $\Delta\mathbf e^{(k+1)}\gets\operatorname{BiCGSTAB}(\mathbf A_{\mathrm{tr}},\mathbf r^{(k)},\mathbf M^{-1})$
  \State $\mathbf e^{(k+1)}\gets\mathbf e^{(k)}+\Delta\mathbf e^{(k+1)}$
  \State Evaluate correction error $\varepsilon_{\mathrm{NL}}^{(k+1)}$ using Eq.~\eqref{eq:transient_implicit_convergence}
  \State \Return updated $\mathcal C^{(k+1)}$
\EndFunction
\Function{TimeStep}{$\mathcal C^n$}
  \State Initialize $\mathbf e^{(0)}\gets\mathbf e^n$ and nonlinear state $\mathcal C^{(0)}$
  \State Use \texttt{lax.cond} to bypass corrections after $\varepsilon_{\mathrm{NL}}<\varepsilon_{\mathrm{NL}}^{\mathrm{tol}}$
  \State $\mathcal C^{(K_{\max})}\gets\texttt{lax.scan}(\textsc{CorrectionStep},\mathcal C^{(0)},K_{\max})$
  \State Recompute $\mathbf e_{\mathrm{eq}}^{n+1}$, $\nabla\mathbf e^{n+1}$, $\mathbf T^{n+1}$, and $\mathbf q^{n+1}$
  \State Record selected time-step outputs in $\mathcal H$
  \State \Return $\mathcal C^{n+1}$ and time-step history output
\EndFunction
\State $(\mathcal C^{N_t},\mathcal H)\gets\texttt{lax.scan}(\textsc{TimeStep},\mathcal C^0,N_t)$
\State \Return $\mathbf e$, $\mathbf T$, $\mathbf q$, and $\mathcal H$
\end{algorithmic}
\end{algorithm}

Compared with forward-Euler integration, backward-Euler integration avoids the same explicit advection and relaxation stability restrictions and therefore permits larger time steps. Nevertheless, $\Delta t$ must still adequately resolve the physical transient dynamics and maintain temporal accuracy. The implicit method is particularly useful for fine spatial discretizations, short phonon relaxation times, and simulations extending over long physical timescales.

\subsection{GPU-Accelerated Implementation in JAX}
\label{sec:gpu_jax}

JAX-Nano is implemented using JAX, a high-performance numerical
computing framework that supports just-in-time (JIT) compilation,
automatic differentiation, and execution on hardware accelerators.
The finite-volume discretization, phonon transport calculations,
and iterative solution procedures are expressed using
JAX-compatible array operations, enabling computationally
intensive kernels to execute on GPUs.

The evaluation of reconstructed gradients, numerical face fluxes,
phonon scattering terms, and macroscopic quantities is formulated
using vectorized operations. JAX's \texttt{jit} transformation
compiles these operations through the Accelerated Linear Algebra (XLA) compiler, reducing
Python execution overhead and enabling optimization of
computational kernels. The steady-state and transient solvers
employ JAX-compatible iterative procedures, including sparse
matrix--vector operations and Krylov-subspace linear solvers.
For implicit transient integration, nested \texttt{lax.scan}
operations organize physical time advancement and nonlinear
correction iterations, while \texttt{lax.cond} provides
convergence-dependent control flow within compiled computations.

A key feature of JAX-Nano is its compatibility with automatic
differentiation. By expressing the governing numerical operations
within the JAX computational framework, sensitivities of simulated
quantities with respect to differentiable model parameters can
be evaluated using automatic differentiation. This capability
provides a foundation for gradient-based parameter estimation,
inverse learning, and optimization of nanoscale heat transport
systems. The differentiable formulation and its verification
are discussed in the following section.

The combination of GPU acceleration, JIT compilation, and
automatic differentiation provides a unified computational
framework for forward phonon transport simulation and
gradient-based inverse analysis.

\subsection{Automatic Differentiation and Inverse Modeling}
\label{sec:automatic_differentiation}

JAX-Nano incorporates automatic differentiation (AD) into the
finite-volume phonon Boltzmann transport equation (BTE)
solver to identify unknown physical and geometric parameters
from temperature observations. The differentiable formulation
connects the parameter-dependent transport model,
steady-state or transient numerical solution, and
temperature-based objective function within the JAX
computational framework. The inverse problems considered
here involve thermalizing boundary temperatures,
phonon relaxation times, characteristic geometric lengths,
and internal heat-generation rates.

Let $\boldsymbol{\theta}$ denote the learnable optimization
variables and $\mathbf p(\boldsymbol{\theta})$ the corresponding
physical parameters. The inverse problem is formulated as
\begin{equation}
\label{eq:inverse_optimization}
\boldsymbol{\theta}^{*}
=
\underset{\boldsymbol{\theta}}{\operatorname{arg\,min}}\,
\mathcal L(\boldsymbol{\theta}),
\end{equation}
where the temperature-based mean-squared loss is defined as
\begin{equation}
\label{eq:inverse_loss}
\mathcal L(\boldsymbol{\theta})
=
\frac{1}{N_{\mathrm{obs}}}
\sum_{i=1}^{N_{\mathrm{obs}}}
\left[
\mathcal H_i\bigl(\mathbf T(\boldsymbol{\theta})\bigr)
-
T_i^{\mathrm{ref}}
\right]^2.
\end{equation}
Here, $\mathbf T(\boldsymbol{\theta})$ denotes the
temperature field predicted by the steady-state or
transient BTE solver, $\mathcal H_i$ is the observation
operator evaluating the predicted temperature at
observation location $i$, and $N_{\mathrm{obs}}$
is the number of temperature observations.
For full-field inverse learning, temperatures from all
$N_c$ computational cells are used, giving
$N_{\mathrm{obs}}=N_c$. For sparse-observation inverse
learning, the objective function is evaluated using
temperatures at selected probe locations.
Configurations with 5 and 16 temperature probes
are considered to assess parameter identification
from limited observations.

For steady-state inverse learning, the objective
function is evaluated using the converged
temperature field. For transient inverse learning,
the time-dependent BTE solver is advanced to the
prescribed observation time, and the predicted
temperature field is compared with the reference
field at the same time. In both cases, the
temperature observations provide the objective
function for gradient-based parameter identification.

For thermalizing boundary-temperature identification,
the prescribed boundary temperature $T_b$ is treated
as a learnable physical parameter. Its dependence
is incorporated into the thermalizing boundary
condition of the finite-volume BTE solver, allowing
the boundary temperature to be inferred from the
observed temperature response.

For relaxation-time identification, each phonon
band $b$ may be assigned an independent positive
scaling factor,
\begin{equation}
\label{eq:inverse_tau}
\tau_b
=
\alpha_{\tau,b}\tau_b^{\mathrm{ref}},
\qquad
\alpha_{\tau,b}
=
\exp(\theta_{\tau,b}),
\end{equation}
where $\tau_b^{\mathrm{ref}}$ denotes the reference
relaxation time and $\theta_{\tau,b}$ is the
unconstrained learnable parameter associated
with phonon band $b$. The exponential
parameterization ensures that the identified
relaxation times remain positive. Alternatively,
a common scaling factor can be used to identify
a global relaxation-time correction. The
parameter dependence is incorporated consistently
into the scattering contribution, equilibrium-energy
evaluation, and transport correction system.
Both individual relaxation-time identification
and simultaneous identification of two relaxation
times are considered.

For geometric identification, a positive
characteristic-length scaling factor is
introduced as
\begin{equation}
\label{eq:inverse_length}
L
=
\alpha_L L^{\mathrm{ref}},
\qquad
\alpha_L
=
\exp(\theta_L),
\end{equation}
where $L^{\mathrm{ref}}$ is the reference
characteristic length and $\theta_L$ is the
unconstrained learnable parameter. The
exponential parameterization ensures that
the identified length remains positive.
The mesh connectivity and topology remain
fixed. Under uniform three-dimensional
geometric scaling, face areas and cell
volumes scale as $\alpha_L^2$ and
$\alpha_L^3$, respectively, so the
finite-volume transport contribution per
unit volume scales as $1/\alpha_L$.
This geometric dependence is incorporated
directly into the transport residual and
corresponding linear correction system,
enabling gradient-based identification
of the characteristic length without
remeshing.

The strength of an internal heat source
is identified by scaling its reference
volumetric heat-generation rate,
\begin{equation}
\label{eq:inverse_heat}
\dot Q
=
\theta_Q \dot Q_{\mathrm{ref}},
\end{equation}
where $\dot Q_{\mathrm{ref}}$ is the reference
volumetric heat-generation rate and
$\theta_Q$ is the learnable source amplitude.
The spatial distribution of the heat source
remains fixed while its magnitude is inferred
from temperature observations.

For multiple spatially distributed heat sources,
the nominal volumetric heat-generation rate
of each source is parameterized as
\begin{equation}
\label{eq:inverse_distributed_heat}
\dot Q_j
=
\theta_j \dot Q_{\mathrm{ref}},
\qquad j=1,\ldots,N_Q,
\end{equation}
where $\theta_j$ is the normalized amplitude
of source $j$. For the $3\times3$ FinFET
array, $N_Q=9$, and all nine source
amplitudes are identified simultaneously
from temperature observations.

The gradients of the objective function
with respect to the learnable parameters
are evaluated using AD. Reverse-mode AD
is employed through JAX's
\texttt{value\_and\_grad} transformation
for single-parameter identification
other than boundary temperature,
simultaneous identification of two phonon
relaxation times, and transient inverse
learning,
\begin{equation}
\label{eq:inverse_gradient}
\mathbf g
=
\nabla_{\boldsymbol{\theta}}
\mathcal L(\boldsymbol{\theta}).
\end{equation}
This transformation evaluates the
temperature-based loss and its gradient
with respect to all learnable parameters.
For transient inverse learning, the
gradient is propagated through the
time-stepping operations of the
parameter-dependent BTE solver,
allowing the observed transient
temperature response to guide
parameter identification.

Forward-mode AD is employed through
JAX's \texttt{jvp} transformation for
boundary-temperature identification
and simultaneous identification of
nine spatially distributed
heat-generation rates.
Each component of the loss gradient
is evaluated using a Jacobian--vector
product,
\begin{equation}
\label{eq:inverse_jvp}
g_j
=
J_{\mathcal L}(\boldsymbol{\theta})
\mathbf e_j
=
\frac{\partial\mathcal L}
{\partial\theta_j},
\qquad j=1,\ldots,N_p,
\end{equation}
where $J_{\mathcal L}$ denotes the Jacobian
of the scalar-valued loss function,
$\mathbf e_j$ is the unit vector
corresponding to the $j$th learnable
parameter, and $N_p$ is the number
of learnable parameters.

For boundary-temperature identification,
$N_p=1$, and the loss gradient is obtained
through a single JVP evaluation.
For the $3\times3$ heat-source array,
$N_p=9$, and the complete loss gradient
is assembled from nine sequential
JVP evaluations.
Processing one parameter direction
at a time limits the number of
tangent directions propagated through
the iterative BTE solver simultaneously,
providing a memory-conscious
differentiation strategy.

The resulting gradients are used
for parameter optimization through
gradient descent, Adam, or
projected gradient descent with
gradient clipping and backtracking
line search, depending on the
inverse problem. For projected
gradient descent, the learnable
parameters are restricted to
prescribed admissible bounds,
while backtracking accepts only
finite, non-increasing objective
values. Both differentiation
approaches operate directly on
the discretized BTE solver,
avoiding the need to derive
separate continuous sensitivity
equations.

Algorithm~\ref{alg:inverse_modeling}
summarizes the general inverse-modeling
workflow for steady-state and transient
BTE simulations. The gradient-evaluation
step employs either reverse-mode AD
or sequential forward-mode JVPs,
with the latter detailed in
Algorithm~\ref{alg:forward_jvp}.
At each optimization iteration,
the learnable variables are mapped
to physical parameters, followed
by forward simulation, objective
evaluation, and gradient-based
updates. The accuracy of the
inferred parameters under
full-field and sparse-temperature
observations is examined in the
numerical results.

\begin{algorithm}
\caption{AD-based inverse modeling in JAX-Nano}
\label{alg:inverse_modeling}
\begin{algorithmic}[1]

\Require Reference temperature observations
$\mathbf T_{\mathrm{obs}}^{\mathrm{ref}}$,
observation operator $\mathcal H$,
initial parameters $\boldsymbol{\theta}^{(0)}$,
optimizer, AD mode, and maximum iterations
$N_{\mathrm{opt}}$

\Ensure Identified physical parameters $\mathbf p^{*}$

\For{$k=0,\ldots,N_{\mathrm{opt}}-1$}

    \State Map $\boldsymbol{\theta}^{(k)}$
    to physical parameters $\mathbf p^{(k)}$

    \State Solve the steady-state BTE or advance
    the transient BTE to the observation time

    \State Obtain the predicted temperature field
    $\mathbf T(\boldsymbol{\theta}^{(k)})$

    \State Evaluate $\mathcal L^{(k)}$
    using Eq.~\eqref{eq:inverse_loss}

    \State Compute $\mathbf g^{(k)}$ using
    reverse-mode AD or Algorithm~\ref{alg:forward_jvp}

    \State Update $\boldsymbol{\theta}^{(k+1)}$
    using the selected optimizer

    \If{termination criterion is satisfied}
        \State \textbf{break}
    \EndIf

\EndFor

\State $\boldsymbol{\theta}^{*}
\gets$ final accepted learnable parameters

\State $\mathbf p^{*}
=\mathbf p(\boldsymbol{\theta}^{*})$

\State \Return $\mathbf p^{*}$

\end{algorithmic}
\end{algorithm}

\begin{algorithm}
\caption{Sequential forward-mode gradient evaluation}
\label{alg:forward_jvp}
\begin{algorithmic}[1]

\Require Learnable parameters
$\boldsymbol{\theta}\in\mathbb R^{N_p}$,
temperature-based loss $\mathcal L$

\Ensure Loss $\mathcal L(\boldsymbol{\theta})$
and gradient $\mathbf g$

\State Evaluate $\mathcal L(\boldsymbol{\theta})$

\State Initialize $\mathbf g\gets\mathbf 0$

\For{$j=1,\ldots,N_p$}

    \State Set $\mathbf e_j$ to the $j$th
    coordinate unit vector

    \State $g_j\gets
    \operatorname{JVP}[\mathcal L]
    (\boldsymbol{\theta},\mathbf e_j)$

\EndFor

\State \Return $\mathcal L(\boldsymbol{\theta}),\mathbf g$

\end{algorithmic}
\end{algorithm}

%%===============================================================================
%	Results
%===============================================================================
\section{Verification and Validation}
The accuracy and robustness of JAX-Nano are assessed
through a series of benchmarks of increasing physical
and computational complexity. Steady-state and transient
phonon transport are first examined using canonical
problems with established analytical and numerical
reference solutions, followed by validation in
three-dimensional FinFET geometries. Systematic
spatial, angular, temporal, and solver convergence
studies are conducted to evaluate the numerical
accuracy and reliability of the proposed framework.
Together, these tests establish the forward predictive
capabilities of JAX-Nano before its application to
gradient-based inverse parameter identification.

\subsection{Steady-State Cross-Plane Phonon Transport}
\label{sec:steady_validation}

\begin{figure}
\begin{center}
\includegraphics [width=1.\columnwidth]{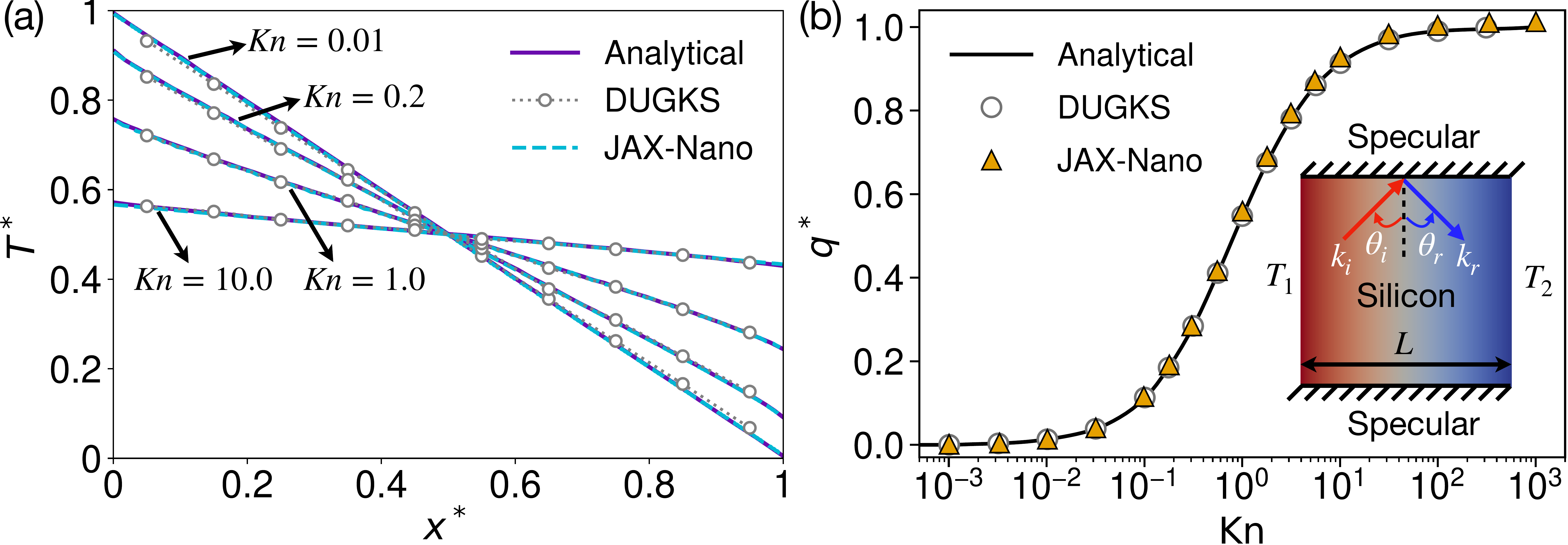}
\end{center}
\caption{Validation of the JAX-Nano steady-state solver for cross-plane phonon
transport across diffusive-to-ballistic transport regimes.
(a) Dimensionless temperature profiles $T^*$ for
$\mathrm{Kn}=0.01$, $0.2$, $1$, and $10$, compared with the analytical
solutions and results from the discrete unified gas kinetic scheme (DUGKS)~\cite{Guo_2016a}.
(b) Dimensionless heat flux $q^*$ as a function of the
Knudsen number, compared with the analytical solution and DUGKS results
from Guo and Xu~\cite{Guo_2016a}.
The inset illustrates the cross-plane transport configuration with prescribed
temperatures $T_1$ and $T_2$ and specular boundaries in the transverse
direction.}
\label{fig:steady_valid}
\end{figure}  

The steady-state implementation of JAX-Nano is first validated using
the canonical cross-plane heat-conduction problem in a dielectric film
considered by Guo and Xu~\cite{Guo_2016a}. The film has a thickness
$L$ in the $x$ direction, with the two boundaries at $x=0$ and $x=L$
maintained at temperatures $T_1$ and $T_2$, respectively. Because the
film and the imposed thermal boundary conditions are uniform in the
two transverse directions, no spatial variation occurs in these
directions, and the problem reduces to one-dimensional transport along
the film thickness. The phonon distribution therefore depends
spatially only on $x$, while retaining its dependence on the phonon
propagation direction. This configuration provides a canonical
benchmark for examining cross-plane phonon transport from the
diffusive to the ballistic regime.

Although the reference problem is one-dimensional in physical space,
it is represented in JAX-Nano using a three-dimensional computational
domain with a quasi-one-dimensional $400\times1\times1$ mesh. The
finite-volume cells are stacked along the cross-plane ($x$) direction,
with a single cell in each of the two transverse directions. Specular
boundary conditions are imposed on the transverse faces. Because the
solution is invariant in the transverse directions, additional cells
in these directions would not provide additional physical information.
This quasi-1D representation preserves the one-dimensional nature of
the benchmark while exercising the same three-dimensional finite-volume
formulation used subsequently for fully three-dimensional device
simulations. A single-band gray phonon model is employed, with the
angular domain discretized using 64 discrete propagation directions.

The dimensionless temperature and spatial coordinate are defined as
\begin{equation}
T^* = \frac{T-T_2}{T_1-T_2},
\qquad
x^* = \frac{x}{L},
\end{equation}
where $T_1$ and $T_2$ denote the temperatures of the hot and cold
boundaries, respectively. The cross-plane heat flux is evaluated from
the discrete angular moment of the phonon energy distribution as
\begin{equation}
q_{x,i}
=
\sum_b \sum_{\alpha=1}^{N_{\Omega}}
w_{\alpha} v_{g,b\alpha,x} e_{ib\alpha},
\end{equation}
where $e_{ib\alpha}$ is the phonon energy distribution in cell $i$,
band $b$, and discrete propagation direction $\alpha$,
$N_{\Omega}$ is the total number of discrete propagation directions,
$w_{\alpha}$ is the corresponding angular quadrature weight, and
$v_{g,b\alpha,x}$ is the cross-plane component of the phonon group
velocity. The heat flux is spatially averaged over the computational
domain and normalized by its ballistic-limit value as
\begin{equation}
q^* = \frac{|\bar{q}_x|}{q_{\infty}},
\qquad
\bar{q}_x = \frac{1}{N_c}\sum_{i=1}^{N_c}q_{x,i},
\qquad
q_{\infty}
= \frac{1}{4} C_v v_g (T_1-T_2),
\end{equation}
where $N_c$ is the number of cells, $C_v$ is the volumetric heat
capacity, and $v_g$ is the phonon group velocity. For the present
single-band gray-model calculations, the boundary temperatures are
set to $T_1=800$ K and $T_2=300$ K. The JAX-Nano predictions are
compared with the corresponding analytical solutions and discrete
unified gas kinetic scheme (DUGKS) results reported by Guo and
Xu~\cite{Guo_2016a}.

Figure~\ref{fig:steady_valid}(a) compares the dimensionless
steady-state temperature profiles for $\mathrm{Kn}=0.01$, $0.2$, $1$,
and $10$, spanning transport from the near-diffusive to the strongly
ballistic regime. At $\mathrm{Kn}=0.01$, the temperature distribution
is close to the diffusive limit and exhibits an approximately linear
variation through the interior of the film. As $\mathrm{Kn}$ increases,
the temperature profile progressively departs from the diffusive
behavior. The interior temperature gradient decreases, while the
temperature jumps associated with boundary nonequilibrium become
increasingly pronounced. At $\mathrm{Kn}=10$, the temperature
distribution becomes nearly uniform in the interior, reflecting the
dominance of ballistic phonon transport.

For all four Knudsen numbers, the JAX-Nano predictions closely reproduce
the analytical solutions and DUGKS results. In particular, the solver
captures both the near-linear temperature distribution in the
diffusive regime and the progressive flattening of the interior
temperature profile as transport becomes increasingly ballistic.
This agreement is noteworthy because the JAX-Nano results are obtained
using the three-dimensional finite-volume implementation on a
quasi-1D mesh rather than a separate one-dimensional solver. The
results therefore verify that the three-dimensional spatial
discretization, angular discretization, and boundary treatment
correctly recover the expected one-dimensional cross-plane transport
behavior.

Figure~\ref{fig:steady_valid}(b) provides a complementary validation
based on the dimensionless heat flux over the substantially broader
range $10^{-3}\leq\mathrm{Kn}\leq10^{3}$. The JAX-Nano predictions
closely follow the analytical solution throughout the entire range
and are consistent with the DUGKS reference results. At small
$\mathrm{Kn}$, the heat flux approaches the diffusive limit, whereas
increasing $\mathrm{Kn}$ produces a continuous transition through the
quasi-ballistic regime toward the ballistic limit. The agreement over
six orders of magnitude in $\mathrm{Kn}$ demonstrates that the
steady-state formulation accurately captures not only the limiting
transport regimes but also the transition between them.

Taken together, the temperature-profile and heat-flux comparisons in
Fig.~\ref{fig:steady_valid} validate both the spatially resolved
temperature field and the overall transport response predicted by
JAX-Nano. The benchmark also confirms that the general
three-dimensional implementation correctly recovers the canonical
one-dimensional solution when applied to a quasi-1D configuration,
providing a baseline for the transient and fully three-dimensional
simulations considered subsequently.

\subsection{Transient Thermal Grating}
\label{sec:ttg_validation}

\begin{figure}
\begin{center}
\includegraphics [width=1.\columnwidth]{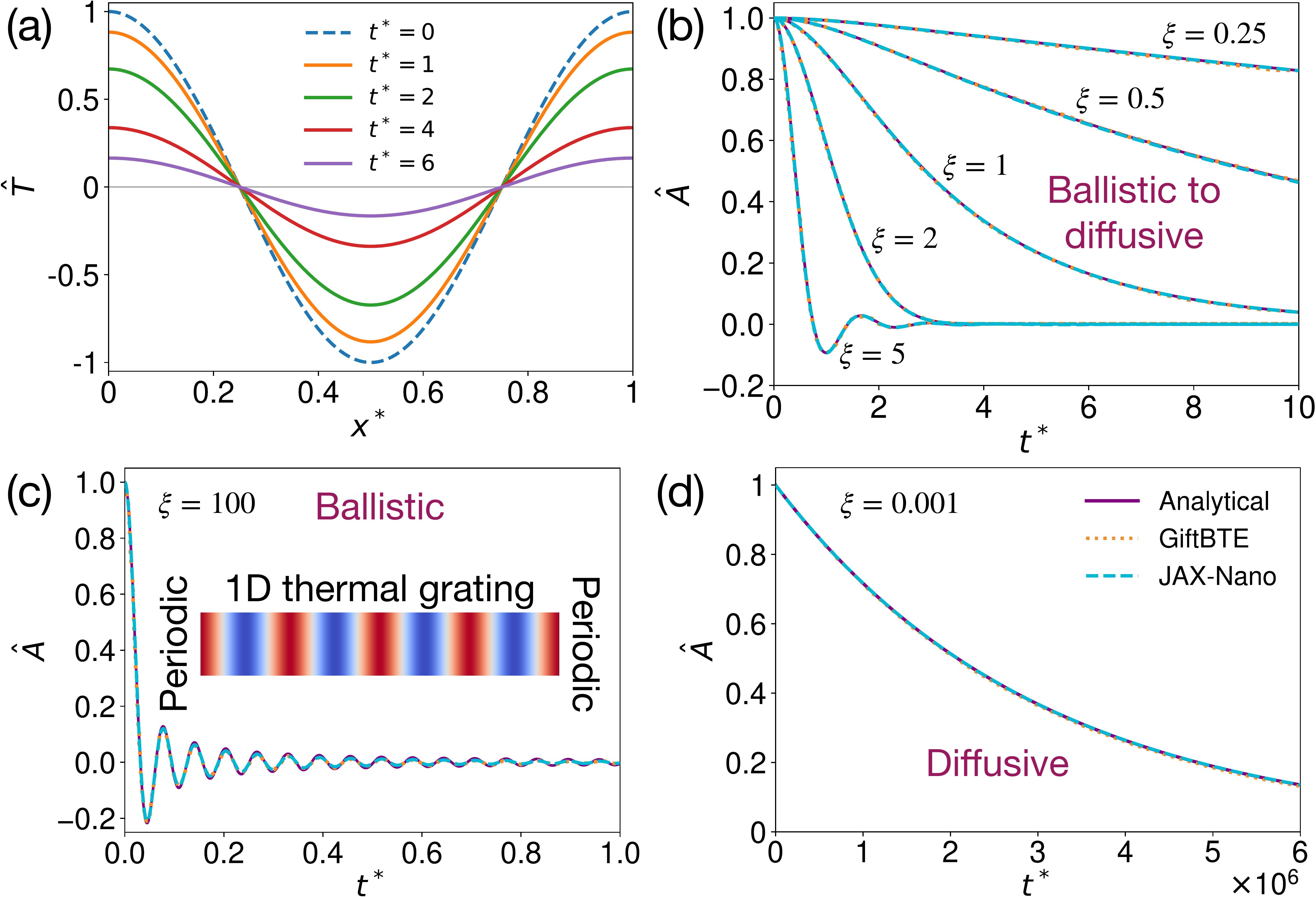}
\end{center}
\caption{Validation of JAX-Nano for one-dimensional transient thermal grating (TTG) transport across ballistic and diffusive regimes. 
(a) Evolution of the dimensionless temperature profile $\hat{T}$ at selected dimensionless times $t^*$ for $\xi=1$, obtained from the analytical solution. 
(b) Decay of the normalized grating amplitude $\hat{A}$ for different values of the transport parameter $\xi$, illustrating the transition from ballistic to diffusive transport. 
(c) Ballistic limit ($\xi=100$), characterized by rapidly damped oscillations of the grating amplitude; the inset illustrates the one-dimensional thermal-grating configuration with periodic boundaries. 
(d) Diffusive limit ($\xi=0.001$), showing monotonic thermal decay. 
Predictions from JAX-Nano are compared with the analytical solution and GiftBTE, demonstrating excellent agreement across the transport regimes.}
\label{fig:transient_valid}
\end{figure} 

The transient capability of JAX-Nano is next assessed using the
transient thermal grating (TTG) problem, which provides a canonical
benchmark for nonequilibrium phonon transport over a broad range of
transport regimes. In this problem, a spatially periodic temperature
perturbation is imposed on an otherwise uniform medium and subsequently
relaxes through phonon transport and scattering. The transient response
depends strongly on the relationship between the phonon mean free path
and the characteristic spatial scale of the thermal grating, making
the TTG problem well suited for examining the transition from
diffusive to ballistic transport.

The initial temperature field is prescribed as a sinusoidal
perturbation,
\begin{equation}
T(x,0)
=
T_0 + \Delta T
\cos\left(\frac{2\pi x}{L}\right),
\end{equation}
where $T_0$ is the reference temperature, $\Delta T$ is the initial
temperature amplitude, and $L$ is the thermal-grating wavelength. The
corresponding grating wave number is $k=2\pi/L$. Periodic boundary
conditions are imposed along the transport direction such that the
computational domain represents one spatial period of the thermal
grating. The temperature perturbation is expressed in dimensionless
form as
\begin{equation}
\hat{T}(x,t)
=
\frac{T(x,t)-T_0}{\Delta T},
\end{equation}
and the temporal relaxation of the grating is characterized by the
normalized amplitude
\begin{equation}
\hat{A}(t)
=
\frac{A(t)}{A(0)}.
\end{equation}
The transport regime is characterized by the dimensionless parameter
\begin{equation}
\xi
=
2\pi\mathrm{Kn}
=
\frac{2\pi\Lambda}{L}
=
k\Lambda,
\end{equation}
where $\Lambda$ is the phonon mean free path and
$\mathrm{Kn}=\Lambda/L$. Thus, $\xi$ measures the phonon mean free
path relative to the characteristic spatial scale of the temperature
variation, $1/k=L/(2\pi)$. The limit $\xi\ll1$ corresponds to
diffusive transport, whereas increasing $\xi$ indicates progressively
stronger quasi-ballistic and ballistic effects. 
The dimensionless time is defined as
\begin{equation}
t^* = \frac{t}{\tau},
\end{equation}
where $\tau$ is the phonon relaxation time. 

The TTG problem is simulated using the same finite-volume phonon BTE
framework employed in the steady-state calculations. Because the
temperature field varies only along the grating direction, the problem
is modeled as one-dimensional transport and discretized using 100
finite-volume cells. A single-band gray phonon model is employed, with
the angular domain discretized using 512 discrete propagation
directions. For the results presented in
Fig.~\ref{fig:transient_valid}, the phonon energy distribution is
advanced in time using the explicit time-integration scheme. This
benchmark examines both the evolution of the spatial temperature field
and the temporal decay of the thermal-grating amplitude over transport
conditions ranging from the strongly diffusive to the strongly
ballistic limits.

Figure~\ref{fig:transient_valid}(a) illustrates the spatial evolution
of the dimensionless temperature profile $\hat{T}$ obtained from the
analytical solution at selected dimensionless times $t^*$. Starting
from the prescribed sinusoidal temperature distribution, the amplitude
of the temperature perturbation decreases with time while retaining
its periodic spatial structure. These profiles illustrate the
characteristic relaxation of the thermal grating and provide a
reference for assessing the numerical solution.

The influence of the transport regime is illustrated in
Fig.~\ref{fig:transient_valid}(b), which shows the normalized grating
amplitude $\hat{A}$ for $\xi=0.25$, $0.5$, $1$, $2$, and $5$. The
transient response changes substantially with $\xi$, reflecting the
competition between phonon scattering and ballistic propagation. At
relatively small $\xi$, the response approaches the smooth decay
characteristic of diffusion-dominated transport. As $\xi$ increases,
ballistic propagation becomes increasingly important, and the
transient response progressively departs from the classical diffusive
behavior. The agreement between the JAX-Nano predictions and the
reference solutions across these intermediate values of $\xi$
demonstrates that the numerical formulation captures the continuous
transition between diffusive, quasi-ballistic, and ballistic
transport.

The strongly ballistic limit is examined in
Fig.~\ref{fig:transient_valid}(c) for $\xi=100$. In this regime, the
phonon mean free path is much larger than the characteristic spatial
scale $1/k$ of the thermal grating. Consequently, the transient
response is no longer described by a simple monotonic diffusive decay.
Instead, the normalized grating amplitude exhibits rapidly damped
oscillations associated with ballistic phonon propagation. JAX-Nano
accurately reproduces both the oscillatory behavior and its decay,
demonstrating its ability to capture the strongly nonequilibrium
transport behavior described by the phonon BTE.

At the opposite limit, Fig.~\ref{fig:transient_valid}(d) considers
$\xi=0.001$. In this regime, the phonon mean free path is much smaller
than the characteristic spatial scale of the thermal grating, and
phonon scattering drives the transport toward the diffusive limit.
The normalized grating amplitude therefore exhibits a smooth,
monotonic decay. The JAX-Nano predictions agree closely with both the
analytical solution and the GiftBTE results, confirming that the
numerical formulation correctly recovers the expected diffusive
behavior.

Taken together, the results in Fig.~\ref{fig:transient_valid}
demonstrate that JAX-Nano accurately resolves transient phonon
transport across a broad range of transport regimes. The same
numerical framework captures the smooth thermal relaxation of the
diffusive limit, the transition through the quasi-ballistic regime,
and the damped oscillatory response characteristic of strongly
ballistic transport. This validation complements the steady-state
cross-plane results in Fig.~\ref{fig:steady_valid} and establishes a
basis for applying JAX-Nano to the three-dimensional device-scale
simulations considered subsequently.

\begin{figure}
\begin{center}
\includegraphics [width=1.\columnwidth]{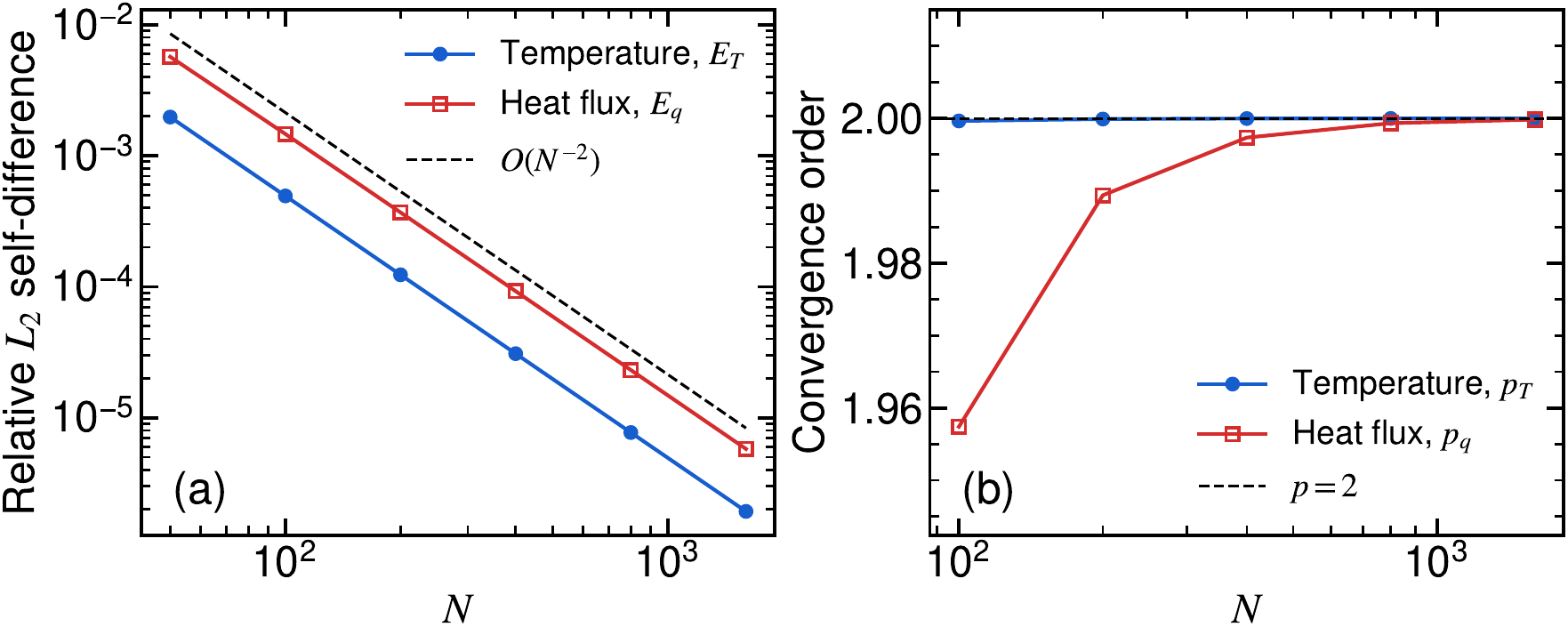}
\end{center}
\caption{Spatial self-convergence of JAX-Nano for the transient
thermal grating (TTG) benchmark at $\xi=100$ and $t^*=0.01$.
(a) Relative volume-weighted $L_2$ self-differences in
temperature and heat flux between successive grid resolutions,
compared with the second-order reference scaling $O(N^{-2})$.
(b) Observed convergence orders for temperature and heat flux
computed from three successive grid resolutions.
Here, $N$ denotes the finest-grid cell count used in each
comparison. Both quantities exhibit approximately
second-order spatial self-convergence.}
\label{fig:L2_error}
\end{figure} 

To further assess the spatial accuracy of JAX-Nano, a systematic grid-refinement study is conducted using a quasi-one-dimensional (quasi-1D) transient thermal grating (TTG) benchmark at $\xi=100$ and $t^*=0.01$. Phonon transport is modeled using the gray Boltzmann transport equation (BTE) with a single phonon band and 512 discrete propagation directions. The initial temperature perturbation varies sinusoidally along the grating direction, while phonon propagation is resolved in three-dimensional angular space. Periodic boundary conditions are imposed at the two ends of the grating direction, whereas specularly reflecting, adiabatic boundary conditions are applied to the four lateral walls. Seven successively refined spatial grids containing $25$, $50$, $100$, $200$, $400$, $800$, and $1600$ finite-volume cells along the grating direction are considered. The same angular quadrature, physical parameters, initial conditions, boundary conditions, and dimensionless evaluation time are maintained across all grid resolutions.

Spatial self-convergence is evaluated by comparing the
temperature and heat-flux solutions obtained on consecutive
grids. The fine-grid solutions are conservatively restricted
onto the corresponding coarse grid using a volume-weighted
averaging operator,
\begin{equation}
\mathcal{R}(\phi_{2N})_i
=
\frac{
V_{2N,2i-1}\phi_{2N,2i-1}
+
V_{2N,2i}\phi_{2N,2i}
}{
V_{2N,2i-1}+V_{2N,2i}
},
\label{eq:ttg_restriction}
\end{equation}
where $\phi$ denotes either the temperature $T$ or the
heat-flux component $q_{\parallel}$ along the grating
direction, $V_{2N,j}$ is the volume of fine-grid cell $j$,
and $\mathcal{R}$ represents the conservative restriction
operator. The spatial grids are nested and aligned such
that each coarse-grid cell corresponds to two adjacent
fine-grid cells.

The volume-weighted $L_2$ norm of a discrete field
$a$ on a grid containing $N$ cells is defined as
\begin{equation}
\|a\|_{2,V_N}
=
\left[
\frac{
\displaystyle\sum_{i=1}^{N}V_{N,i}a_i^2
}{
\displaystyle\sum_{i=1}^{N}V_{N,i}
}
\right]^{1/2},
\label{eq:ttg_l2_norm}
\end{equation}
where $V_{N,i}$ denotes the volume of coarse-grid
cell $i$.

The relative volume-weighted $L_2$ self-difference
for temperature is calculated as
\begin{equation}
E_T(N,2N)
=
\frac{
\left\|
T_N-\mathcal{R}(T_{2N})
\right\|_{2,V_N}
}{
\left\|
T_N-\overline{T}_N
\right\|_{2,V_N}
},
\label{eq:ttg_relative_temperature}
\end{equation}
where the volume-averaged temperature is
\begin{equation}
\overline{T}_N
=
\frac{
\displaystyle\sum_{i=1}^{N}V_{N,i}T_{N,i}
}{
\displaystyle\sum_{i=1}^{N}V_{N,i}
}.
\label{eq:ttg_mean_temperature}
\end{equation}
Normalizing by the temperature perturbation rather than
the absolute temperature is appropriate for the TTG
problem, in which the spatially varying thermal response
is superimposed on a uniform reference temperature.

Similarly, the relative volume-weighted $L_2$
self-difference for the heat flux is defined as
\begin{equation}
E_q(N,2N)
=
\frac{
\left\|
q_{\parallel,N}
-
\mathcal{R}(q_{\parallel,2N})
\right\|_{2,V_N}
}{
\left\|
q_{\parallel,N}
\right\|_{2,V_N}
}.
\label{eq:ttg_relative_heatflux}
\end{equation}

These measures quantify differences between numerical
solutions obtained at successive grid resolutions,
rather than errors relative to an exact solution.
Both quantities are dimensionless, allowing the
temperature and heat-flux self-differences to be
compared directly.

The observed spatial convergence order is estimated
using three successive grid resolutions according to
\begin{equation}
p_{\phi}(N)
=
\log_2
\left[
\frac{
E_{\phi}(N/4,N/2)
}{
E_{\phi}(N/2,N)
}
\right],
\label{eq:ttg_convergence_order}
\end{equation}
where $\phi\in\{T,q\}$ and $N$ denotes the finest-grid
cell count in each three-grid comparison.
For an asymptotically convergent spatial discretization
of order $p$, the relative self-differences are
expected to decrease approximately as $N^{-p}$
under uniform grid refinement, provided that
temporal and angular discretization errors are
sufficiently controlled.

Figure~\ref{fig:L2_error}(a) presents the relative
volume-weighted $L_2$ self-differences for temperature
and heat flux as functions of spatial resolution.
Both quantities decrease systematically with grid
refinement and closely follow the second-order
reference scaling $O(N^{-2})$.

The corresponding observed convergence orders are
presented in Fig.~\ref{fig:L2_error}(b).
The temperature convergence order remains close to
two across the examined resolutions, while the
heat-flux convergence order approaches two as the
grid is refined. These results demonstrate approximately
second-order spatial self-convergence for both the
temperature field and the heat flux in the strongly
ballistic TTG regime.

Together with the steady-state and transient benchmark
comparisons presented in Figs.~\ref{fig:steady_valid}
and \ref{fig:transient_valid}, respectively, the spatial
self-convergence results provide complementary evidence
of the accuracy and numerical consistency of the
JAX-Nano phonon transport solver.

\subsection{Three-Dimensional Phonon Transport:
Verification and Validation}
\label{sec:3d_cube_validation}
To evaluate the accuracy and robustness of JAX-Nano
for three-dimensional phonon transport, a series of
numerical studies is conducted in a cubic silicon domain
with a side length of $L=100~\mathrm{nm}$.
Thermalizing boundary conditions are imposed at
the bottom ($y=0$) and top ($y=L$) surfaces,
maintained at 300 and 800 K, respectively,
while the remaining four faces are subject to
either diffuse or specular phonon reflection.
Both steady-state and transient transport are
investigated, with particular attention to the
influence of mesh structure, numerical reconstruction
schemes, and angular and spectral discretization.
The numerical predictions are systematically compared
with those obtained using GiftBTE, which employs
a second-order spatial discretization based on
upwind reconstruction with least-squares gradient
evaluation. For steady-state calculations, GiftBTE
uses a synthetic iterative scheme to accelerate
convergence. These comparisons assess the accuracy,
numerical consistency, and robustness of JAX-Nano
across the investigated configurations.

Unless otherwise specified, all JAX-Nano simulations
employ second-order upwind reconstruction with the
Barth--Jespersen (BJ) slope limiter.
Alternative reconstruction schemes, including central
reconstruction and unlimited second-order upwind
reconstruction, are considered only in the numerical
scheme comparison.

\subsubsection{Steady-State Transport with Diffuse Boundary Conditions}

\begin{figure}
\begin{center}
\includegraphics [width=1.\columnwidth]{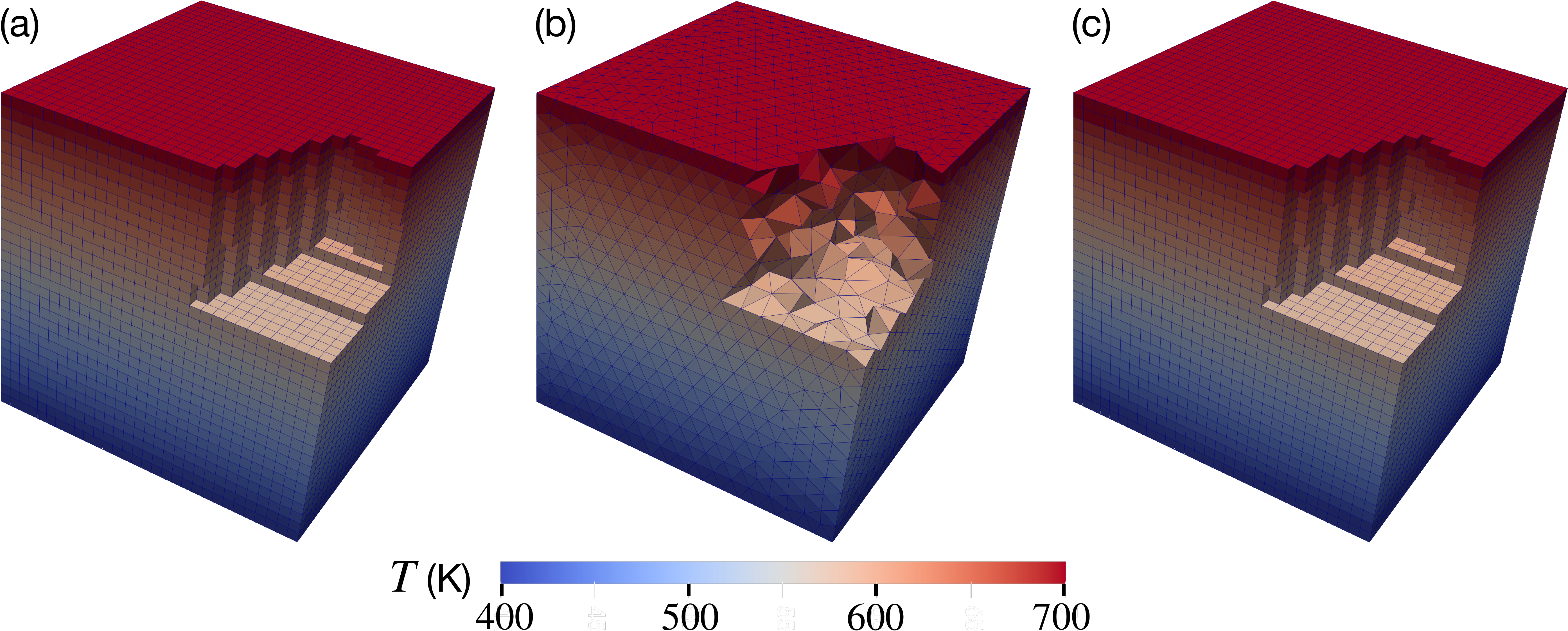}
\end{center}
\caption{Comparison of three-dimensional steady-state
temperature distributions in a cubic silicon domain with
thermalizing boundaries at $y=0$ and $y=L$
and diffuse reflection boundaries at the remaining four faces, computed
using (a) JAX-Nano on a structured mesh,
(b) JAX-Nano on an unstructured mesh, and
(c) GiftBTE on the same structured mesh as in (a).
The structured mesh contains $30^3=27,000$
computational cells, while the unstructured mesh
contains $22,789$ cells.
The same numerical framework is employed for both JAX-Nano
simulations to assess the influence of mesh structure.
Cutaway views reveal the interior temperature distributions,
with an identical color scale used for all three cases.
The close agreement demonstrates the consistency of
JAX-Nano predictions across structured and unstructured
meshes and their agreement with the GiftBTE reference.}
\label{fig:3D_cube_T}
\end{figure} 

\begin{figure}
\begin{center}
\includegraphics [width=1.\columnwidth]{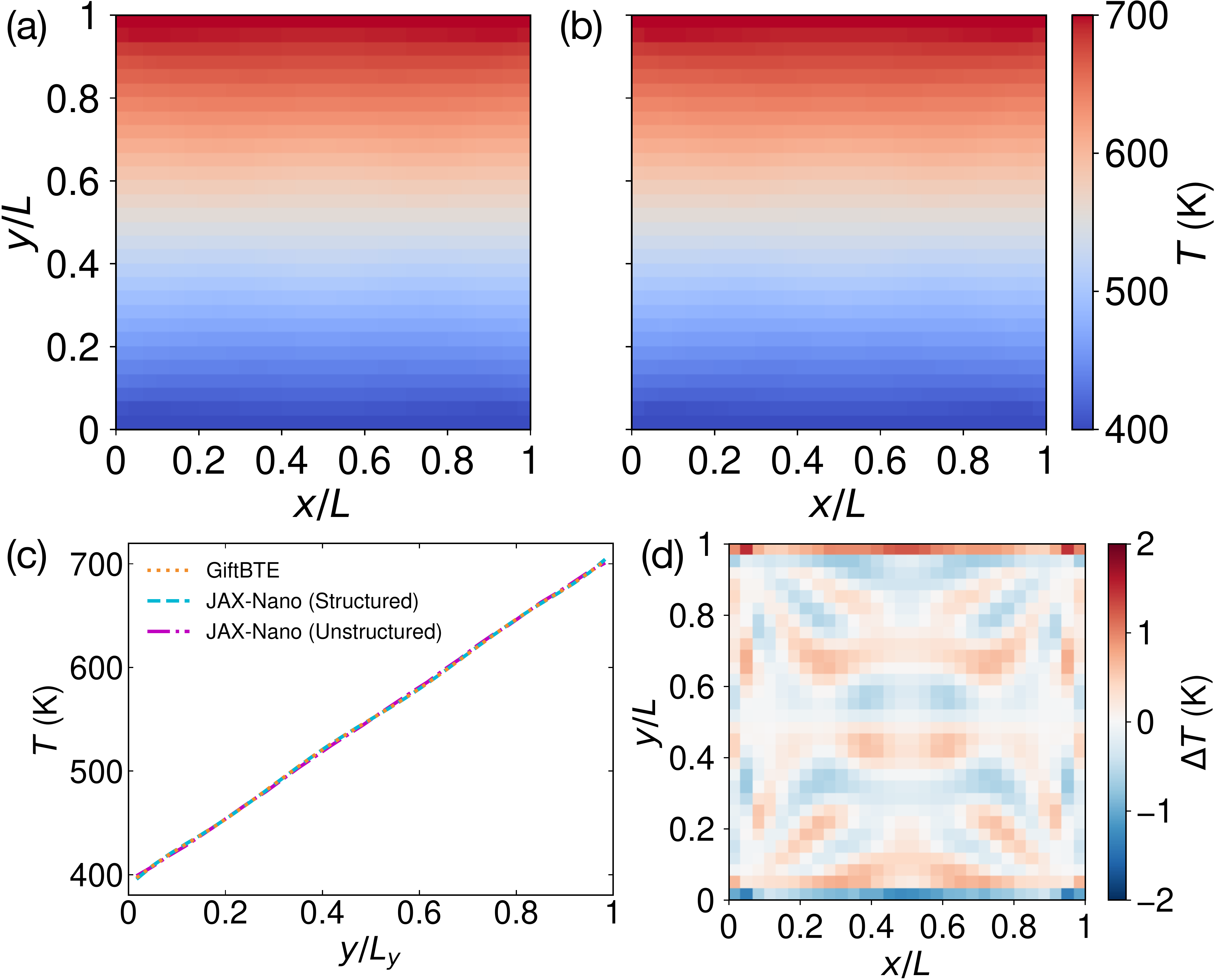}
\end{center}
\caption{Comparison of steady-state temperature distributions
in a three-dimensional cubic silicon domain with thermalizing
boundary conditions at $y=0$ and $y=L$ and diffuse
reflection at the remaining four faces.
(a) Temperature distribution on the central cross-sectional
plane predicted by GiftBTE on a structured mesh.
(b) Corresponding temperature distribution predicted by
JAX-Nano on the same structured mesh.
(c) Centerline temperature profiles obtained using GiftBTE
and JAX-Nano on structured and unstructured meshes.
The centerline is defined by $x=z=L/2$, with $y$
extending from the bottom (300 K) to the top (800 K)
surfaces, both subject to thermalizing boundary conditions.
(d) Local temperature difference
$\Delta T = T_{\mathrm{JAX\mbox{-}Nano}} -
T_{\mathrm{GiftBTE}}$ between the structured JAX-Nano
and GiftBTE solutions on the central cross-sectional plane.
The close agreement demonstrates the consistency of
JAX-Nano predictions on structured and unstructured meshes
with the GiftBTE reference solution.}
\label{fig:3D_cube_T_slice}
\end{figure} 

\begin{figure}
\begin{center}
\includegraphics [width=1.\columnwidth]{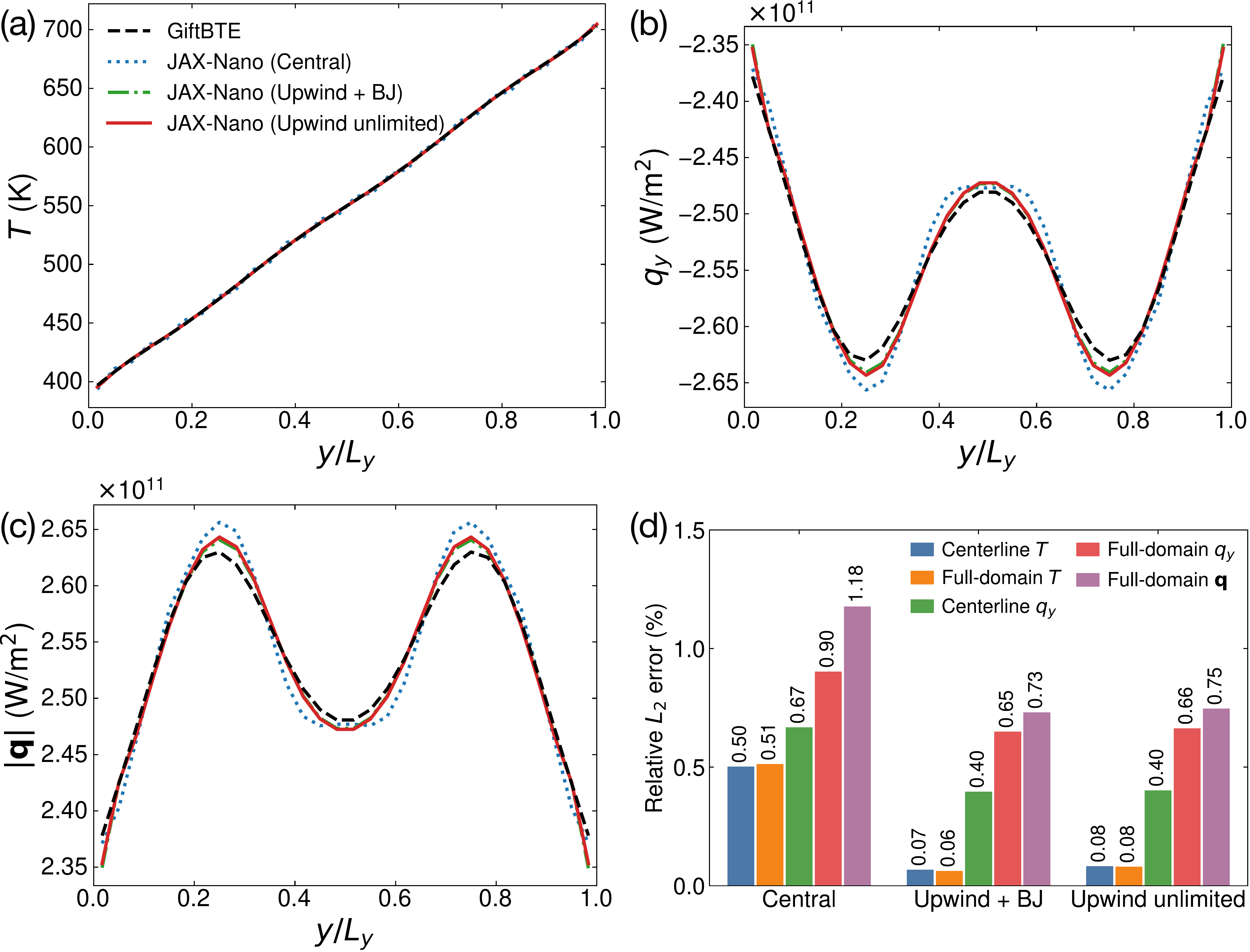}
\end{center}
\caption{Comparison of steady-state temperature and heat-flux predictions obtained using GiftBTE and three spatial reconstruction schemes implemented in JAX-Nano for a three-dimensional cubic silicon domain with thermalizing boundary conditions at $y=0$ and $y=L$ and diffuse reflection at the remaining four faces. The structured mesh contains $30^3=27{,}000$ computational cells. (a) Temperature $T$ along the domain centerline. (b) Signed heat-flux component $q_y$ along the centerline. (c) Heat-flux magnitude $|\mathbf{q}|$ along the centerline. (d) Relative $L_2$ errors in the centerline temperature, full-domain temperature, centerline $q_y$, full-domain $q_y$, and full-domain heat-flux vector $\mathbf{q}$. The three JAX-Nano schemes comprise central reconstruction, second-order upwind reconstruction with the Barth--Jespersen (BJ) limiter, and unlimited second-order upwind reconstruction. The centerline is defined by $x=z=L/2$, with $y$ varying from the bottom to the top thermalizing boundary. GiftBTE serves as the reference solution for all relative-error calculations.
}
\label{fig:flux_diffuse}
\end{figure} 

We first consider steady-state phonon transport
in a cubic silicon domain with thermalizing boundary
conditions imposed on the top ($800$~K) and
bottom ($300$~K) surfaces, while diffuse reflection
conditions are applied to the four lateral surfaces.
The resulting temperature gradient drives phonon
transport between the hot and cold boundaries.
The phonon BTE is discretized using 64 angular
directions and a single spectral band.
JAX-Nano simulations are performed on both structured
and unstructured meshes to examine the influence
of mesh structure on the numerical predictions.
The steady-state temperature and heat-flux distributions
are compared with those obtained using GiftBTE,
which serves as the numerical reference.

Figure~\ref{fig:3D_cube_T} presents the
three-dimensional steady-state temperature
distributions obtained using JAX-Nano on
structured and unstructured meshes, together
with the GiftBTE reference solution.
The second-order upwind reconstruction with
the Barth--Jespersen (BJ) limiter is employed
for both JAX-Nano simulations.
The cutaway visualizations reveal the interior
temperature fields and the spatial variation
between the hot and cold surfaces.
All three simulations exhibit similar temperature
distributions, with a gradual transition from
the high-temperature region near the hot boundary
to the low-temperature region near the cold boundary.
Despite differences in mesh connectivity and
cell geometry, the structured and unstructured
JAX-Nano solutions exhibit consistent temperature
patterns and agree closely with GiftBTE.
These results demonstrate the capability of
JAX-Nano to handle three-dimensional phonon
transport on different mesh configurations.

A more detailed quantitative comparison is
provided in Fig.~\ref{fig:3D_cube_T_slice}.
Figures~\ref{fig:3D_cube_T_slice}(a) and
\ref{fig:3D_cube_T_slice}(b) show the temperature
distributions on the central cross-sectional
plane obtained using GiftBTE and JAX-Nano,
respectively, on the same structured mesh. Both solutions exhibit nearly
identical spatial temperature variations.
The centerline temperature profiles in
Fig.~\ref{fig:3D_cube_T_slice}(c) further
confirm the agreement among GiftBTE and the
structured and unstructured JAX-Nano solutions,
with the three curves nearly overlapping
throughout the domain.
To examine local discrepancies,
Fig.~\ref{fig:3D_cube_T_slice}(d) presents
the temperature difference between the
structured JAX-Nano and GiftBTE solutions.
The differences remain approximately within
$\pm 2$~K over the central cross-sectional
plane, which is small compared with the
imposed temperature difference of $500$~K.
The spatially varying discrepancies indicate
minor differences between the numerical
solutions, while the overall temperature
distribution is consistently reproduced.

In addition to the spatial temperature distributions,
the influence of numerical reconstruction schemes on
the accuracy of temperature and heat-flux predictions
is investigated using the same three-dimensional cubic
configuration. The structured domain is discretized into
$30\times30\times30=27{,}000$ computational cells.
Three reconstruction schemes are considered in JAX-Nano:
central reconstruction, second-order upwind reconstruction
with the Barth--Jespersen (BJ) limiter, and unlimited
second-order upwind reconstruction. The numerical
predictions are compared with GiftBTE through centerline
profiles and relative $L_2$ errors evaluated along the
centerline and over the entire computational domain,
as shown in Fig.~\ref{fig:flux_diffuse}.

Figure~\ref{fig:flux_diffuse}(a) compares the steady-state
temperature profiles along the centerline connecting the
two thermalizing boundaries. All three JAX-Nano
reconstruction schemes closely reproduce the GiftBTE
temperature distribution, with nearly overlapping
profiles throughout the domain.
Figures~\ref{fig:flux_diffuse}(b) and
\ref{fig:flux_diffuse}(c) present the signed heat-flux
component $q_y$ and the heat-flux magnitude
$|\mathbf{q}|$, respectively.
All three schemes capture the overall spatial variation
observed in the GiftBTE reference solution, including
the two local minima and the intermediate maximum in
$q_y$, and the corresponding extrema in the heat-flux
magnitude. However, noticeable differences arise in
the predicted heat-flux distributions.
The central reconstruction exhibits larger deviations
from GiftBTE, particularly near the local extrema,
whereas both upwind reconstruction schemes produce
profiles that more closely follow the reference solution.
The BJ-limited and unlimited upwind schemes yield
nearly overlapping distributions, indicating that
the limiter introduces only minor differences in
the predicted heat flux for this configuration.

To quantify these discrepancies, the relative $L_2$
error is defined as
\begin{equation}
\varepsilon_{L_2}
=
\frac{
\left\|u_{\mathrm{JAX\text{-}Nano}}
-u_{\mathrm{GiftBTE}}\right\|_2
}{
\left\|u_{\mathrm{GiftBTE}}\right\|_2
}
\times 100\%,
\end{equation}
where $u$ represents the temperature $T$, the signed
heat-flux component $q_y$, or the heat-flux vector
$\mathbf{q}$, evaluated over the specified set of
computational cells. For the heat-flux vector, the
$L_2$ norm includes all three Cartesian components.

Figure~\ref{fig:flux_diffuse}(d) summarizes the relative
$L_2$ errors for the three reconstruction schemes.
For central reconstruction, the centerline and
full-domain temperature errors are $0.50\%$ and
$0.51\%$, respectively, while the corresponding errors
for the BJ-limited upwind reconstruction decrease to
$0.07\%$ and $0.06\%$. The unlimited upwind
reconstruction yields similarly small temperature
errors of $0.08\%$ for both comparisons.

For the heat-flux component $q_y$, central reconstruction
produces relative errors of $0.67\%$ along the centerline
and $0.90\%$ over the full domain. These errors decrease
to $0.40\%$ and $0.65\%$, respectively, with BJ-limited
upwind reconstruction, and to $0.40\%$ and $0.66\%$
with unlimited upwind reconstruction.
The full-domain heat-flux vector errors are $1.18\%$,
$0.73\%$, and $0.75\%$ for central, BJ-limited upwind,
and unlimited upwind reconstruction, respectively.

Overall, these results demonstrate that JAX-Nano accurately reproduces three-dimensional steady-state phonon transport on both structured and unstructured meshes, with temperature distributions closely matching the GiftBTE reference solution. The small temperature discrepancies across the central cross-sectional plane and the nearly overlapping centerline profiles confirm the consistency of the numerical predictions across different mesh configurations. The reconstruction scheme comparison further reveals that, although all three schemes capture the overall temperature and heat-flux distributions, second-order upwind reconstruction provides substantially improved accuracy over central reconstruction, particularly for heat-flux predictions. The BJ-limited and unlimited upwind schemes exhibit comparable accuracy, with the BJ-limited formulation achieving slightly smaller full-domain heat-flux errors while providing additional control over reconstructed values through slope limiting. These findings establish the accuracy and numerical consistency of JAX-Nano for the investigated three-dimensional steady-state configuration and support the use of BJ-limited second-order upwind reconstruction in subsequent phonon transport simulations.

\subsubsection{Steady-State Transport with Specular Boundary Conditions}

\begin{figure}
\begin{center}
\includegraphics [width=1.\columnwidth]{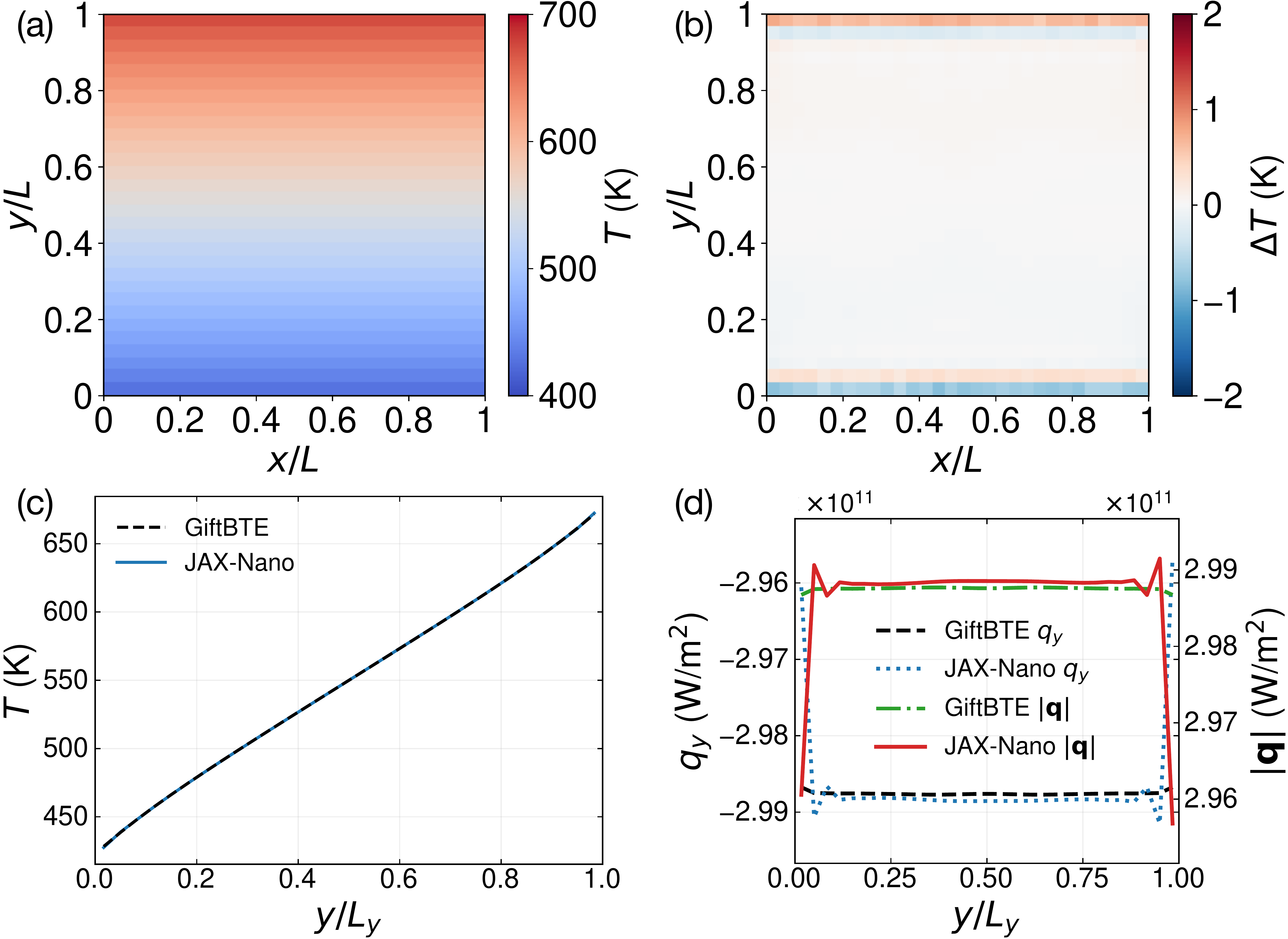}
\end{center}
\caption{Validation of JAX-Nano against GiftBTE for three-dimensional
phonon transport in a silicon cube with thermalizing boundaries
at $y=0$ and $y=L_y$ and specular reflection at the remaining
four faces.
(a) Temperature distribution predicted by JAX-Nano on the central
$x$--$y$ plane ($z=L/2$).
(b) Temperature difference,
$\Delta T = T_{\mathrm{JAX\text{-}Nano}} -
T_{\mathrm{GiftBTE}}$,
on the same plane.
(c) Comparison of temperature profiles along the vertical
centerline ($x=z=L/2$).
(d) Comparison of the signed heat-flux component $q_y$
(left axis) and heat-flux magnitude $|\mathbf{q}|$
(right axis) along the same centerline.
The results demonstrate close agreement between JAX-Nano
and GiftBTE in both temperature and heat flux, with small
discrepancies near the boundaries.}
\label{fig:3D_cube_T_slice_specular}
\end{figure} 

To further evaluate the treatment of phonon-boundary
interactions in JAX-Nano, steady-state transport in a three-dimensional silicon cube with
specular reflection at the four lateral faces is
investigated using a structured mesh containing
$30^3=27,000$ computational cells,
with $64$ discrete phonon propagation directions
and a single spectral band. 
The thermalizing boundary conditions
at the bottom and top surfaces remain unchanged
from the diffuse-reflection case, allowing the
influence of the lateral boundary treatment to
be examined under identical thermal driving
conditions.

Figure~\ref{fig:3D_cube_T_slice_specular}
compares the temperature and heat-flux predictions
obtained using JAX-Nano and GiftBTE.
As shown in Fig.~\ref{fig:3D_cube_T_slice_specular}(a),
the temperature varies predominantly along the
$y$ direction, with negligible variation in the
lateral direction. The temperature difference
between JAX-Nano and GiftBTE remains small
throughout the interior, with localized deviations
near the thermalizing boundaries, as illustrated
in Fig.~\ref{fig:3D_cube_T_slice_specular}(b).
The centerline temperature profiles in
Fig.~\ref{fig:3D_cube_T_slice_specular}(c)
exhibit nearly indistinguishable predictions
from the two solvers.

Figure~\ref{fig:3D_cube_T_slice_specular}(d)
further compares the signed heat-flux component
$q_y$ and the heat-flux magnitude $|\mathbf{q}|$
along the vertical centerline.
Both quantities remain approximately uniform
throughout the interior, with localized variations
near the thermalizing boundaries.
The negative values of $q_y$ indicate heat
transport from the higher-temperature upper
surface toward the lower-temperature bottom
surface. Although small differences in heat-flux
magnitude are observed between JAX-Nano and
GiftBTE, the overall profiles exhibit close
agreement. These results demonstrate the
capability of JAX-Nano to reproduce steady-state
phonon transport with specular boundary reflection
in three-dimensional geometries.

\subsubsection{Transient Transport with Diffuse and Specular Boundary Conditions}

\begin{figure}
\begin{center}
\includegraphics [width=1.\columnwidth]{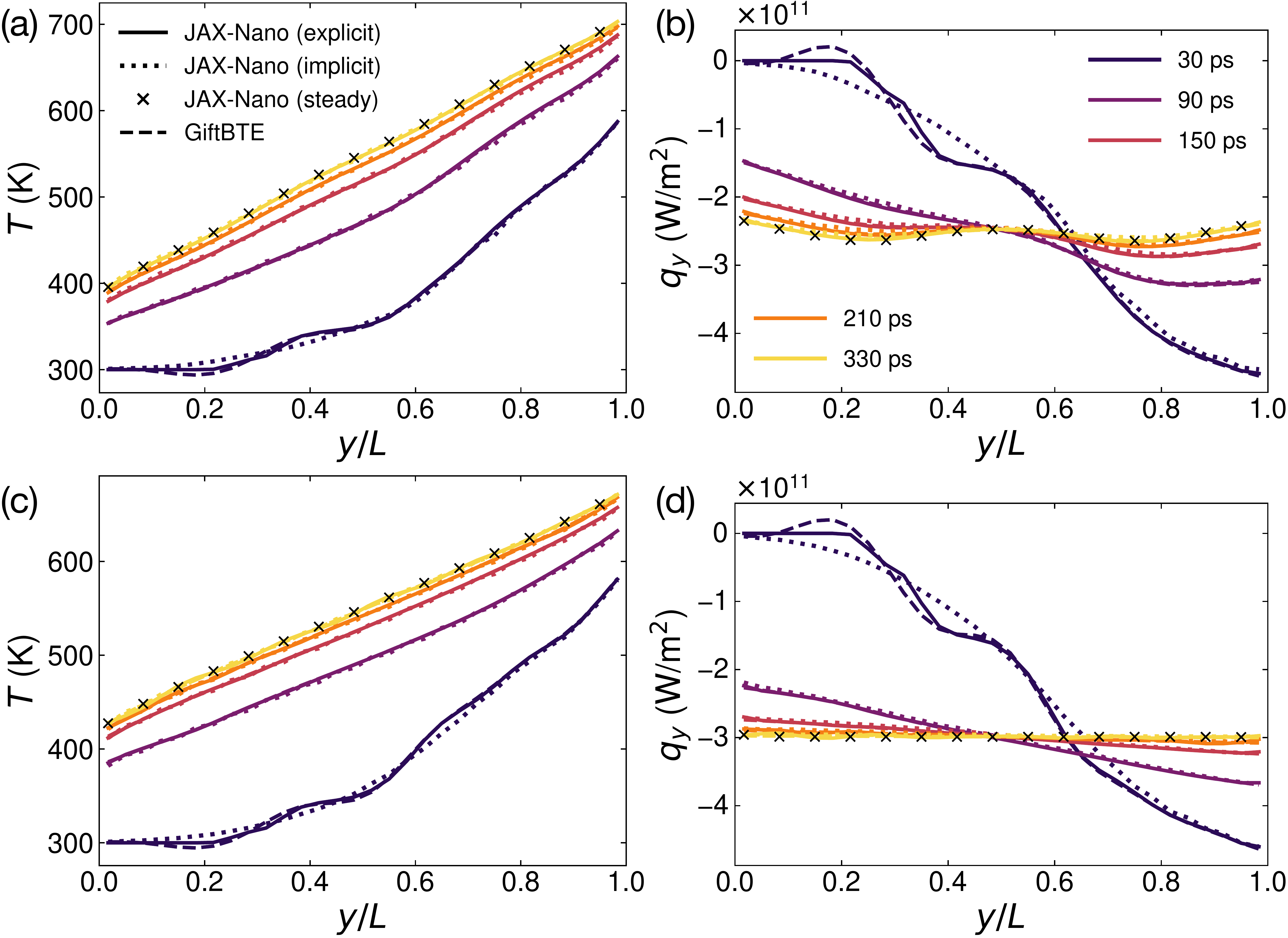}
\end{center}
\caption{Comparison of transient temperature and heat-flux distributions
predicted by JAX-Nano using explicit and implicit time integration
and by GiftBTE in a three-dimensional silicon cube.
(a) Temperature and (b) $y$-direction heat flux under diffuse
boundary conditions; (c) temperature and (d) $y$-direction heat
flux under specular boundary conditions.
The top ($y=L$) and bottom ($y=0$) surfaces are maintained at
$800~\mathrm{K}$ and $300~\mathrm{K}$, respectively, with a
domain length of $L=100~\mathrm{nm}$.
Profiles are extracted along the central line at
$t=30$, $90$, $150$, $210$, and $330~\mathrm{ps}$.
Solid, dotted, and dashed lines represent JAX-Nano explicit,
JAX-Nano implicit, and GiftBTE solutions, respectively,
while black crosses denote the JAX-Nano steady-state solutions.
All simulations employ $27{,}000$ spatial cells and
$64$ discrete phonon propagation directions.}
\label{fig:transient_cube_explicit_implicit}
\end{figure} 

To evaluate the transient transport capabilities of JAX-Nano,
phonon heat transport is simulated in a three-dimensional silicon cube with a side length of $L=100~\mathrm{nm}$. The top
($y=L$) and bottom ($y=0$) surfaces are maintained at
$800~\mathrm{K}$ and $300~\mathrm{K}$, respectively.
Both diffuse and specular phonon boundary-scattering conditions
are considered. The computational domain is discretized into
$27{,}000$ spatial cells, with $64$ discrete phonon propagation
directions and a single spectral band. The explicit and implicit time-integration schemes
employ time-step sizes of $\Delta t=3.0\times10^{-13}~\mathrm{s}$
and $\Delta t=1.5\times10^{-12}~\mathrm{s}$, respectively.
The implicit scheme therefore permits a time step five times
larger than that used in the explicit simulations.

Figure~\ref{fig:transient_cube_explicit_implicit} compares
the transient temperature and $y$-direction heat-flux
distributions obtained using the explicit and implicit
formulations of JAX-Nano with the corresponding GiftBTE
solutions obtained using explicit time integration. Profiles are extracted along the central line
at $t=30$, $90$, $150$, $210$, and $330~\mathrm{ps}$.
Figures~\ref{fig:transient_cube_explicit_implicit}(a) and (b)
present the temperature and heat-flux distributions under
diffuse boundary conditions, while panels (c) and (d)
show the corresponding results under specular boundary
conditions.

For both boundary-scattering conditions, the explicit
and implicit JAX-Nano formulations exhibit overall
agreement with GiftBTE, reproducing the transient evolution
of the temperature and heat-flux distributions.
The two time-integration schemes capture similar transport
behavior despite their different time-step sizes.
Differences between the numerical solutions are more
noticeable at earlier times, particularly in the heat-flux
profiles. As the simulation progresses, the temperature
and heat-flux distributions evolve toward the corresponding
steady-state solutions, indicated by black crosses.
These comparisons demonstrate the capability of JAX-Nano
to capture transient phonon transport under different
boundary-scattering conditions using both explicit and
implicit time integration.

\subsubsection{Nongray Phonon Transport with Multiple Spectral Bands}

\begin{figure}
\begin{center}
\includegraphics [width=1.\columnwidth]{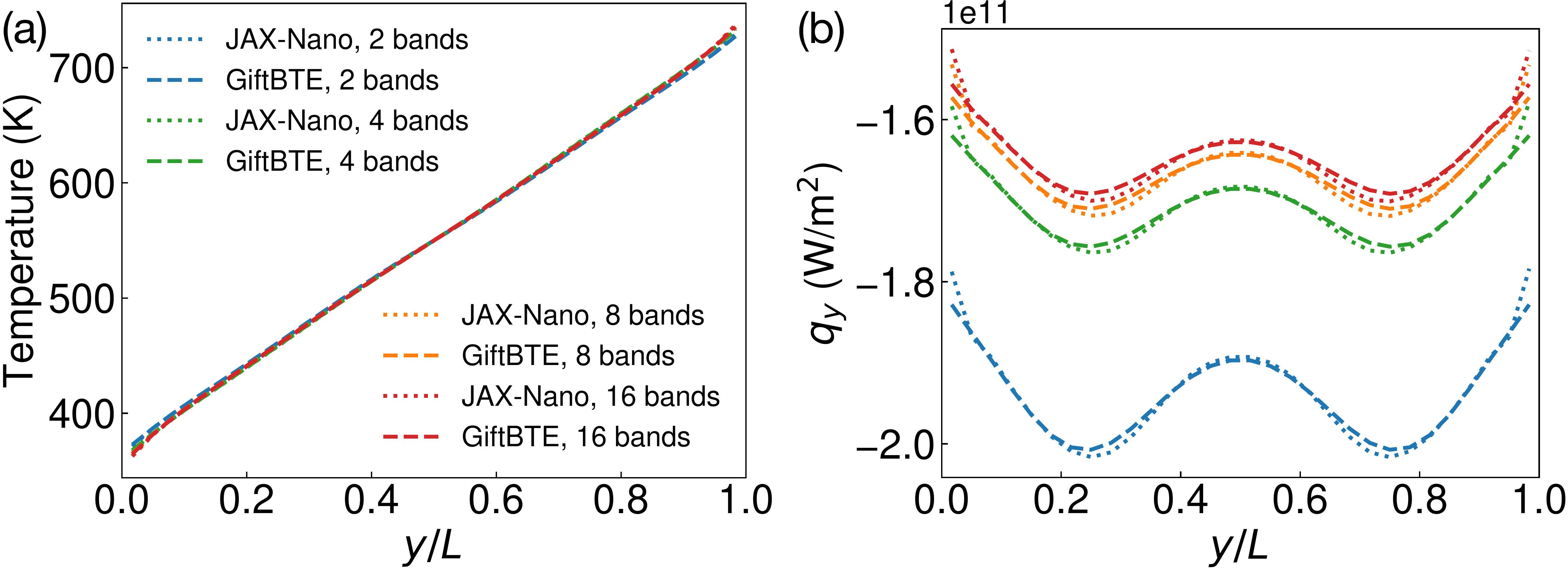}
\end{center}
\caption{Comparison of steady-state nongray phonon transport
predictions obtained using JAX-Nano and GiftBTE for
different spectral resolutions ($N_b=2,4,8,16$, where $N_b$ denotes the number
of spectral bands).
(a) Temperature and (b) $y$-direction heat-flux
distributions along the centerline of a three-dimensional
silicon cube with a side length of $L=100~\mathrm{nm}$.
The bottom ($y=0$) and top ($y=L$) surfaces are maintained
at $300~\mathrm{K}$ and $800~\mathrm{K}$, respectively,
while diffuse phonon reflection is imposed on the four
lateral faces. All simulations employ $27{,}000$
spatial cells and $64$ discrete phonon propagation
directions. Dotted and dashed lines represent JAX-Nano
and GiftBTE predictions, respectively.}
\label{fig:nongray_T_flux}
\end{figure}

To validate the capability of JAX-Nano to simulate nongray phonon
transport, steady-state simulations are performed with different
numbers of spectral bands, $N_b=2,4,8,$ and $16$. The computational
domain is a three-dimensional silicon cube with a side length of
$L=100~\mathrm{nm}$. The bottom and top surfaces are maintained at
$300~\mathrm{K}$ and $800~\mathrm{K}$, respectively, while diffuse
phonon reflection is imposed on the four lateral surfaces. The
computational domain is discretized into $27{,}000$ spatial cells,
with $64$ discrete phonon propagation directions. The JAX-Nano
results are compared with those obtained using GiftBTE for the
same spectral, spatial, and angular resolutions.

Figure~\ref{fig:nongray_T_flux} compares the steady-state
temperature and $y$-direction heat-flux distributions
along the centerline of the silicon cube for different
numbers of spectral bands. As shown in
Fig.~\ref{fig:nongray_T_flux}(a), the temperature
distributions exhibit weak sensitivity to spectral
resolution, with nearly overlapping profiles obtained
using 2--16 spectral bands. The JAX-Nano predictions
closely agree with the corresponding GiftBTE results
throughout the domain, indicating consistent temperature
predictions across the investigated spectral
discretizations.

In contrast, the heat-flux distributions shown in
Fig.~\ref{fig:nongray_T_flux}(b) exhibit a more
pronounced dependence on spectral resolution.
The two-band configuration produces a substantially
different heat-flux profile compared with the
higher-band configurations. As the number of spectral
bands increases, the differences between successive
heat-flux predictions generally decrease, with the
8- and 16-band solutions exhibiting relatively close
agreement. This behavior suggests a trend toward
spectral convergence of the heat-flux predictions. The stronger
dependence of heat flux on spectral resolution reflects
its sensitivity to the representation of phonon
transport properties across different spectral bands.
Despite these variations, JAX-Nano closely reproduces
the corresponding GiftBTE heat-flux distributions
at each spectral resolution, including their spatial
variations along the centerline.

Overall, these results demonstrate that JAX-Nano
consistently reproduces the steady-state nongray
phonon transport predictions of GiftBTE across
the investigated spectral resolutions. The observed
reduction in heat-flux differences with increasing
spectral resolution further suggests convergence
toward a spectrally resolved solution. Together with
the preceding single-band validation cases, these
findings support the accuracy and applicability of
JAX-Nano for three-dimensional nongray phonon
transport involving multiple spectral bands.

\subsubsection{Angular Resolution and Convergence}
\label{sec:angular_convergence}

\begin{figure}
\begin{center}
\includegraphics [width=1.\columnwidth]{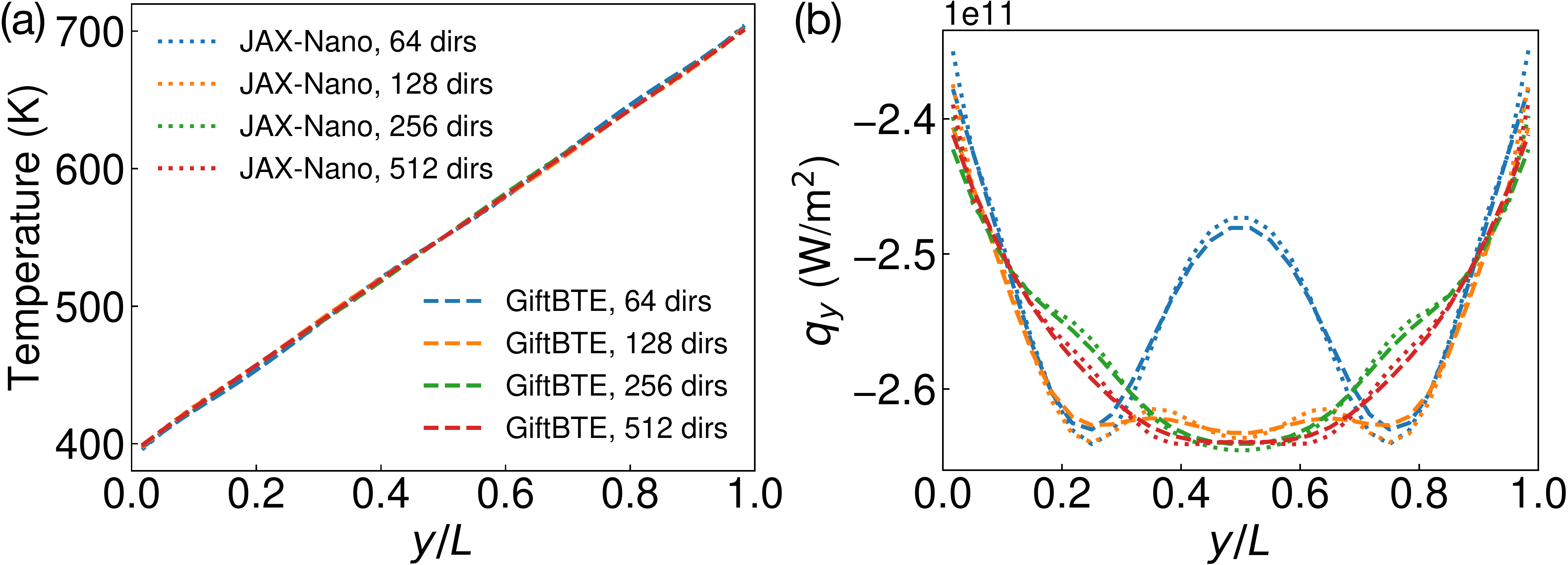}
\end{center}
\caption{Angular-resolution dependence of steady-state phonon transport
predictions obtained using JAX-Nano and GiftBTE.
(a) Temperature and (b) $y$-direction heat flux along the centerline
of a three-dimensional silicon cube with a side length of
$L=100$~nm.
Four angular resolutions are considered:
$N_\Omega=64$, 128, 256, and 512, where $N_\Omega$ denotes
the number of discrete phonon propagation directions.
The bottom ($y=0$) and top ($y=L$) surfaces are maintained
at 300 and 800~K, respectively, while diffuse phonon reflection
is imposed on the four lateral surfaces.
All simulations employ 27,000 spatial cells and a single
spectral band.
Dotted and dashed lines represent JAX-Nano and GiftBTE
predictions, respectively.}
\label{fig:gray_T_flux_dirs}
\end{figure}

The influence of angular discretization on the accuracy and
convergence of JAX-Nano is investigated by varying the number
of discrete phonon propagation directions. Four angular
resolutions, $N_\Omega=64$, 128, 256, and 512, are considered
for the three-dimensional silicon cube described above.
All simulations employ 27,000 spatial cells, a single
spectral band, and second-order upwind reconstruction with
the Barth--Jespersen limiter in JAX-Nano.
The bottom and top surfaces are maintained at 300 and
800~K, respectively, while diffuse phonon reflection is
imposed on the four lateral surfaces. The corresponding
GiftBTE solutions are included for comparison to assess
the consistency between the two numerical implementations.

Figure~\ref{fig:gray_T_flux_dirs} compares the steady-state
temperature and $y$-direction heat-flux distributions along
the centerline of the cubic domain for different angular
resolutions. As shown in Fig.~\ref{fig:gray_T_flux_dirs}(a),
the temperature profiles exhibit weak sensitivity to angular
resolution. The predictions obtained using 64--512 directions
nearly overlap, and JAX-Nano closely reproduces the
corresponding GiftBTE results across the entire domain.
This indicates that the temperature distribution is
relatively insensitive to angular discretization over
the investigated range.

In contrast, the heat-flux distributions in
Fig.~\ref{fig:gray_T_flux_dirs}(b) exhibit a more pronounced
dependence on angular resolution. At $N_\Omega=64$, the
$y$-direction heat flux displays substantial spatial
variation, including a relatively pronounced local maximum
near the center of the domain. Increasing the angular
resolution to 128 directions significantly modifies the
heat-flux profile, while the predictions obtained using
256 and 512 directions become progressively closer.
These results indicate that the heat flux requires finer
angular discretization than temperature to achieve
comparable numerical convergence.

The different convergence behavior of temperature and
heat flux can be attributed to their distinct dependence
on the angular phonon distribution. Temperature is
determined by an angularly averaged energy moment,
whereas heat flux involves a directionally weighted
moment of the phonon distribution. Consequently, angular
discretization errors that have a relatively small
influence on temperature may produce more noticeable
differences in heat-flux predictions. This sensitivity
is particularly relevant in the presence of diffuse
boundary scattering, which redistributes phonon energy
among propagation directions.

Despite the variations associated with angular resolution,
JAX-Nano and GiftBTE predict closely matching heat-flux
profiles at each corresponding angular discretization.
The agreement between the two solvers, together with
the decreasing differences between successive angular
resolutions, demonstrates the consistency of the
JAX-Nano angular discretization and its convergence
behavior for three-dimensional steady-state phonon
transport. The results also emphasize the importance
of assessing heat-flux convergence independently of
temperature convergence when selecting the angular
resolution for phonon BTE simulations.

\section{Three-Dimensional FinFET Thermal Transport}

\subsection{Single-Fin Transistor}

\begin{figure}
\begin{center}
\includegraphics [width=1.\columnwidth]{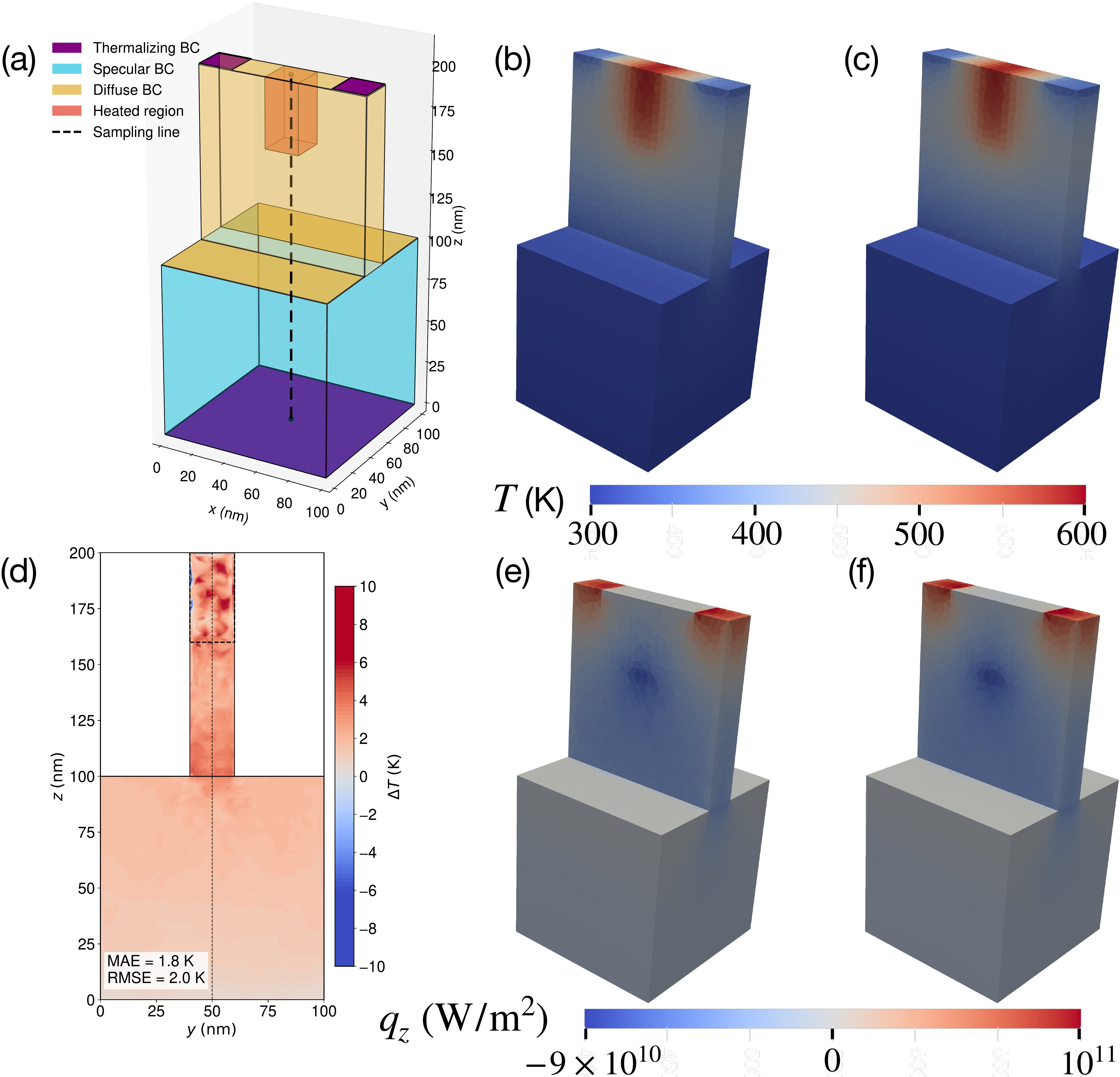}
\end{center}
\caption{
Three-dimensional phonon Boltzmann transport simulation in a silicon FinFET geometry and comparison between JAX-Nano and GiftBTE.
(a) Computational geometry, boundary conditions, heated region, and vertical sampling line. 
(b,c) Three-dimensional temperature distributions predicted by JAX-Nano and GiftBTE, respectively.
(d) Temperature difference between the two solvers on the central $yz$ plane, with a mean absolute error (MAE) of 1.8 K and a root-mean-square error (RMSE) of 2.0 K. 
The vertical dashed line indicates the fin centerline
($y=50$ nm), while the horizontal dashed line marks the
lower boundary of the heated region ($z=160$ nm).
(e,f) Three-dimensional vertical heat-flux distributions, $q_z$, predicted by JAX-Nano and GiftBTE, respectively.
}
\label{fig:finfet_3d_validation}
\end{figure}

\begin{figure}
\begin{center}
\includegraphics [width=.92\columnwidth]{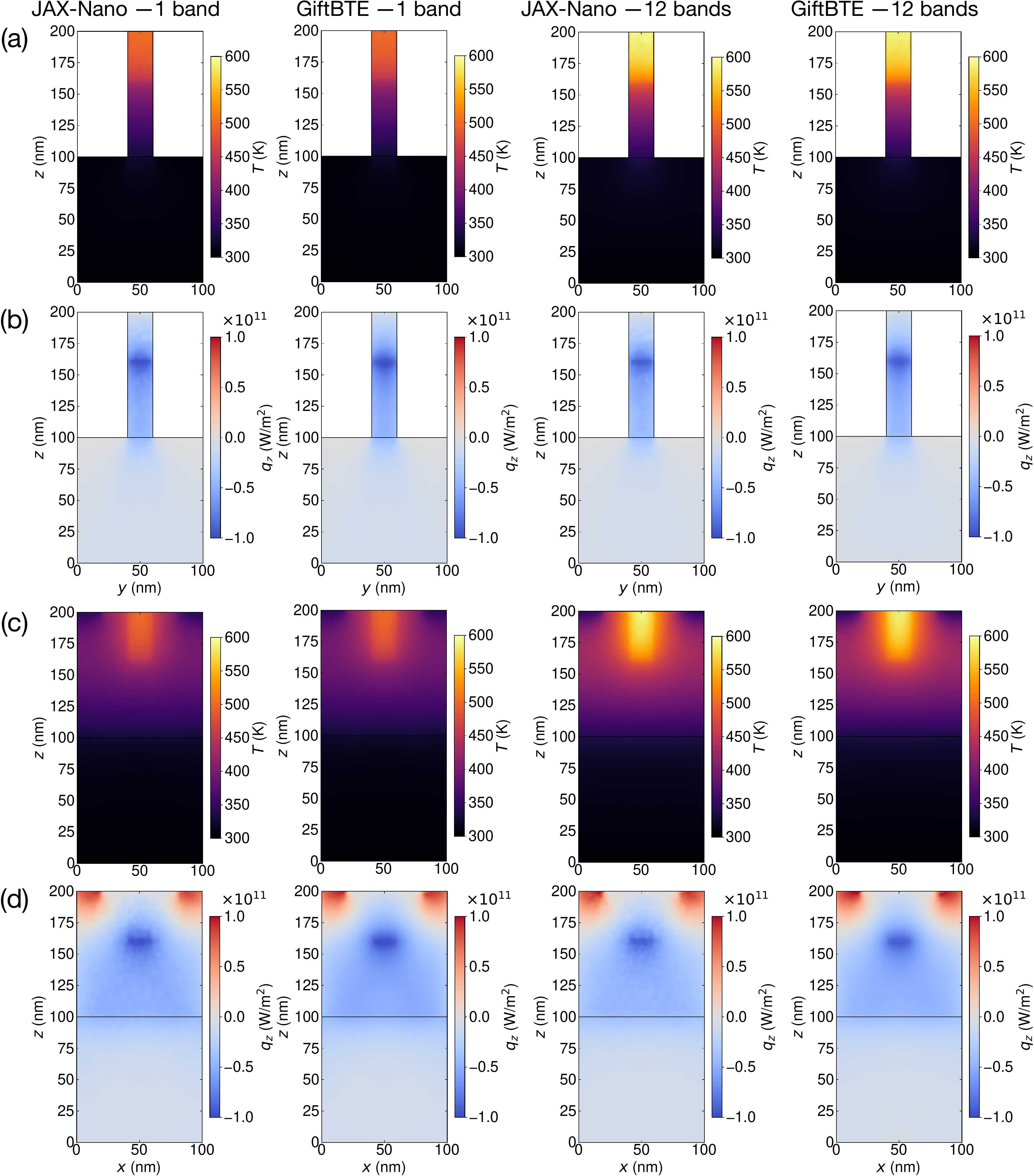}
\end{center}
\caption{Comparison of temperature and vertical heat-flux distributions
predicted by JAX-Nano and GiftBTE for three-dimensional phonon
Boltzmann transport in a silicon FinFET geometry using an unstructured
mesh with 45,804 cells and 128 discrete directions.
Results are presented for single-band and 12-band phonon models.
The four rows show
(a) temperature $T$ on the central $yz$ plane ($x=50$ nm),
(b) vertical heat flux $q_z$ on the same plane,
(c) temperature $T$ on the central $xz$ plane ($y=50$ nm), and
(d) vertical heat flux $q_z$ on the same plane.
}
\label{fig:xz_yz_T_flux}
\end{figure}

\begin{figure}
\begin{center}
\includegraphics [width=1.\columnwidth]{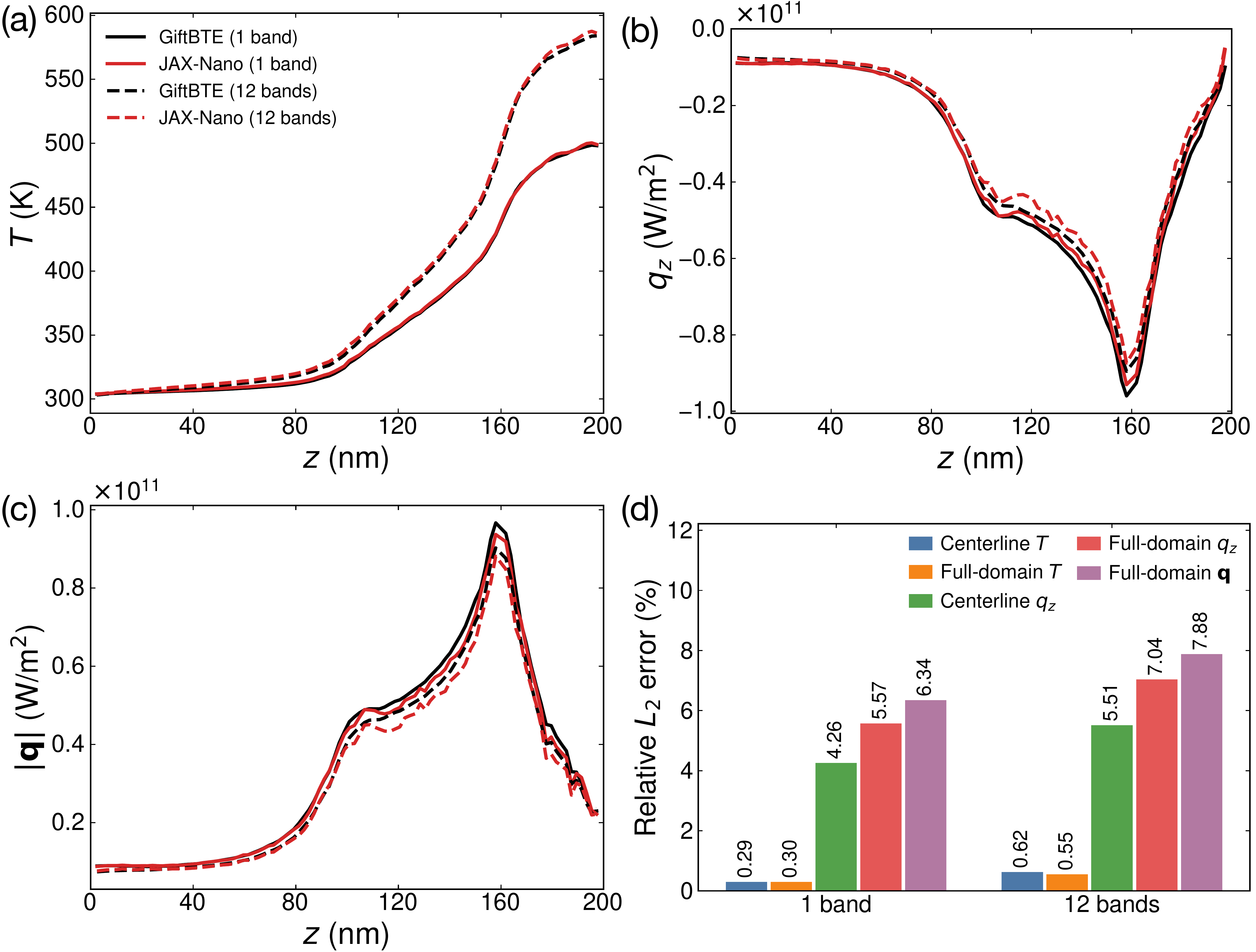}
\end{center}
\caption{
Comparison of JAX-Nano and GiftBTE predictions for phonon
Boltzmann transport in a three-dimensional silicon FinFET geometry
using an unstructured mesh with 45,804 cells and 128
discrete directions.
Results are presented for single-band and 12-band phonon models.
(a) Temperature profiles $T$ along the vertical centerline.
(b) Vertical heat-flux profiles $q_z$ along the same centerline.
(c) Heat-flux magnitude profiles $|\mathbf{q}|$ along the centerline.
Solid and dashed lines represent the single-band and 12-band
models, respectively, while black and red lines denote GiftBTE
and JAX-Nano predictions.
(d) Relative $L_2$ errors between JAX-Nano and GiftBTE
for centerline temperature, full-domain temperature,
centerline vertical heat flux, full-domain vertical heat flux,
and full-domain heat-flux vector.
}
\label{fig:finfet_temperature_heatflux_validation}
\end{figure}

To evaluate the capability of JAX-Nano to simulate phonon transport
in complex three-dimensional geometries, we consider a silicon
single-fin field-effect transistor (FinFET), as illustrated in
Fig.~\ref{fig:finfet_3d_validation}(a).
The computational domain consists of a rectangular substrate
supporting a vertically oriented silicon fin with a localized
heat-generation region near its upper portion.
The boundary conditions and heat-source strength follow those
adopted by Hu et al.~\cite{Hu_2024a}.
Thermalizing boundary conditions are imposed on surfaces in
contact with the metal electrodes (source and drain) and on
substrate boundaries away from the heated region, with the
boundary temperature fixed at $T=300$ K to represent a
room-temperature environment.
Diffusely reflecting boundary conditions are prescribed on
surfaces in contact with the dielectric layer, while the
remaining boundaries are treated as specularly reflecting
to represent symmetry between neighboring devices.
A uniform total volumetric heat-generation rate of
$\sum_{b=1}^{N_b}\dot{Q}_b
=10^{19}\ \mathrm{W\,m^{-3}}$
is applied within the localized hot spot
$\Omega_{\mathrm{hot}}=
[40,60]\times[40,60]\times[160,200]\ \mathrm{nm}^3$,
where $(x,y,z)\in\Omega_{\mathrm{hot}}$.
The computational domain is discretized using an unstructured tetrahedral
mesh containing 45,804 cells, and the angular domain is
represented by 128 discrete directions.
Both single-band and 12-band phonon transport models are
considered to examine the influence of spectral resolution
on the predicted thermal response.

Figure~\ref{fig:finfet_3d_validation}(b,c) compares the
three-dimensional temperature distributions predicted by
JAX-Nano and GiftBTE.
Both solvers capture a localized high-temperature region
near the upper center of the silicon fin, resulting from
the prescribed volumetric heat generation.
The temperature decreases toward the upper corners of the
fin, consistent with the cooling effect of the adjacent
thermalizing boundaries maintained at 300 K.
A substantial temperature gradient is also observed along
the fin height, with the underlying substrate remaining
relatively close to the prescribed boundary temperature.
The corresponding vertical heat-flux distributions,
shown in Fig.~\ref{fig:finfet_3d_validation}(e,f),
exhibit similar spatial patterns, indicating consistent
predictions of heat transport from the heated region
toward the substrate.

To quantify the agreement between the two numerical solutions,
Fig.~\ref{fig:finfet_3d_validation}(d) presents the temperature
difference on the central $yz$ plane.
The difference remains relatively small compared with the overall
temperature variation, with a mean absolute error (MAE) of
1.8 K and a root-mean-square error (RMSE) of 2.0 K.
These results indicate that JAX-Nano reproduces the temperature
distribution predicted by GiftBTE with good accuracy for the
considered three-dimensional FinFET configuration.

A more detailed comparison of the spatial temperature and
heat-flux distributions is presented in
Fig.~\ref{fig:xz_yz_T_flux}.
The first two rows show the temperature and vertical heat-flux
distributions on the central $yz$ plane ($x=50$ nm), while
the last two rows show the corresponding fields on the central
$xz$ plane ($y=50$ nm).
For both single-band and 12-band models, JAX-Nano and GiftBTE
produce closely matching spatial distributions.
The localized volumetric heat generation produces a pronounced
high-temperature region near the upper center of the silicon fin.
In contrast, substantially lower temperatures are observed
near the upper corners of the fin, consistent with the
thermalizing boundary conditions maintained at 300 K on the
adjacent surfaces.
This localized cooling, together with the internal heat
generation, results in strong temperature gradients near
the upper portion of the fin.
The temperature gradually decreases toward the underlying
substrate, which remains relatively close to the prescribed
boundary temperature.
Compared with the single-band model, the 12-band model
predicts a higher peak temperature, highlighting the
influence of spectral phonon properties on thermal transport
in nanoscale structures.

The corresponding vertical heat-flux distributions reveal
the spatial redistribution of thermal energy within the
FinFET geometry.
Negative values of $q_z$ dominate within the fin, indicating
net downward heat transport toward the substrate.
Localized variations in the magnitude and direction of
$q_z$ are observed near the heated region and the upper
fin boundaries, reflecting the combined effects of
volumetric heat generation, boundary scattering, and
three-dimensional phonon transport.
These variations are particularly pronounced near regions
of strong temperature gradients.
Despite differences in the spectral representations,
JAX-Nano reproduces the principal heat-flux features
predicted by GiftBTE for both band models, including
the spatial distribution of downward heat transport
and localized heat-flux variations near the upper fin.

Figure~\ref{fig:finfet_temperature_heatflux_validation}(a--c)
provides a quantitative comparison of temperature, vertical
heat flux, and heat-flux magnitude along the vertical centerline
of the silicon FinFET.
The profiles are obtained by spatial averaging within a
cylindrical sampling region centered at $x=y=50$ nm,
followed by interpolation along the $z$ direction.
The temperature profiles predicted by JAX-Nano and GiftBTE
show close agreement for both single-band and 12-band models.
The temperature increases from the substrate toward the
heated region in the upper portion of the fin, reflecting
the localized volumetric heat generation and the
redistribution of thermal energy through phonon transport.
The 12-band model predicts higher temperatures than the
single-band model, consistent with the spatial distributions
shown in Fig.~\ref{fig:xz_yz_T_flux}.

The vertical heat-flux profiles in
Fig.~\ref{fig:finfet_temperature_heatflux_validation}(b)
exhibit pronounced negative values within the fin, with
a peak magnitude near $z=160$ nm, approximately corresponding
to the lower boundary of the heated region.
These negative values indicate downward heat transport
toward the substrate.
The heat-flux magnitude profiles in
Fig.~\ref{fig:finfet_temperature_heatflux_validation}(c)
show corresponding variations along the centerline.
Both solvers reproduce the principal features of these
profiles, including the locations of strong heat flux
associated with the heated region.
The discrepancies are more pronounced in the heat-flux
predictions than in the temperature profiles, particularly
near regions of strong spatial variation.

The relative $L_2$ errors between JAX-Nano and GiftBTE
are summarized in
Fig.~\ref{fig:finfet_temperature_heatflux_validation}(d).
For the single-band model, the centerline and full-domain
temperature errors are 0.29\% and 0.30\%, respectively,
compared with 0.62\% and 0.55\% for the 12-band model.
The corresponding errors for centerline $q_z$,
full-domain $q_z$, and the full-domain heat-flux vector
are 4.26\%, 5.57\%, and 6.34\% for the single-band model,
and 5.51\%, 7.04\%, and 7.88\% for the 12-band model.
The larger heat-flux discrepancies indicate that the
heat-flux field provides a more stringent assessment
of agreement between the two solvers than the
temperature field.
Nevertheless, the overall agreement demonstrates that
JAX-Nano reproduces the principal thermal transport
characteristics predicted by GiftBTE for both
spectral models.

Overall, this validation demonstrates that JAX-Nano can reproduce
both the macroscopic temperature response and the spatially
resolved heat-flux characteristics of a three-dimensional silicon
FinFET using an unstructured computational mesh.
The close agreement with GiftBTE for both single-band and
12-band models supports the applicability of the proposed
GPU-accelerated phonon BTE framework to complex nanoscale
thermal transport problems.

\subsection{Double-Fin Transistor}

\begin{figure}
\begin{center}
\includegraphics[width=1.\columnwidth]{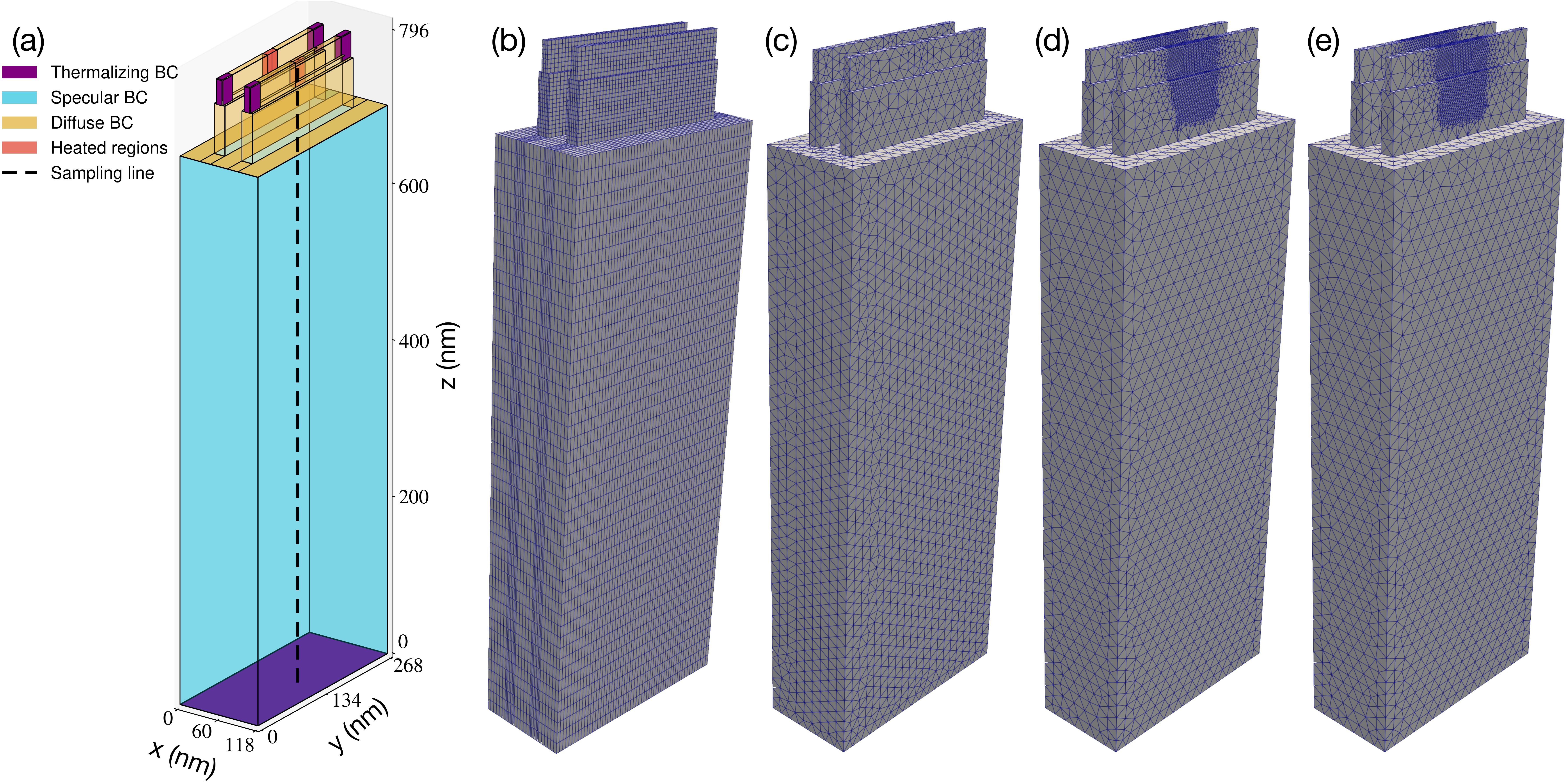}
\end{center}
\caption{
Geometry, boundary conditions, and computational meshes
for the two-fin silicon FinFET configuration.
(a) Three-dimensional device geometry showing the
thermalizing, specular, and diffuse boundary conditions,
localized volumetric heat-generation regions, and
the sampling line used for quantitative comparisons.
(b) Structured hexahedral mesh used by JAX-Nano
(99,694 cells).
(c) Coarse unstructured tetrahedral mesh used by JAX-Nano
(33,685 cells).
(d) Fine unstructured tetrahedral mesh used by JAX-Nano
(54,839 cells).
(e) Unstructured tetrahedral mesh used by GiftBTE
(55,040 cells).
The computational meshes resolve the same device
geometry while differing in cell topology and
spatial resolution.
}
\label{fig:twofinfet_geometry_mesh}
\end{figure}

\begin{figure}
\begin{center}
\includegraphics[width=1.\columnwidth]{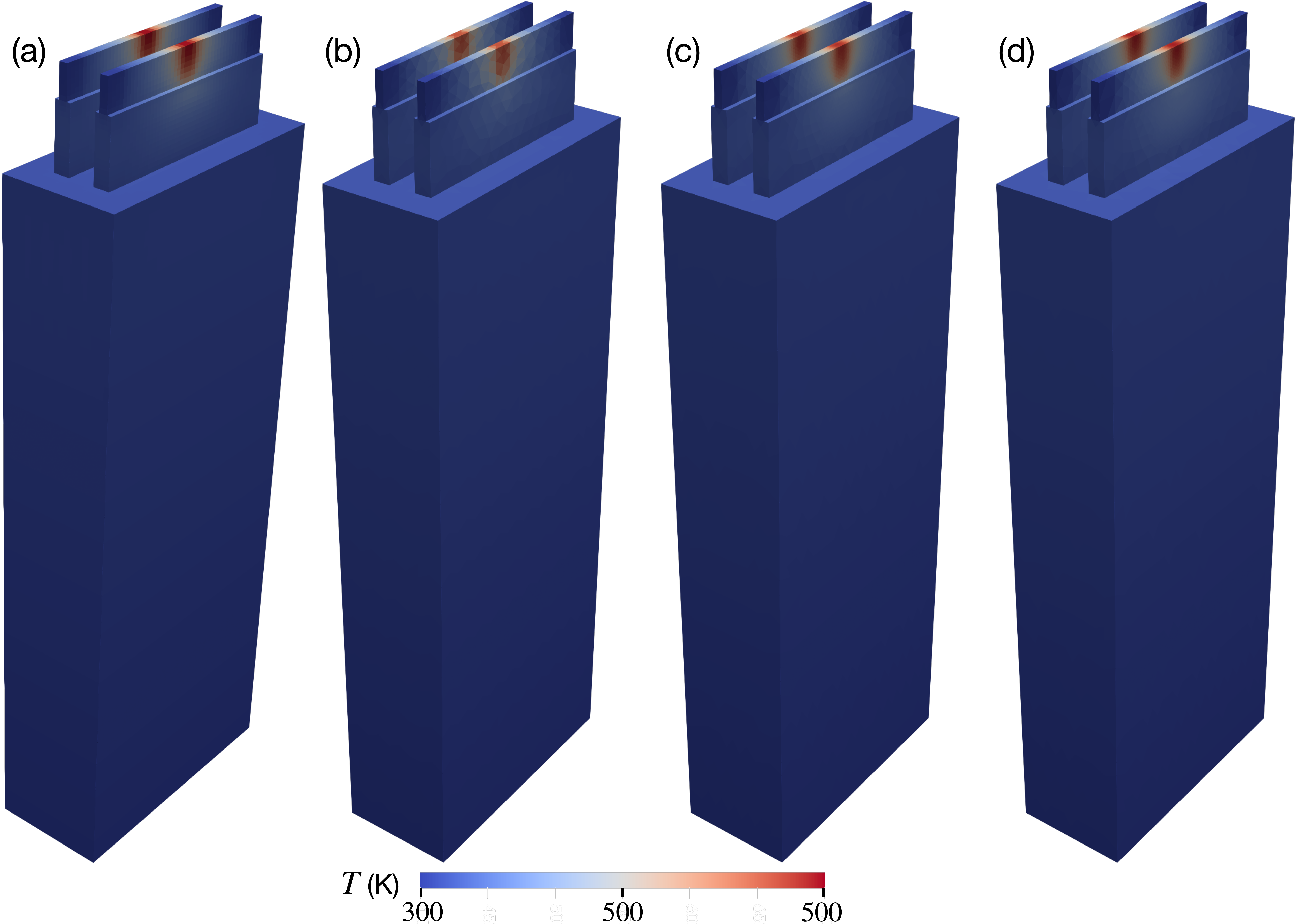}
\end{center}
\caption{
Comparison of steady-state temperature distributions
in the two-fin silicon FinFET configuration.
(a) JAX-Nano solution obtained using the structured
hexahedral mesh (99,694 cells).
(b) JAX-Nano solution obtained using the coarse
unstructured tetrahedral mesh (33,685 cells).
(c) JAX-Nano solution obtained using the fine
unstructured tetrahedral mesh (54,839 cells).
(d) GiftBTE solution obtained using the unstructured
tetrahedral mesh (55,040 cells).
All simulations employ the same silicon device geometry
and thermal boundary conditions.
The temperature fields are displayed using a common
color scale to facilitate direct comparison of the
spatial temperature distributions and localized
hot spots near the heated regions.
}
\label{fig:twofinfet_temperature}
\end{figure}

\begin{figure}
\begin{center}
\includegraphics[width=1.\columnwidth]{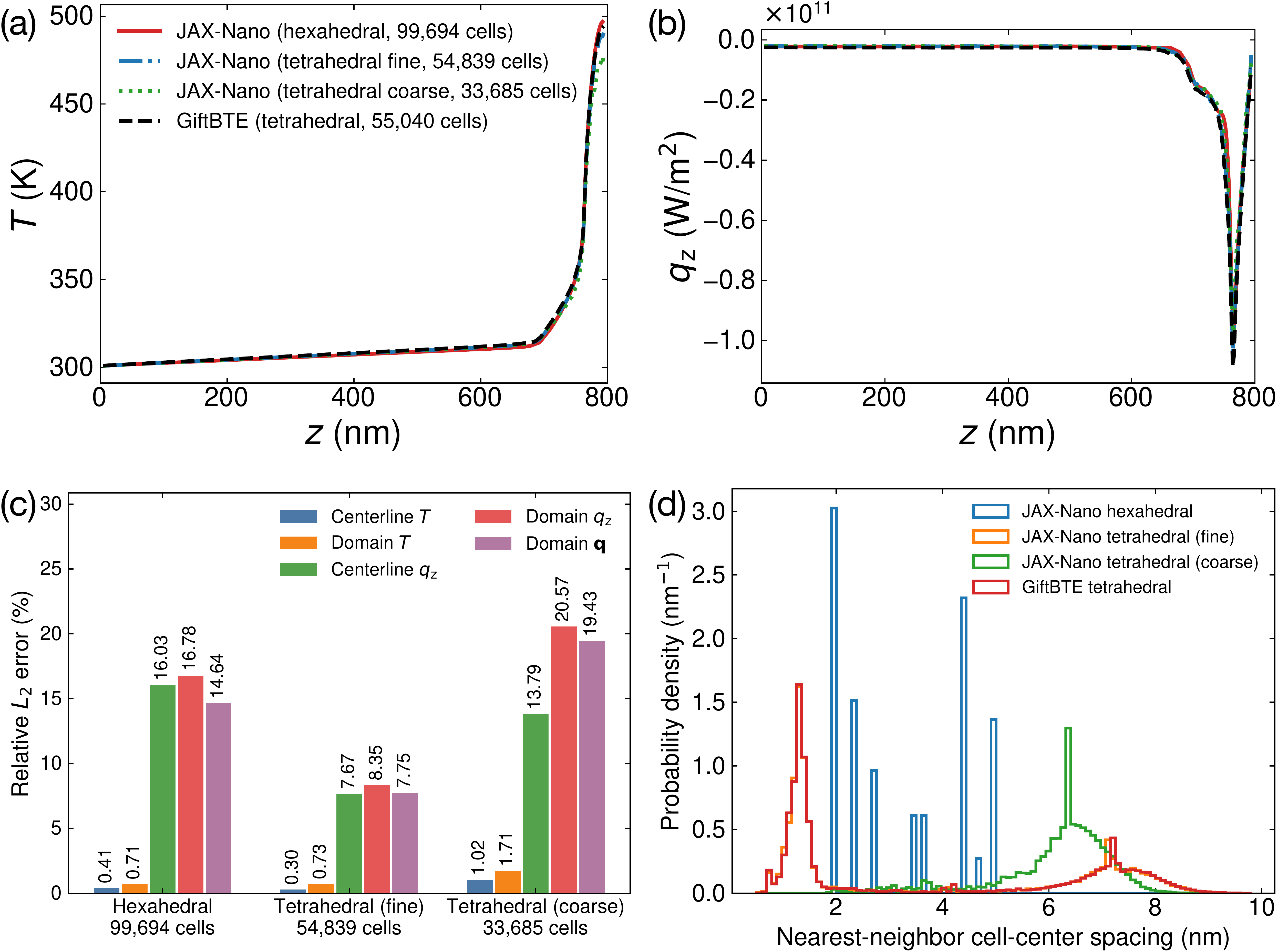}
\end{center}
\caption{
Steady-state validation of JAX-Nano against GiftBTE
for the two-fin silicon FinFET configuration using different
computational meshes.
(a) Temperature profiles along the sampling line
for the structured hexahedral, fine tetrahedral,
and coarse tetrahedral JAX-Nano meshes, compared
with the GiftBTE tetrahedral solution.
(b) Corresponding axial heat-flux profiles, $q_z$,
along the same sampling line.
(c) Relative $L_2$ errors in the centerline and
full-domain temperature, axial heat flux, and
full-domain heat-flux vector, using the GiftBTE
solution as the reference.
(d) Normalized distributions of nearest-neighbor
cell-center spacing for the four computational
meshes, illustrating differences in local spatial
resolution and mesh nonuniformity.
The fine tetrahedral meshes used by JAX-Nano and
GiftBTE exhibit closely matching spacing
distributions.
}
\label{fig:twofinfet_steady_validation}
\end{figure}

\begin{figure}
\begin{center}
\includegraphics[width=1.\columnwidth]{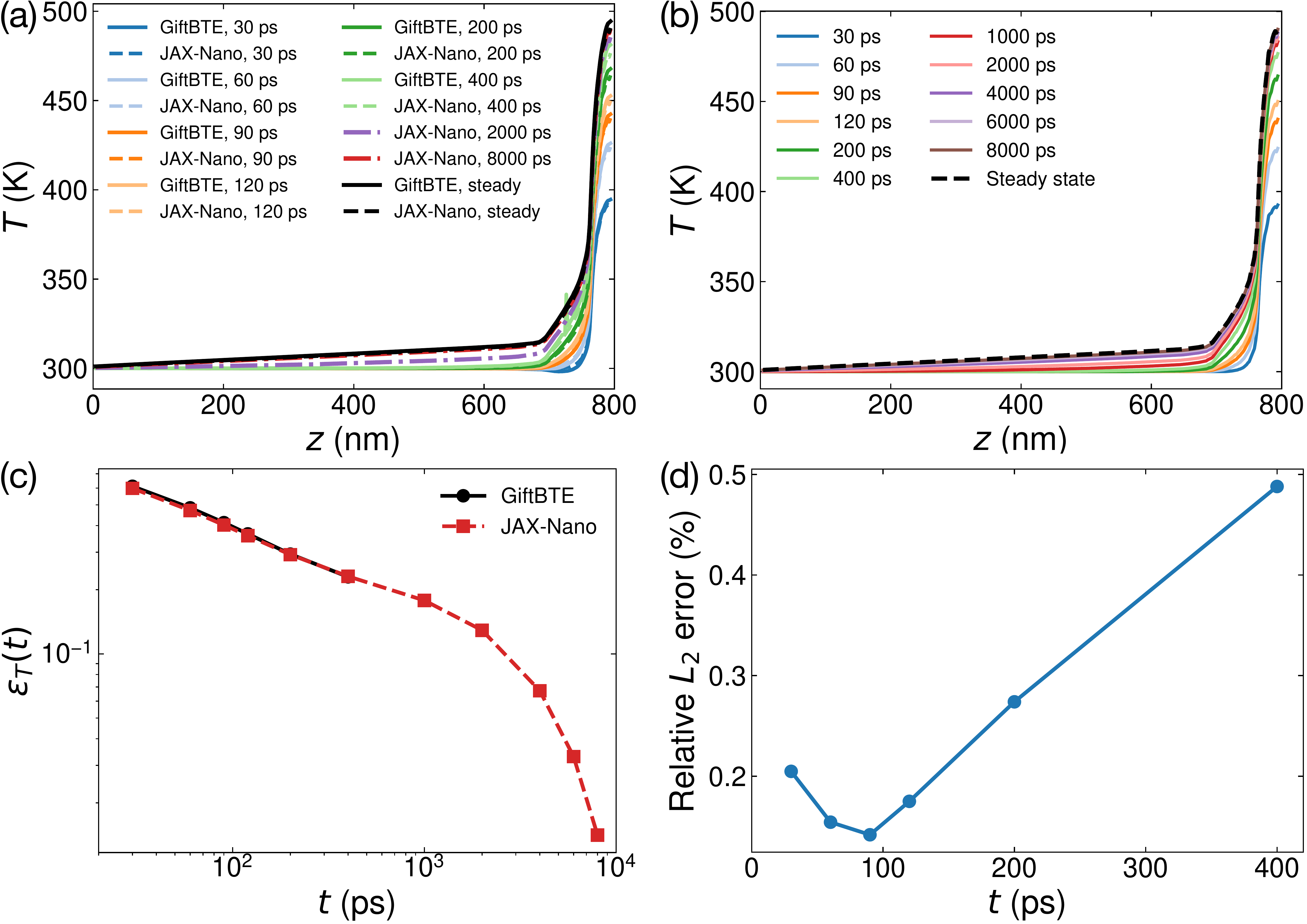}
\end{center}
\caption{
Transient thermal response and approach to steady state
in the two-fin silicon FinFET configuration.
(a) Comparison of temperature profiles predicted
by JAX-Nano and GiftBTE along the sampling line
at selected times from 30 to 400 ps, together
with additional JAX-Nano profiles at 2000 and
8000 ps and the corresponding steady-state solutions.
(b) Evolution of the JAX-Nano temperature profile
from 30 to 8000 ps, illustrating the progressive
development of the temperature field toward
steady state.
(c) Temporal evolution of the relative $L_2$
deviation, $\varepsilon_T(t)$, of each transient
temperature profile from its corresponding
steady-state solution, normalized by the
steady-state temperature rise above the
reference temperature of 300 K.
(d) Relative $L_2$ error between JAX-Nano and
GiftBTE transient temperature profiles at
corresponding simulation times, expressed
as a percentage.
The transient comparisons employ the fine
tetrahedral meshes containing 54,839 cells
for JAX-Nano and 55,040 cells for GiftBTE.
}
\label{fig:twofinfet_transient_validation}
\end{figure}

\begin{table}[t]
\centering
\caption{
Mesh characteristics of the structured and unstructured
discretizations employed in the two-fin FinFET simulations.
The local mesh spacing $h_i$ is defined as the Euclidean
distance between the center of cell $i$ and its nearest
neighboring cell center,
$h_i=\min_{j\ne i}\|\mathbf{x}_i-\mathbf{x}_j\|_2$,
where $\mathbf{x}_i$ denotes the centroid of cell $i$.
The quantities $h_{\min}$, $h_{\mathrm{med}}$,
$h_{\mathrm{mean}}$, $h_{95}$, and $h_{\max}$
represent the minimum, median, mean, 95th-percentile,
and maximum values of $h_i$, respectively.
These statistics characterize the distribution of
cell-center spacings across each computational mesh,
including the effects of nonuniform spatial refinement.
All spacing values are reported in nanometers.
}
\label{tab:twofinfet_mesh}
\begin{tabular}{lrrrrrr}
\hline
Mesh & Cells
& $h_{\min}$
& $h_{\mathrm{med}}$
& $h_{\mathrm{mean}}$
& $h_{95}$
& $h_{\max}$ \\
\hline
JAX-Nano hexahedral
& 99,694 & 2.000 & 2.667 & 3.247 & 5.000 & 5.000 \\

JAX-Nano tetrahedral (fine)
& 54,839 & 0.508 & 1.507 & 3.717 & 8.022 & 9.603 \\

JAX-Nano tetrahedral (coarse)
& 33,685 & 1.861 & 6.376 & 6.166 & 7.418 & 9.071 \\

GiftBTE tetrahedral
& 55,040 & 0.520 & 1.508 & 3.723 & 8.033 & 9.782 \\
\hline
\end{tabular}
\end{table}

The applicability of JAX-Nano to three-dimensional
nanoscale thermal transport is further investigated
using a two-fin silicon field-effect transistor
(FinFET) configuration. This benchmark involves
localized volumetric heat generation, multiple
phonon boundary-scattering conditions, and
computational meshes with different topologies
and spatial resolutions. Both steady-state and
transient phonon transport are considered, and
the numerical predictions of JAX-Nano are
compared with those obtained using GiftBTE.
All simulations employ a single phonon band
and 64 discrete angular directions.

Figure~\ref{fig:twofinfet_geometry_mesh}(a)
illustrates the three-dimensional device geometry,
boundary conditions, localized heat-generation
regions, and sampling line used for quantitative
comparisons. The silicon device consists of two
fins connected to a common substrate.
A uniform volumetric heat-generation rate of
$10^{19}\ \mathrm{W\,m^{-3}}$ is applied within
the localized heated regions.
The bottom surface of the substrate and
the two corner regions of each fin are
maintained at a reference temperature of
$300~\mathrm{K}$ through thermalizing
boundary conditions. Specular reflection
is imposed on the lateral surfaces of
the substrate, while diffuse reflection
is prescribed on the remaining external
surfaces, as illustrated in the figure.
The resulting thermal transport problem involves
spatially nonuniform temperature and heat-flux
fields arising from the localized energy input
and phonon scattering at device boundaries.

To assess the influence of mesh topology and
spatial resolution, four computational meshes
are considered, as shown in
Figs.~\ref{fig:twofinfet_geometry_mesh}(b)--(e).
JAX-Nano is applied to a structured hexahedral
mesh containing 99,694 cells, a coarse
unstructured tetrahedral mesh containing
33,685 cells, and a fine unstructured
tetrahedral mesh containing 54,839 cells.
GiftBTE employs an unstructured tetrahedral
mesh containing 55,040 cells.
For the two fine unstructured tetrahedral meshes,
local refinement is introduced near the
heat-generation regions to better resolve
the associated temperature gradients and
heat-flux variations.
All four meshes represent the same physical
device geometry but differ in cell connectivity
and spatial resolution.

The mesh characteristics are summarized in
Table~\ref{tab:twofinfet_mesh}.
The local mesh spacing is defined using the
distance between each cell centroid and its
nearest neighboring cell centroid,
\begin{equation}
h_i =
\min_{j\neq i}
\left\|
\mathbf{x}_i-\mathbf{x}_j
\right\|_2,
\end{equation}
where $\mathbf{x}_i$ denotes the centroid
of cell $i$.
The minimum, median, mean, 95th-percentile,
and maximum values of $h_i$ are reported
to characterize the spatial resolution
and nonuniformity of each mesh.

The structured hexahedral mesh has a mean
cell-center spacing of 3.247 nm and a
maximum spacing of 5.000 nm.
The coarse tetrahedral mesh has a median
spacing of 6.376 nm, whereas the fine
tetrahedral mesh has a substantially smaller
median spacing of 1.507 nm.
The corresponding GiftBTE mesh has a median
spacing of 1.508 nm.
Furthermore, the mean spacings of the fine
JAX-Nano and GiftBTE tetrahedral meshes are
3.717 and 3.723 nm, respectively.
These closely matched spacing statistics
provide a basis for comparing the numerical
solutions while reducing differences
associated with mesh resolution.

Figure~\ref{fig:twofinfet_temperature}
compares the steady-state temperature
distributions obtained using the four
computational meshes.
All solutions exhibit similar overall
thermal behavior, with elevated temperatures
concentrated near the localized heat-generation
regions and decreasing toward the thermalizing
substrate boundary.
The JAX-Nano solutions obtained using both
structured and unstructured meshes reproduce
the principal features of the GiftBTE
temperature distribution.
The fine tetrahedral JAX-Nano solution
exhibits particularly close agreement
with the GiftBTE reference, whereas the
coarse tetrahedral solution shows more
noticeable local differences.
These comparisons demonstrate that
JAX-Nano can resolve the steady-state
temperature distribution in a complex
three-dimensional silicon device using
different mesh topologies.

A quantitative comparison of the steady-state
solutions is presented in
Fig.~\ref{fig:twofinfet_steady_validation}.
For the centerline comparisons, temperature
and axial heat-flux profiles are extracted
along the same physical sampling line,
extending from $z=0$ to $z=796$ nm.
The profiles are evaluated at 400 uniformly
distributed sampling locations using
piecewise-linear interpolation based on
the surrounding cell centers, without
nearest-neighbor extrapolation.
Only sampling locations with valid
interpolated values for all four datasets
are retained in the centerline error
calculations.

Figure~\ref{fig:twofinfet_steady_validation}(a)
shows the temperature profiles along the
sampling line.
The JAX-Nano solutions exhibit similar
spatial temperature variations to the
GiftBTE reference.
The fine tetrahedral mesh provides close
agreement throughout the sampling line,
while the structured hexahedral and
coarse tetrahedral meshes exhibit
somewhat larger local deviations.

The corresponding axial heat-flux
profiles, $q_z$, are compared in
Fig.~\ref{fig:twofinfet_steady_validation}(b).
Compared with the temperature profiles,
the heat-flux distributions exhibit
greater sensitivity to the underlying
mesh discretization.
This behavior reflects the dependence
of heat flux on the directional moment
of the phonon distribution function,
which may be more sensitive to spatial
discretization errors than the scalar
temperature field.
Among the three JAX-Nano meshes, the
fine tetrahedral discretization provides
the closest overall agreement with
the GiftBTE heat-flux distribution.

To quantify the numerical differences,
five relative $L_2$ error metrics are
considered: centerline temperature,
full-domain temperature, centerline
axial heat flux, full-domain axial
heat flux, and full-domain heat-flux
vector.

For the full-domain comparisons,
the GiftBTE temperature cell centers
are selected as the common evaluation
points.
The JAX-Nano temperature and heat-flux
fields are mapped onto these locations
using inverse-distance-weighted (IDW)
interpolation with eight nearest neighbors
and an inverse-distance exponent of two.
The GiftBTE heat-flux field is similarly
interpolated when its sampling coordinates
differ from those of the temperature field.
This procedure enables direct comparison
of the numerical solutions on a common
set of spatial locations despite
differences in mesh connectivity
and spatial resolution.

For a scalar or vector field $\phi$,
the relative discrete $L_2$ error is
defined as
\begin{equation}
\varepsilon_{\phi}
=
\frac{
\left(
\displaystyle\sum_{i\in\mathcal{V}}
\left\|
\phi_i^{\mathrm{JAX-Nano}}
-
\phi_i^{\mathrm{GiftBTE}}
\right\|_2^2
\right)^{1/2}
}{
\left(
\displaystyle\sum_{i\in\mathcal{V}}
\left\|
\phi_i^{\mathrm{GiftBTE}}
\right\|_2^2
\right)^{1/2}
}
\times 100\%,
\end{equation}
where $\mathcal{V}$ denotes the set
of valid common evaluation points.
For scalar quantities, the inner norm
reduces to the absolute value.
The errors are evaluated using discrete,
unweighted sampling-point values rather
than volume-integrated norms.

To avoid unsupported extrapolation,
an interpolated point is accepted only
when its distance from the nearest
source cell center does not exceed
three times the corresponding local
source-cell spacing.
Full-domain errors are reported only
when at least 95\% of the common
evaluation points satisfy the
interpolation validity criteria.
For heat-flux comparisons, the validity
conditions must be satisfied by both
the JAX-Nano and GiftBTE fields.
The centerline errors are evaluated
separately using the corresponding
linearly interpolated profiles.

Figure~\ref{fig:twofinfet_steady_validation}(c)
summarizes the relative $L_2$ errors.
For the fine tetrahedral mesh, the
centerline and full-domain temperature
errors are approximately 0.30\% and
0.73\%, respectively.
The corresponding centerline axial
heat-flux and full-domain axial
heat-flux errors are approximately
7.67\% and 8.35\%, while the
full-domain heat-flux vector error
is approximately 7.75\%.

For the structured hexahedral mesh,
the centerline and full-domain
temperature errors are approximately
0.41\% and 0.71\%, respectively,
while the heat-flux errors are larger.
The coarse tetrahedral mesh produces
centerline and full-domain temperature
errors of approximately 1.02\% and
1.71\%, respectively, accompanied
by larger heat-flux discrepancies.

These results indicate that temperature
predictions remain comparatively
consistent across the different meshes,
whereas heat-flux predictions exhibit
greater sensitivity to spatial
discretization.
The fine tetrahedral mesh provides
the best overall agreement with
GiftBTE, particularly for the
heat-flux quantities.
The remaining discrepancies may
reflect the combined effects of
mesh topology, local spatial
resolution, interpolation, and
differences in the numerical
discretizations employed by
the two solvers.

The normalized nearest-neighbor
cell-center spacing distributions
are compared in
Fig.~\ref{fig:twofinfet_steady_validation}(d).
The fine tetrahedral meshes used
by JAX-Nano and GiftBTE exhibit
nearly overlapping distributions,
consistent with the mesh statistics
reported in Table~\ref{tab:twofinfet_mesh}.
In contrast, the coarse tetrahedral
mesh exhibits a different spacing
distribution associated with its
lower spatial resolution.
The structured hexahedral mesh
shows a distinct distribution
reflecting its structured
cell arrangement.
These comparisons provide additional
context for interpreting the
mesh-dependent differences in
the steady-state solutions.

The transient thermal response is
subsequently investigated using the
fine unstructured tetrahedral meshes
containing 54,839 cells for JAX-Nano
and 55,040 cells for GiftBTE.
Both solvers employ one phonon band,
64 discrete angular directions, and
explicit time integration, consistent
with the spatial and angular
discretizations used in the steady-state
calculations.

The JAX-Nano transient simulations
use a time step of
$\Delta t=3.0\times10^{-13}\ \mathrm{s}$,
whereas GiftBTE employs
$\Delta t=1.2\times10^{-13}\ \mathrm{s}$.
For the present configuration,
larger time steps were observed
to cause numerical instability
in GiftBTE.
The two solvers therefore use
their respective stable time steps,
and the transient temperature
profiles are compared at matching
physical times rather than at
identical iteration counts.

Figure~\ref{fig:twofinfet_transient_validation}(a)
compares the transient temperature profiles
predicted by JAX-Nano and GiftBTE along the
sampling line at 30, 60, 90, 120, 200,
and 400 ps.
The two solvers predict similar overall
temperature distributions throughout the
common simulation interval.
A localized spike is observed in the
GiftBTE temperature profile at $t=400$ ps,
whereas the corresponding JAX-Nano profile
remains smooth.
Due to the substantial computational cost
of extending the GiftBTE simulations to
longer physical times, the long-time
transient analysis is performed using
JAX-Nano.
Additional JAX-Nano results at 2000 and
8000 ps illustrate the longer-time
thermal response.
At $t=8000$ ps, the JAX-Nano temperature
profile closely approaches its corresponding
steady-state solution, indicating convergence
toward the stationary temperature distribution.
The corresponding steady-state solutions
are also shown for reference.

Figure~\ref{fig:twofinfet_transient_validation}(b)
presents the evolution of the
JAX-Nano temperature profile
over the extended simulation interval.
At early times, the temperature
distribution remains relatively
close to the initial reference state.
As time progresses, localized
heat generation produces a
progressive temperature rise,
while phonon transport redistributes
energy through the silicon device.
The transient temperature profiles
gradually approach the corresponding
steady-state distribution.

To quantify the approach toward
steady state, the normalized
relative $L_2$ deviation is
defined as
\begin{equation}
\varepsilon_T(t)
=
\frac{
\left[
\displaystyle\sum_{i=1}^{N_s}
\left(
T_i(t)-T_{i,\mathrm{ss}}
\right)^2
\right]^{1/2}
}{
\left[
\displaystyle\sum_{i=1}^{N_s}
\left(
T_{i,\mathrm{ss}}-T_{\mathrm{ref}}
\right)^2
\right]^{1/2}
},
\end{equation}
where $N_s$ denotes the number
of valid sampling points,
$T_{i,\mathrm{ss}}$ is the
corresponding steady-state
temperature, and
$T_{\mathrm{ref}}=300\ \mathrm{K}$.
The norms are evaluated along
the sampling line using the
same valid spatial locations.
Each transient solution is
compared against the steady-state
solution obtained using the
same solver.

Figure~\ref{fig:twofinfet_transient_validation}(c)
shows the temporal evolution
of $\varepsilon_T(t)$ for
JAX-Nano and GiftBTE.
Both solvers exhibit decreasing
deviations from their respective
steady-state solutions over the
common simulated interval.
The extended JAX-Nano results
demonstrate continued convergence
toward the steady-state temperature
distribution over longer physical
times.

Finally,
Fig.~\ref{fig:twofinfet_transient_validation}(d)
presents the relative $L_2$ error between
the JAX-Nano and GiftBTE transient temperature
profiles at matching physical times.
The cross-solver error is defined as
\begin{equation}
\varepsilon_{T,\mathrm{cross}}(t)
=
\frac{
\left[
\displaystyle\sum_{i\in\mathcal{S}(t)}
\left(
T_i^{\mathrm{JAX-Nano}}(t)
-
T_i^{\mathrm{GiftBTE}}(t)
\right)^2
\right]^{1/2}
}{
\left[
\displaystyle\sum_{i\in\mathcal{S}(t)}
\left(
T_i^{\mathrm{GiftBTE}}(t)
\right)^2
\right]^{1/2}
}
\times 100\%,
\end{equation}
where $\mathcal{S}(t)$ denotes the set of
valid common sampling points at time $t$.
The error quantifies the discrepancy between
the two transient predictions, rather than
their individual deviations from steady state.
As shown in
Fig.~\ref{fig:twofinfet_transient_validation}(d),
the two solvers exhibit relatively close
agreement at early times, followed by an
increase in the cross-solver error as the
transient temperature field develops.
The larger discrepancy at later times is
consistent with the localized differences
observed in the temperature profiles,
including the spike in the GiftBTE solution
at $t=400$ ps.
Nevertheless, the two solvers reproduce
similar overall transient thermal behavior
over the common simulation interval.

The combined steady-state and
transient comparisons demonstrate
the applicability of JAX-Nano
to three-dimensional phonon
transport in complex silicon
device geometries.
The results show that JAX-Nano
can accommodate structured and
unstructured computational meshes,
reproduce the principal temperature
and heat-flux characteristics
predicted by GiftBTE, and capture
the temporal evolution toward
steady state.
The comparisons also highlight
the importance of spatial
resolution for accurate heat-flux
predictions and consistent
interpolation procedures for
cross-mesh numerical validation.

\subsection{Array of FinFETs}

\begin{figure}
\begin{center}
\includegraphics[width=1.\columnwidth]{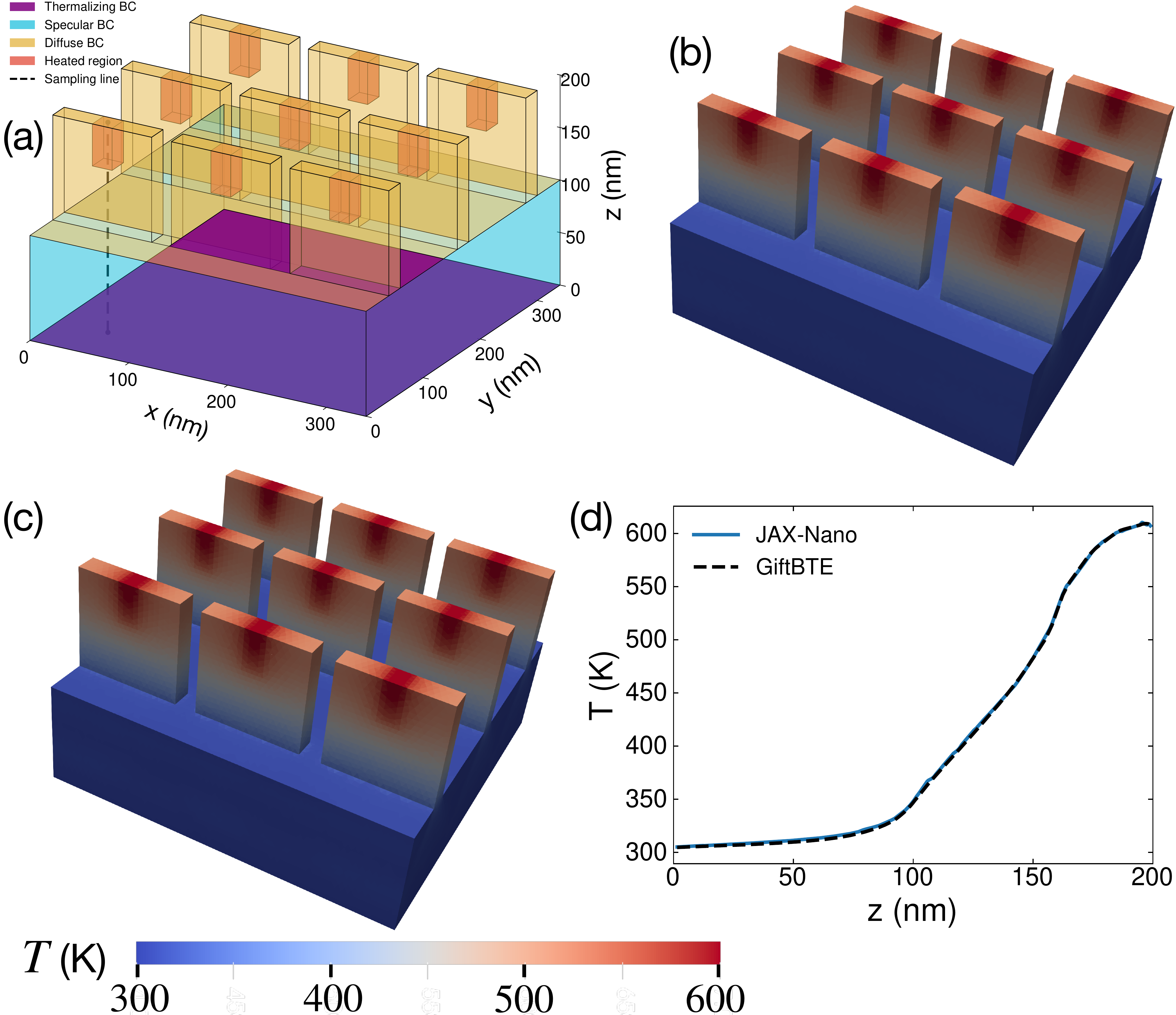}
\end{center}
\caption{
Steady-state phonon thermal transport in a three-dimensional
$3\times3$ silicon FinFET array.
(a) Schematic of the computational geometry and boundary conditions,
including thermalizing boundaries, specular boundaries, diffuse boundaries, and localized volumetric
heat-generation regions.
The black dashed line indicates the sampling line at
$x=50~\mathrm{nm}$ and $y=50~\mathrm{nm}$,
extending from $z=0$ to $200~\mathrm{nm}$.
(b) Steady-state temperature distribution predicted by JAX-Nano.
(c) Corresponding temperature distribution predicted by GiftBTE.
(d) Comparison of temperature profiles along the sampling line
obtained using JAX-Nano and GiftBTE.
Both simulations employ 313,345 computational cells,
one phonon band, and 64 angular directions.
}
\label{fig:FinFET_Array_Validation}
\end{figure}

\begin{figure}
\begin{center}
\includegraphics[width=1.\columnwidth]{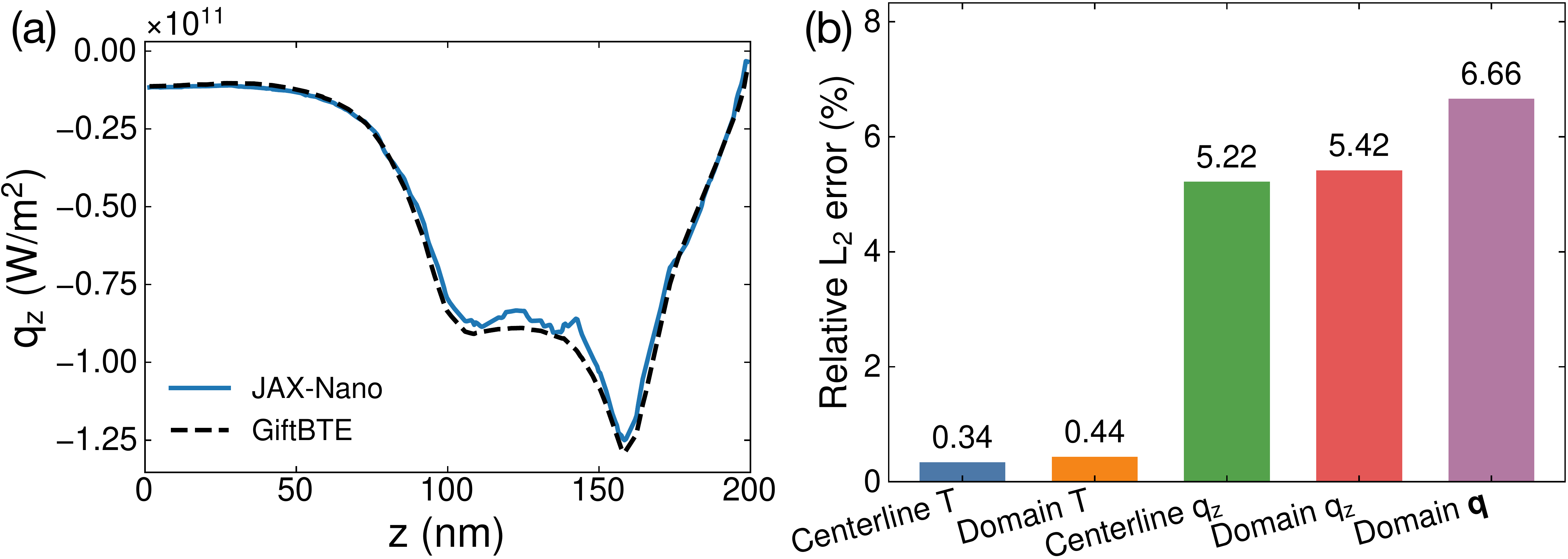}
\end{center}
\caption{
Steady-state heat-flux validation for a three-dimensional
$3\times3$ silicon FinFET array.
(a) Comparison of the axial heat-flux component $q_z$
along the sampling line at $x=50~\mathrm{nm}$ and
$y=50~\mathrm{nm}$, extending from $z=0$ to
$200~\mathrm{nm}$, as predicted by JAX-Nano
(solid blue line) and GiftBTE (dashed black line).
(b) Relative $L_2$ errors between JAX-Nano and GiftBTE
for centerline temperature, full-domain temperature,
centerline axial heat flux, full-domain axial heat flux,
and full-domain heat-flux vector, yielding
$0.34\%$, $0.44\%$, $5.22\%$, $5.42\%$,
and $6.66\%$, respectively.
}
\label{fig:FinFET_Array_HeatFlux_Error_Validation}
\end{figure}

To further evaluate the capability of JAX-Nano to simulate
phonon transport in complex three-dimensional geometries,
we consider a $3\times3$ silicon FinFET array, as illustrated
in Fig.~\ref{fig:FinFET_Array_Validation}(a).
Compared with the single-fin configuration considered
previously, the array introduces multiple spatially
distributed heat-generation regions within a larger
computational domain.
The geometry consists of nine vertically oriented silicon
fins arranged on a common substrate.
Each fin contains a localized heat-generation region
near its upper portion.
The substrate has dimensions of
$340\times340\times100~\mathrm{nm}^3$,
while each fin measures
$100\times20\times100~\mathrm{nm}^3$.
The fins are arranged with a center-to-center spacing
of $120~\mathrm{nm}$ in both horizontal directions,
and the total height of the computational domain
is $200~\mathrm{nm}$.

The boundary conditions are similar to those employed
in the single-fin configuration, with an important
modification to the upper fin boundaries.
Thermalizing boundary conditions are imposed on the
bottom surface of the substrate, with the temperature
fixed at $T=300~\mathrm{K}$.
Specularly reflecting boundary conditions are prescribed
on the lateral surfaces of the substrate,
while diffusely reflecting boundary conditions are
applied to the exposed substrate surface and all
exposed fin surfaces.
Unlike the single-fin configuration, the upper corners
of the fins are also treated as diffusely reflecting
rather than thermalizing boundaries.
Consequently, the upper fin surfaces are not maintained
at a prescribed temperature, and heat removal occurs
through phonon transport toward the thermalizing
boundary at the bottom of the substrate.
A uniform total volumetric heat-generation rate of
$\sum_{b=1}^{N_b}\dot{Q}_b
=10^{19}~\mathrm{W\,m^{-3}}$
is applied within each localized heated region.
For the fin indexed by $(i,j)$,
where $i,j\in\{0,1,2\}$,
the heated region is defined as
\begin{equation}
\Omega_{\mathrm{hot}}^{(i,j)}
=
[40+120i,60+120i]
\times
[40+120j,60+120j]
\times
[160,200]~\mathrm{nm}^3.
\end{equation}
The computational domain is discretized using
an unstructured tetrahedral mesh containing 313,345 cells,
and the angular domain is represented by
64 discrete directions.
A single-band phonon transport model is considered
for both JAX-Nano and GiftBTE.

Figure~\ref{fig:FinFET_Array_Validation}(b,c)
compares the three-dimensional steady-state temperature
distributions predicted by JAX-Nano and GiftBTE.
Both solvers predict localized high-temperature regions
near the upper portions of the nine fins,
corresponding to the prescribed volumetric
heat-generation regions.
The temperature decreases from the heated regions
toward the underlying substrate, which remains
relatively close to the prescribed boundary temperature.
Unlike the single-fin configuration, where the
thermalizing upper corners provide additional
local cooling, the diffusely reflecting fin surfaces
in the present array do not impose a fixed temperature.
As a result, the temperature field is primarily
governed by internal heat generation, phonon transport,
boundary scattering, and heat removal through
the substrate.
The spatial temperature distributions predicted
by JAX-Nano and GiftBTE exhibit close agreement,
including the localized hot spots and the temperature
gradients extending from the fins into the substrate.

A quantitative comparison of the temperature profiles
along the vertical sampling line is presented in
Fig.~\ref{fig:FinFET_Array_Validation}(d).
The sampling line is located at
$x=y=50~\mathrm{nm}$ and extends from
$z=0$ to $200~\mathrm{nm}$,
passing through the substrate and one of the fins.
The profiles are obtained by interpolating the
cell-centered numerical solutions onto a common
set of sampling points.
Both solvers predict a gradual temperature increase
within the substrate, followed by a more pronounced
increase along the fin toward the heated region.
The temperature rises rapidly near the lower boundary
of the heated region at $z=160~\mathrm{nm}$,
reaching approximately $610~\mathrm{K}$
near the upper end of the fin.
The absence of a thermalizing boundary at the
fin top allows the temperature to remain elevated
throughout the heated region.
The JAX-Nano and GiftBTE temperature profiles
nearly overlap over the entire sampling line,
demonstrating close agreement in the predicted
local thermal response.

The corresponding vertical heat-flux profiles
are compared in
Fig.~\ref{fig:FinFET_Array_HeatFlux_Error_Validation}(a).
Negative values of $q_z$ dominate along the
sampling line, indicating net downward heat
transport from the heated region toward the substrate.
The magnitude of the downward heat flux increases
substantially within the fin and reaches a pronounced
maximum near $z=160~\mathrm{nm}$,
approximately corresponding to the lower boundary
of the heat-generation region.
Above this location, the magnitude of $q_z$
decreases sharply toward the fin top.
This behavior reflects the redistribution of
thermal energy generated within the upper fin
and its transport toward the underlying substrate.
Both JAX-Nano and GiftBTE reproduce the principal
features of the vertical heat-flux profile,
including the direction of heat transport,
the location of the maximum downward heat flux,
and the rapid variation near the heated region.
Local differences are observed between the two
predictions, particularly in regions where
the heat flux varies sharply with position.
These discrepancies are more pronounced than
those observed in the temperature profiles.

The relative $L_2$ errors between JAX-Nano and
GiftBTE are summarized in
Fig.~\ref{fig:FinFET_Array_HeatFlux_Error_Validation}(b).
The centerline and full-domain temperature errors
are $0.34\%$ and $0.44\%$, respectively,
demonstrating close agreement between the
two solvers in predicting the temperature field.
The corresponding errors for centerline $q_z$,
full-domain $q_z$, and the full-domain heat-flux
vector are $5.22\%$, $5.42\%$, and $6.66\%$,
respectively.
The full-domain comparisons are performed by
interpolating the numerical solutions onto common
cell-centered evaluation points.
As observed in the single-fin validation,
the heat-flux errors are larger than the
temperature errors, indicating that the
heat-flux field provides a more stringent
assessment of agreement between the two solvers.
Nevertheless, the close temperature agreement
and consistent heat-flux distributions demonstrate
that JAX-Nano reproduces the principal thermal
transport characteristics predicted by GiftBTE
for the three-dimensional FinFET array.

Overall, this validation extends the single-fin
analysis to a larger three-dimensional configuration
containing multiple heat-generating nanoscale structures.
The agreement between JAX-Nano and GiftBTE in both
temperature and heat-flux predictions demonstrates
the capability of the proposed GPU-accelerated
phonon BTE framework to simulate steady-state
thermal transport in complex device arrays
with multiple localized heat sources and
diffusely reflecting fin boundaries.

% ============================================================
\section{Gradient-Based Inverse Parameter Identification}
\label{sec:inverse_identification}
% ============================================================

Having validated the forward predictive capability
of JAX-Nano for three-dimensional phonon transport,
we next investigate its application to gradient-based
inverse parameter identification.
Unlike the forward problems considered in the
preceding sections, where the material properties,
geometric parameters, boundary conditions, and
heat-generation rates are prescribed, the inverse
problems seek to recover selected unknown parameters
from reference temperature distributions.
The differentiable implementation of JAX-Nano
enables the sensitivity of the thermal response
to these parameters to be evaluated through
automatic differentiation, thereby allowing
the unknown parameters to be optimized using
gradient-based methods.

We consider both single-parameter and
multi-parameter identification problems.
The single-parameter studies involve boundary
temperature, domain size, volumetric
heat-generation rate, and phonon relaxation time.
The multi-parameter studies consider the
simultaneous identification of mode-resolved
relaxation times and nine spatially distributed
volumetric heat-generation rates in a
$3\times3$ FinFET array.

\subsection{Inverse Problem Formulation}
\label{subsec:inverse_formulation}

To evaluate the inverse-modeling capabilities of JAX-Nano,
a series of parameter-identification problems is considered
using temperature fields generated by the forward phonon
Boltzmann transport equation (BTE) solver as reference data.
The objective is to recover unknown physical parameters
from their influence on the predicted temperature distribution.

The inverse problems include the identification of
thermalizing boundary temperatures, phonon relaxation
times, characteristic geometric lengths, and internal
heat-generation rates. Both single-parameter and
simultaneous multi-parameter identification are
investigated. In addition, inverse learning from
sparse temperature observations and transient
temperature responses is examined.

For each inverse problem, reference temperature data
are generated using prescribed physical parameters.
The learnable parameters are initialized at values
different from their reference values and subsequently
updated using gradients obtained through automatic
differentiation. The temperature-based objective
function and gradient-evaluation procedures follow
the formulation described in
Section~\ref{sec:automatic_differentiation}.

The inverse-learning performance is assessed by
comparing the recovered parameters with their
prescribed reference values and examining the
convergence of the objective function. These
numerical experiments evaluate the ability of
JAX-Nano to recover physical parameters under
different parameter dimensions, observation
configurations, and temporal conditions.

\subsection{Single-Parameter Identification}
\label{subsec:single_parameter}

We first investigate the identification of individual
physical parameters to evaluate the accuracy and
robustness of gradient-based inverse learning in
JAX-Nano. The considered parameters include the
thermalizing boundary temperature, phonon relaxation
time, characteristic geometric length, and internal
heat-generation rate.

For each case, a reference temperature field is
obtained from the forward BTE simulation using
a prescribed parameter value. Starting from a
different initial estimate, the unknown parameter
is iteratively updated to minimize the discrepancy
between the predicted and reference temperatures.
The loss gradient is evaluated using automatic
differentiation through the numerical BTE solver.

These experiments examine whether the differentiable
solver can recover parameters governing distinct
physical mechanisms, including boundary thermal
excitation, phonon scattering, geometric transport
scaling, and internal heat generation. The resulting
parameter trajectories and loss histories are used
to assess convergence toward the reference values
and the accuracy of the recovered parameters.

\subsubsection{Domain Size}
\label{sec:inverse_domain_size}

\paragraph{Steady-State Inverse Learning}

\begin{figure}
\begin{center}
\includegraphics[width=1.\columnwidth]{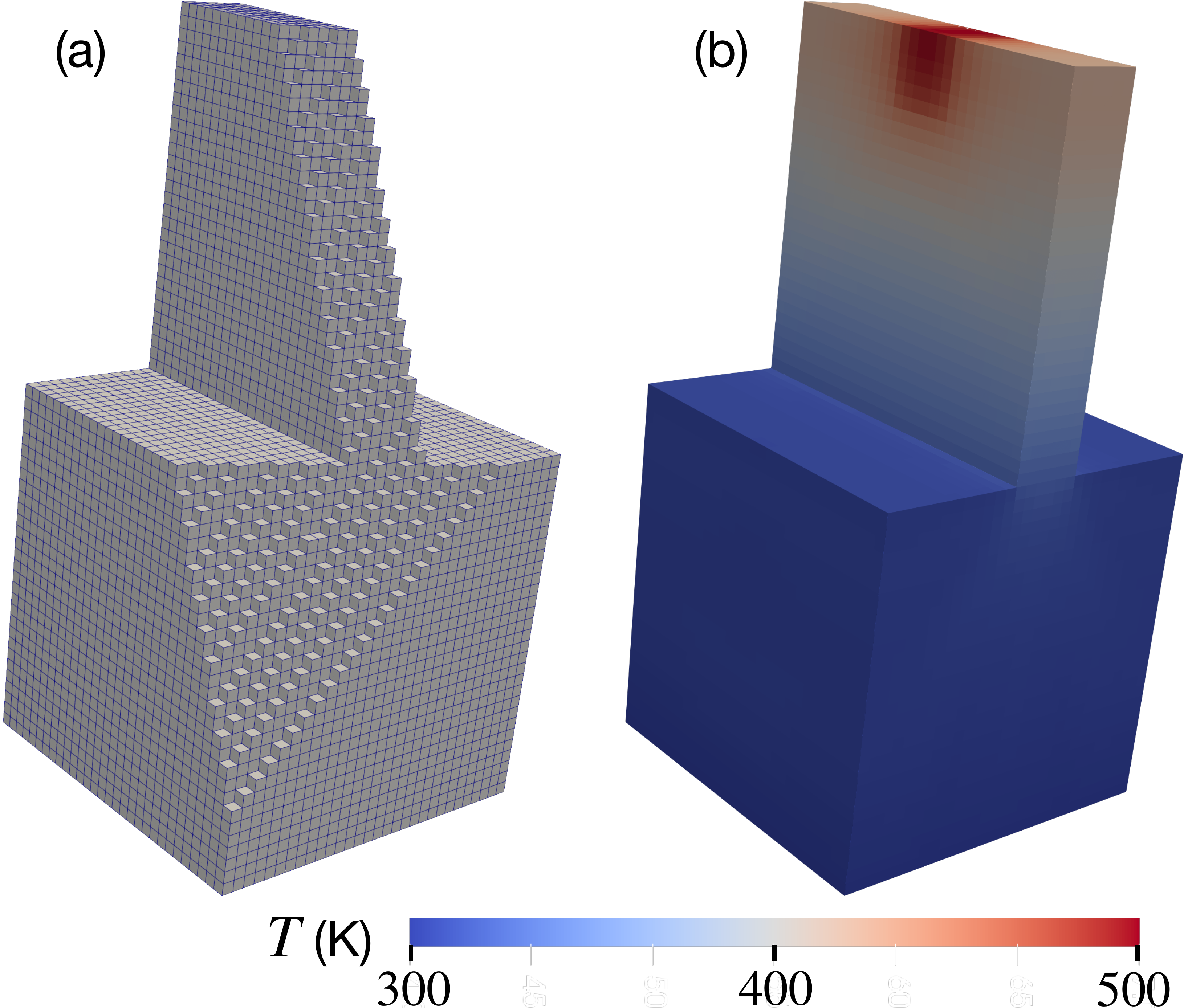}
\end{center}
\caption{
Geometry, computational mesh, and steady-state temperature
distribution of a single-fin silicon FinFET device.
(a) Three-dimensional structured hexahedral mesh consisting
of 32,400 cells. A thermalizing boundary condition is imposed
at the bottom surface of the substrate, while diffuse boundary
conditions are applied to all remaining external surfaces.
(b) Steady-state temperature field predicted by JAX-Nano,
illustrating localized heating near the top of the fin and
the resulting temperature distribution throughout the fin
and underlying substrate. 
}
\label{fig:single_finfet_geometry_temperature}
\end{figure}

\begin{figure}
\begin{center}
\includegraphics[width=1.\columnwidth]{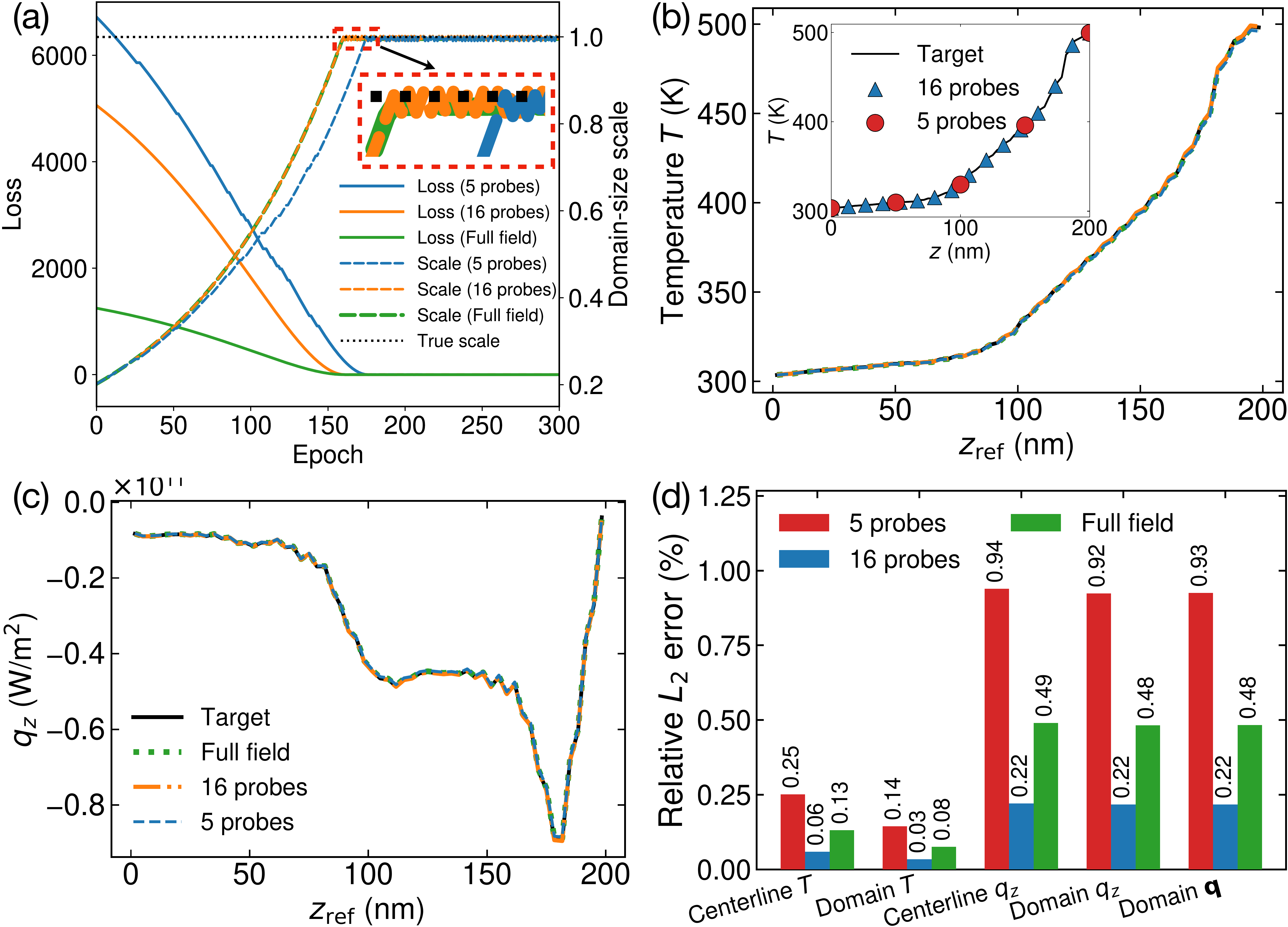}
\end{center}
\caption{
Inverse identification of the geometric scale of a single-fin
silicon FinFET device using differentiable phonon Boltzmann transport simulations
with three temperature-observation configurations:
5 probes, 16 probes, and full-field measurements.
(a) Evolution of the inverse-learning loss and recovered
domain-size scale over training epochs. The dotted horizontal
line indicates the true scale of unity, while the enlarged
inset highlights small-amplitude oscillations in the recovered
scales for the 5- and 16-probe configurations near convergence.
(b) Comparison of the target and reconstructed temperature
profiles along the vertical centerline, with the inset
showing the spatial distribution of the 5 and 16
temperature probes.
(c) Comparison of the corresponding vertical heat-flux
profiles, $q_z$, along the centerline.
(d) Relative $L_2$ errors in the centerline temperature,
full-domain temperature, centerline vertical heat flux,
full-domain vertical heat flux, and full-domain heat-flux
vector for the three observation configurations.
The errors are evaluated using corresponding cell indices
without spatial interpolation.
}
\label{fig:single_finfet_inverse_geometry}
\end{figure}

We first investigate the capability of JAX-Nano to identify
an unknown geometric scale through differentiable phonon
Boltzmann transport simulations. The inverse problem is
formulated using a single-fin silicon FinFET device, as
illustrated in
Fig.~\ref{fig:single_finfet_geometry_temperature}.
The computational domain is discretized using a structured
hexahedral mesh consisting of 32,400 cells.
A thermalizing boundary condition at 300~K is imposed at
the bottom surface of the substrate, while diffuse boundary
conditions are applied to all remaining external surfaces.
The phonon Boltzmann transport equation is discretized
using a single phonon band and 64 discrete propagation
directions.
Each forward BTE evaluation employs 200 iterations to obtain a converged steady-state temperature field for inverse learning. 

A localized volumetric heat source is prescribed near
the top of the silicon fin within the region
\begin{equation}
\Omega_{\mathrm{hot}}(s)
=
[40s,60s]\times[40s,60s]\times[180s,200s]
\ \mathrm{nm}^3,
\end{equation}
where $s$ denotes the dimensionless geometric scale factor.
A uniform volumetric heat-generation rate of
$\dot{Q}=10^{19}~\mathrm{W\,m^{-3}}$
is imposed within this region and is zero elsewhere.
The heat-source dimensions and location scale consistently
with the computational geometry, while the volumetric
heat-generation rate remains fixed.
The resulting steady-state temperature field ranges
from approximately 300 to 500~K for the reference geometry.

The inverse problem seeks to recover a dimensionless
geometric scale factor, $s$, which uniformly scales the
spatial coordinates of the computational domain according to
\begin{equation}
    \mathbf{x}(s) = s\,\mathbf{x}_{\mathrm{ref}},
\end{equation}
where $\mathbf{x}_{\mathrm{ref}}$ denotes the reference
geometry. The target geometry corresponds to
$s_{\mathrm{true}}=1$, whereas the inverse-learning procedure
is initialized with $s_0=0.2$. The geometric scale is
optimized by minimizing the discrepancy between the
predicted and target steady-state temperature observations,
with gradients computed through automatic differentiation
using \texttt{jax.value\_and\_grad}.

To examine the influence of measurement availability,
three observation configurations are considered:
5 temperature probes, 16 temperature probes, and
full-field temperature measurements. The discrete probes
are distributed along the vertical centerline of the
FinFET device, while the full-field configuration uses
temperature observations throughout the computational
domain. All three inverse problems employ the same
forward phonon BTE solver and differ only in the number
of temperature observations used to construct the
inverse-learning objective.

Figure~\ref{fig:single_finfet_inverse_geometry}(a)
presents the evolution of the inverse-learning loss
and the recovered geometric scale. For all three
observation configurations, the loss decreases as
the inferred scale approaches the true value of unity.
The enlarged inset reveals small-amplitude oscillations
in the recovered geometric scales for the 5- and 16-probe
configurations near convergence, whereas the full-field
configuration exhibits comparatively stable convergence.
Despite these oscillations, the recovered scales remain
close to the true value, demonstrating that the geometric
scale can be identified from a limited number of temperature
measurements without requiring full-field observations.

The reconstructed steady-state temperature profiles
along the vertical centerline are compared with the
target solution in
Fig.~\ref{fig:single_finfet_inverse_geometry}(b).
The corresponding vertical heat-flux profiles,
$q_z$, are shown in
Fig.~\ref{fig:single_finfet_inverse_geometry}(c).
The agreement in both temperature and heat flux
indicates that the inferred geometry reproduces
not only the observed thermal response but also
the associated heat-transport characteristics.
Notably, heat flux is not directly included in
the inverse-learning objective, providing an
additional assessment of reconstruction accuracy.

To quantify the reconstruction errors, the relative
$L_2$ norm is defined as
\begin{equation}
    \varepsilon_{L_2}(\phi)
    =
    \frac{
        \left\|
            \phi_{\mathrm{pred}}
            -
            \phi_{\mathrm{target}}
        \right\|_2
    }{
        \left\|
            \phi_{\mathrm{target}}
        \right\|_2
    }
    \times 100\%,
\end{equation}
where $\phi_{\mathrm{pred}}$ and
$\phi_{\mathrm{target}}$ denote the reconstructed
solution obtained using the inferred geometric scale
and the reference solution corresponding to
$s_{\mathrm{true}}=1$, respectively.
The variable $\phi$ represents temperature,
the vertical heat-flux component, or the complete
heat-flux vector. For vector-valued quantities,
the $L_2$ norm includes all three Cartesian
components. Five error metrics are evaluated:
centerline temperature, full-domain temperature,
centerline vertical heat flux, full-domain vertical
heat flux, and full-domain heat-flux vector.
The errors are computed using corresponding
cell indices in the reference and reconstructed
meshes, without spatial interpolation.

As shown in
Fig.~\ref{fig:single_finfet_inverse_geometry}(d),
all three observation configurations yield relatively
small reconstruction errors in both temperature and
heat flux. The 16-probe configuration achieves the
lowest errors across all five metrics, followed by
the full-field and 5-probe configurations.
In particular, the 16-probe configuration yields
relative $L_2$ errors of approximately 0.03\%
for the full-domain temperature and 0.22\%
for the full-domain heat-flux vector.
Although the 5-probe configuration exhibits
larger reconstruction errors, its full-domain
temperature and heat-flux vector errors remain
below 0.2\% and 1.0\%, respectively.
These results demonstrate that accurate geometric
parameter identification and reconstruction of
the associated thermal fields can be achieved
using sparse temperature measurements.

\paragraph{Transient Inverse Learning}

\begin{figure}
\begin{center}
\includegraphics[width=1.\columnwidth]{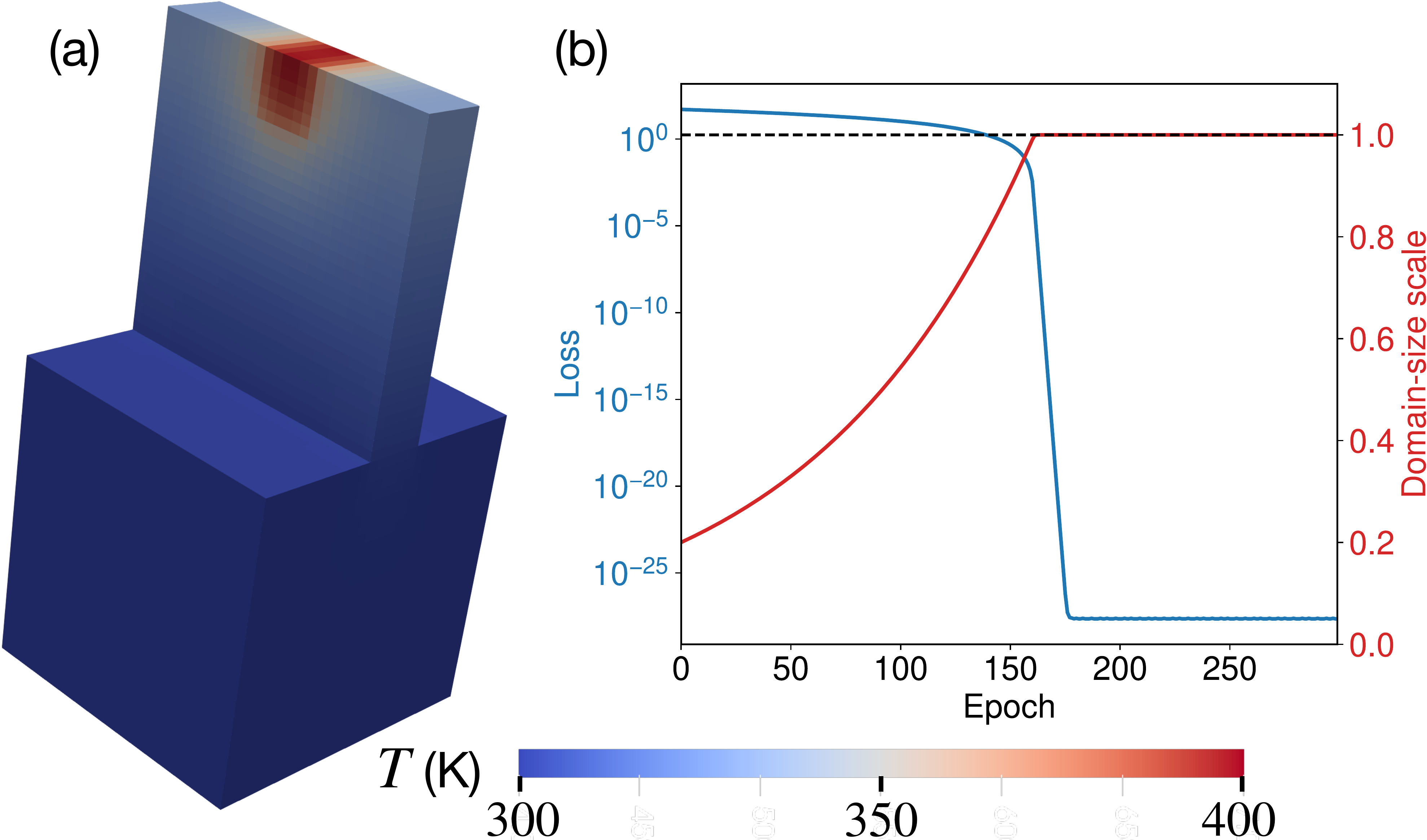}
\end{center}
\caption{
Transient inverse learning of the geometric scale for a
single-fin silicon FinFET using JAX-Nano.
The transient BTE solver employs 1000 time steps with a
time-step size of $\Delta t=10^{-13}~\mathrm{s}$.
(a) Transient temperature distribution for the reference
geometry at $t=100~\mathrm{ps}$.
(b) Evolution of the temperature-based loss and learned domain-size scale during optimization. The black dashed line indicates
the true geometric scale, $s=1$.
}
\label{fig:inverse_domain_size_transient}
\end{figure}

The geometric inverse-learning problem is further
extended to transient phonon transport using the same
single-fin silicon FinFET geometry, computational mesh,
boundary conditions, volumetric heat source, phonon
band, and angular discretization as those employed
in the steady-state case.
The transient BTE is solved using an explicit
time-integration scheme. 
Each forward BTE evaluation employs 1000 time steps
with a fixed time-step size of
$\Delta t=10^{-13}~\mathrm{s}$,
corresponding to a final simulation time of
$t=100~\mathrm{ps}$.
The dimensionless geometric scale factor $s$ is
treated as the unknown parameter and optimized
through reverse-mode automatic differentiation
of the transient BTE solver.

Figure~\ref{fig:inverse_domain_size_transient}(a)
shows the transient temperature distribution
for the reference geometry ($s=1$) at
$t=100~\mathrm{ps}$.
The corresponding inverse-learning convergence
history is presented in
Fig.~\ref{fig:inverse_domain_size_transient}(b),
where the temperature-based loss and learned
geometric scale are plotted against the
optimization epoch.
Starting from an initial scale of $s=0.2$,
the learned geometric scale increases
progressively toward the prescribed
reference value of $s=1$, reaching this
value at approximately 160 epochs.
The temperature-based loss subsequently
decreases sharply and reaches a numerical
floor at approximately 175 epochs.

The recovered geometric scale agrees with
the reference value to the reported numerical
precision.
The final temperature-based mean-squared loss
reaches approximately
$8.08\times10^{-27}$,
corresponding to a relative $L_2$ temperature
error of approximately
$2.95\times10^{-14}\%$.
These results demonstrate that JAX-Nano
can accurately propagate parameter sensitivities
through transient phonon transport simulations
and recover the geometric scale from
noise-free synthetic temperature observations.

\subsubsection{Boundary Temperature}

\begin{figure}
\begin{center}
\includegraphics[width=1.\columnwidth]{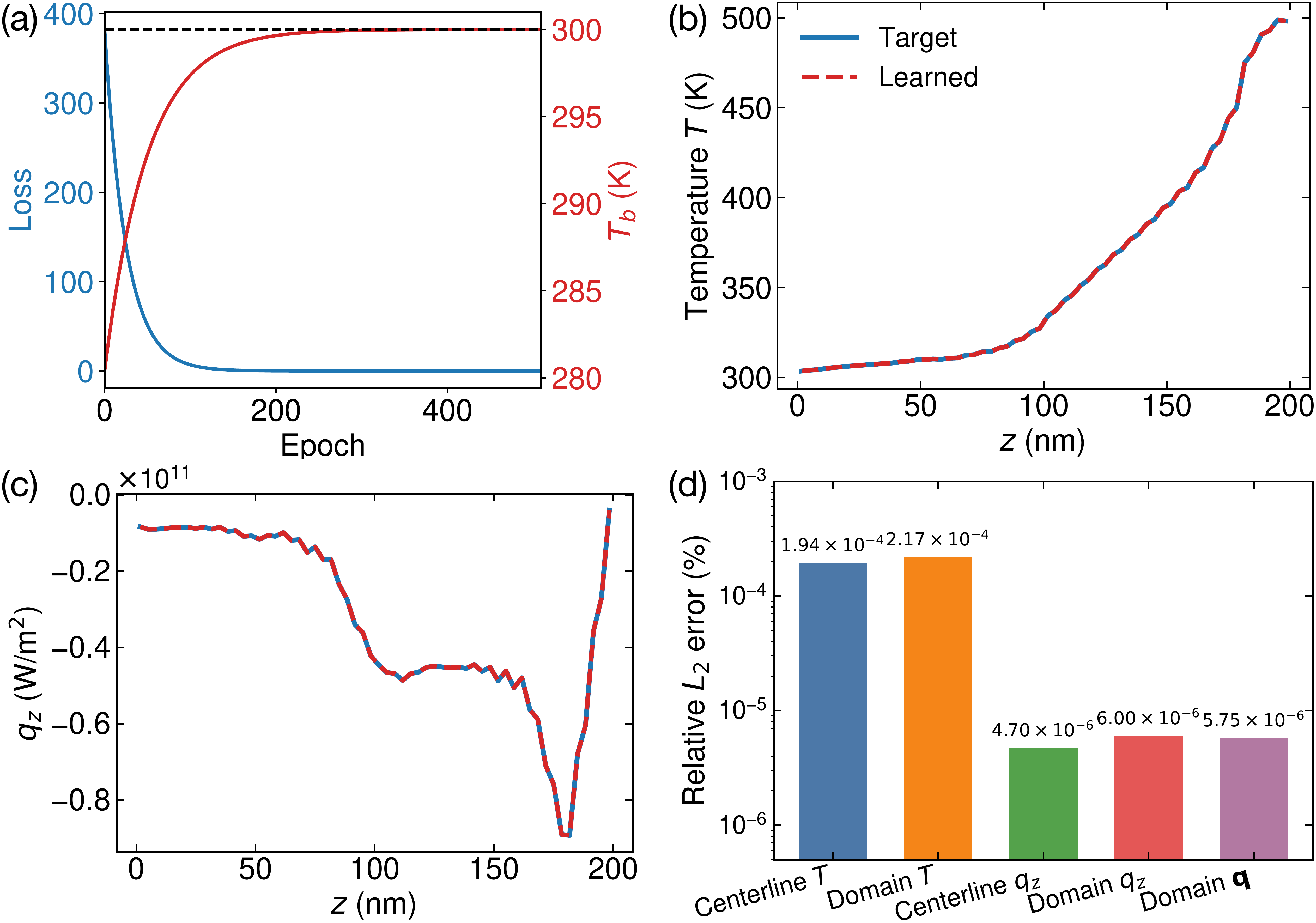}
\end{center}
\caption{
Inverse identification of the thermalizing boundary temperature
in a single-fin silicon FinFET using JAX-Nano.
(a) Evolution of the loss function and learned boundary temperature
$T_b$ during optimization. The dashed horizontal line indicates
the target boundary temperature of $300~\mathrm{K}$.
(b) Comparison of the target and reconstructed temperature
profiles along the $z$-direction centerline.
(c) Corresponding comparison of the vertical heat-flux
component $q_z$.
(d) Relative $L_2$ errors in the centerline and full-domain
temperature and heat-flux fields, including the full heat-flux
vector $\mathbf{q}$. 
}
\label{fig:finfet_inverse_Tb}
\end{figure}

We investigate the inverse identification of the thermalizing
boundary temperature in the single-fin silicon FinFET configuration.
The target boundary temperature is prescribed as
$T_b^{\mathrm{true}}=300~\mathrm{K}$, while the initial estimate
is set to $T_b^{(0)}=280~\mathrm{K}$.
The objective is to recover the unknown boundary temperature
by minimizing the discrepancy between the predicted and target
steady-state temperature fields.
Gradients of the loss function with respect to $T_b$ are
computed using forward-mode automatic differentiation
through the JAX-Nano solver.

Figure~\ref{fig:finfet_inverse_Tb}(a) presents the evolution
of the loss function and the inferred boundary temperature
during optimization.
The loss decreases rapidly during the initial iterations
and subsequently approaches zero.
Meanwhile, the inferred boundary temperature increases
monotonically from $280~\mathrm{K}$ toward the prescribed
value of $300~\mathrm{K}$.
After 500 optimization epochs, the recovered temperature
reaches approximately $299.993~\mathrm{K}$,
demonstrating accurate identification of the unknown
thermal boundary condition.

To assess the accuracy of the reconstructed solution,
Figs.~\ref{fig:finfet_inverse_Tb}(b) and
\ref{fig:finfet_inverse_Tb}(c) compare the target and
reconstructed temperature and vertical heat-flux profiles,
respectively, along the $z$-direction centerline.
The reconstructed profiles closely reproduce the target
solutions throughout the computational domain,
including regions exhibiting pronounced temperature
gradients and variations in heat flux.

A quantitative assessment is provided in
Fig.~\ref{fig:finfet_inverse_Tb}(d), where the relative
$L_2$ errors are evaluated for both the centerline
profiles and the full-domain fields.
The centerline and full-domain temperature errors are
$1.9354\times10^{-4}\%$ and
$2.1650\times10^{-4}\%$, respectively.
The corresponding errors in the vertical heat-flux
component are $4.70\times10^{-6}\%$ and
$6.00\times10^{-6}\%$, while the full-domain heat-flux
vector exhibits an error of $5.75\times10^{-6}\%$.
These results demonstrate that the inferred boundary
temperature accurately reproduces both the temperature
distribution and the heat-flux field.

Overall, this example demonstrates the capability of
JAX-Nano to identify an unknown thermal boundary condition
through gradient-based inverse learning while maintaining
high accuracy in the reconstructed steady-state solution.

\subsubsection{Volumetric Heat-Generation Rate}

\begin{figure}
\begin{center}
\includegraphics[width=1.\columnwidth]{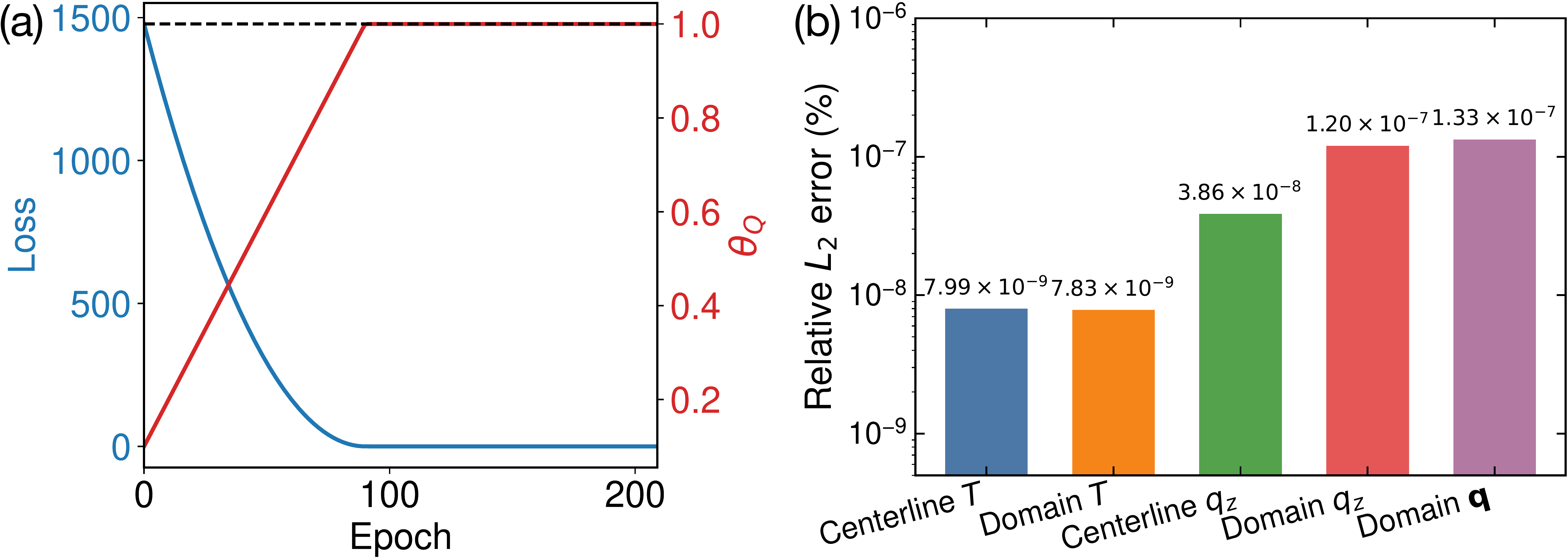}
\end{center}
\caption{
Inverse identification of the volumetric heat-generation rate
in a single-fin silicon FinFET using JAX-Nano.
(a) Evolution of the loss function and learned source amplitude
$\theta_Q$ during optimization.
The dashed horizontal line indicates the reference value
$\theta_Q=1$.
(b) Relative $L_2$ errors in the centerline temperature,
full-domain temperature, centerline vertical heat flux $q_z$,
full-domain vertical heat flux $q_z$, and full-domain
heat-flux vector $\mathbf{q}$.
}
\label{fig:finfet_inverse_heat_generation}
\end{figure}

We investigate the inverse identification of the
volumetric heat-generation rate in the single-fin FinFET.
Following the parameterization introduced in
Eq.~\eqref{eq:inverse_heat}, the source amplitude
$\theta_Q$ is treated as the unknown parameter,
while the spatial distribution of heat generation
remains fixed.
The target temperature field is generated using
$\theta_Q=1$, and the inverse optimization is
initialized with $\theta_Q=0.1$.
The source amplitude is updated by minimizing the
mean squared difference between the predicted and
target temperature fields, with gradients computed
using reverse-mode automatic differentiation
through the BTE solver.

Figure~\ref{fig:finfet_inverse_heat_generation}(a)
shows the evolution of the loss function and the
learned source amplitude during optimization.
The loss decreases as $\theta_Q$ approaches its
reference value, and the recovered parameter
converges toward $\theta_Q=1$.
This demonstrates the capability of JAX-Nano
to identify the magnitude of an internal heat source
from temperature observations.

The accuracy of the reconstructed thermal fields
is further evaluated in
Fig.~\ref{fig:finfet_inverse_heat_generation}(b).
The relative $L_2$ errors are reported for the
centerline temperature, full-domain temperature,
centerline vertical heat flux $q_z$,
full-domain vertical heat flux $q_z$,
and full-domain heat-flux vector $\mathbf{q}$.
All five errors remain below
$1.4\times10^{-7}\%$,
indicating excellent agreement between the
reconstructed and reference solutions.
Importantly, although only temperature observations
are used in the inverse optimization,
the reconstructed heat-flux fields also exhibit
very small errors.
These results demonstrate that JAX-Nano
can accurately recover the amplitude of an
internal volumetric heat source while maintaining
consistency between the reconstructed temperature
and heat-flux fields.

\subsubsection{Phonon Relaxation Time}

\begin{figure}
\begin{center}
\includegraphics[width=1.\columnwidth]{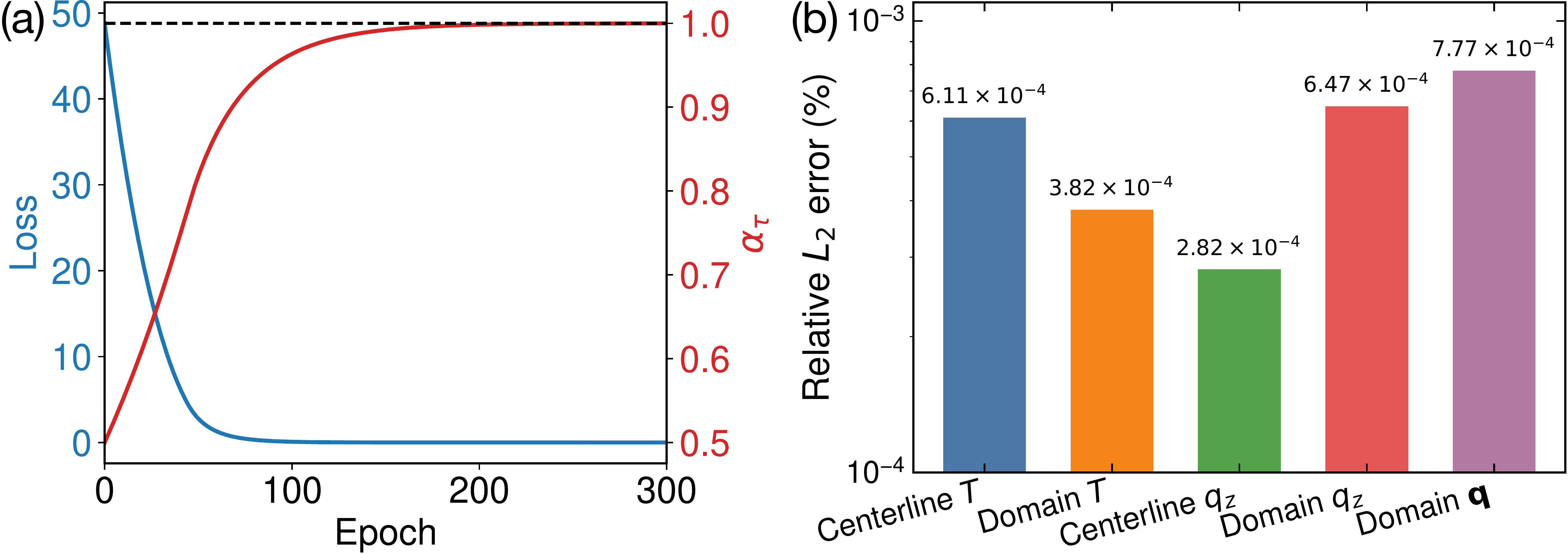}
\end{center}
\caption{
Inverse identification of the phonon relaxation-time scaling factor
in a single-fin silicon FinFET using JAX-Nano.
(a) Evolution of the loss function and learned relaxation-time
scaling factor $\alpha_\tau$ during optimization.
The dashed horizontal line indicates the reference value
$\alpha_\tau=1$.
(b) Relative $L_2$ errors in the centerline temperature,
full-domain temperature, centerline vertical heat flux $q_z$,
full-domain vertical heat flux $q_z$, and full-domain
heat-flux vector $\mathbf{q}$.
}
\label{fig:finfet_inverse_relaxation_time}
\end{figure}

We investigate the inverse identification of the phonon
relaxation time in the single-fin FinFET.
Following the parameterization introduced in
Eq.~\eqref{eq:inverse_tau}, a global relaxation-time
scaling factor $\alpha_\tau$ is treated as the unknown
parameter, such that
$\tau_b=\alpha_\tau\tau_b^{\mathrm{ref}}$.
The target temperature field is generated using
$\alpha_\tau=1$, while the inverse optimization is
initialized with $\alpha_\tau=0.5$.
The scaling factor is identified by minimizing the
mean squared difference between the predicted and
target temperature fields using gradient descent
with a learning rate of $0.01$.
The loss-function gradients are computed using
reverse-mode automatic differentiation through
the BTE solver.

Figure~\ref{fig:finfet_inverse_relaxation_time}(a)
shows the evolution of the loss function and the
learned relaxation-time scaling factor during optimization.
The loss decreases rapidly during the initial optimization
epochs, while $\alpha_\tau$ increases monotonically
toward its reference value.
The recovered scaling factor gradually approaches
$\alpha_\tau=1$, demonstrating the capability of
JAX-Nano to identify the phonon relaxation time
from temperature observations.

The accuracy of the reconstructed thermal fields
is further evaluated in
Fig.~\ref{fig:finfet_inverse_relaxation_time}(b).
The relative $L_2$ errors are reported for the
centerline temperature, full-domain temperature,
centerline vertical heat flux $q_z$,
full-domain vertical heat flux $q_z$,
and full-domain heat-flux vector $\mathbf{q}$.
The corresponding errors are
$6.11\times10^{-4}\%$,
$3.82\times10^{-4}\%$,
$2.82\times10^{-4}\%$,
$6.47\times10^{-4}\%$, and
$7.77\times10^{-4}\%$, respectively.
All five errors remain below $8\times10^{-4}\%$,
indicating close agreement between the reconstructed
and reference solutions.
Notably, although only temperature observations
are employed in the inverse optimization,
the reconstructed heat-flux fields also exhibit
small errors.
These results demonstrate that JAX-Nano can
accurately identify a global phonon relaxation-time
scaling factor while preserving consistency between
the reconstructed temperature and heat-flux fields.

\subsection{Multi-Parameter Identification}
\label{subsec:multi_parameter}

We further investigate the capability of JAX-Nano for
simultaneous identification of multiple unknown physical
parameters in nanoscale thermal transport. Compared with
single-parameter inverse learning, multi-parameter
identification introduces additional challenges arising
from parameter coupling, nonuniform sensitivities, and
potentially ill-conditioned optimization landscapes.
To assess the applicability of the differentiable phonon
BTE framework to such problems, two representative
inverse-learning scenarios are considered. The first
involves simultaneous identification of mode-resolved
phonon relaxation times, whereas the second focuses on
recovering spatially distributed heat-generation rates
in a $3\times3$ FinFET array. These examples examine
the capability of JAX-Nano to identify both material
transport properties and spatially varying thermal
source parameters through gradient-based inverse learning.

\subsubsection{Mode-Resolved Relaxation Times}
\label{subsubsec:mode_resolved_relaxation}

\begin{figure}
\begin{center}
\includegraphics[width=1.\columnwidth]{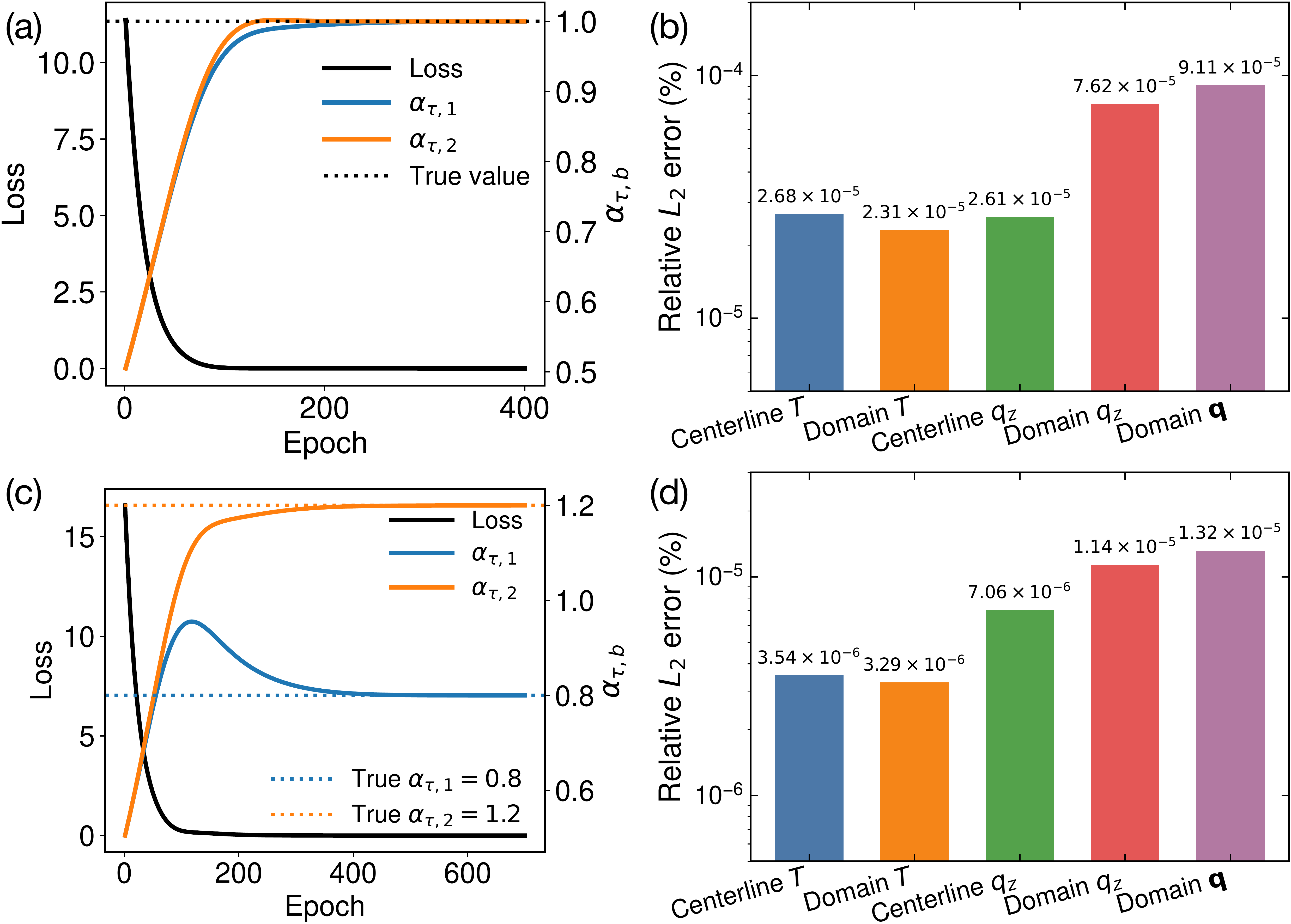}
\end{center}
\caption{
Inverse learning of band-dependent phonon relaxation times
in a three-dimensional single-fin silicon FinFET geometry using JAX-Nano.
(a) Evolution of the loss function and learned relaxation-time
scaling factors for identical reference values,
$\alpha_{\tau,1}=\alpha_{\tau,2}=1.0$.
(b) Relative $L_2$ errors in the reconstructed temperature
and heat-flux fields for the case in (a), evaluated
along the centerline and over the computational domain.
(c) Evolution of the loss function and learned relaxation-time
scaling factors for unequal reference values,
$\alpha_{\tau,1}=0.8$ and $\alpha_{\tau,2}=1.2$.
(d) Corresponding relative $L_2$ errors in the reconstructed
temperature and heat-flux fields for the case in (c).
Solid curves represent the optimization histories,
while dotted horizontal lines indicate the prescribed
reference relaxation-time scaling factors.
}
\label{fig:finfet_inverse_relaxation_time_2modes}
\end{figure}

We first consider the simultaneous identification of
two mode-dependent phonon relaxation times in a
three-dimensional FinFET geometry. The relaxation
times are parameterized using independent scaling
factors,
\begin{equation}
    \tau_b
    =
    \alpha_{\tau,b}
    \tau_b^{\mathrm{ref}},
    \qquad b=1,2,
\end{equation}
where $\tau_b^{\mathrm{ref}}$ denotes the reference
relaxation time of phonon mode $b$, and
$\alpha_{\tau,b}$ represents the corresponding
unknown scaling factor. The objective is to
simultaneously recover both scaling factors from
a prescribed target temperature field.

The inverse problem is formulated by minimizing
the mean-squared difference between the predicted
and target temperature distributions,
\begin{equation}
    \mathcal{L}
    \left(
        \boldsymbol{\alpha}_{\tau}
    \right)
    =
    \frac{1}{N_c}
    \sum_{i=1}^{N_c}
    \left[
        T_i
        \left(
            \boldsymbol{\alpha}_{\tau}
        \right)
        -
        T_i^{\mathrm{target}}
    \right]^2,
\end{equation}
where
$\boldsymbol{\alpha}_{\tau}
=
[\alpha_{\tau,1},\alpha_{\tau,2}]^{\mathrm{T}}$,
$N_c$ is the total number of computational cells,
$T_i$ denotes the temperature predicted by
JAX-Nano, and $T_i^{\mathrm{target}}$ is the
corresponding target temperature.

The gradients of the loss function with respect
to the unknown relaxation-time parameters are
computed using reverse-mode automatic differentiation
through the phonon BTE solver.
Both parameters are optimized simultaneously
using the Adam optimizer with a learning rate
of $0.01$. The initial scaling factors are
set to $\alpha_{\tau,1}=\alpha_{\tau,2}=0.5$.
The inverse-learning experiments are conducted
using the three-dimensional FinFET geometry
discretized into 32,400 computational cells.

Two cases are considered with identical numerical
settings but different reference scaling factors.
In the first case, both scaling factors are set
to unity,
$\alpha_{\tau,1}=\alpha_{\tau,2}=1.0$,
corresponding to the original mode-dependent
relaxation times. In the second case, unequal
scaling factors,
$\alpha_{\tau,1}=0.8$ and
$\alpha_{\tau,2}=1.2$,
are prescribed to examine the simultaneous
recovery of independently modified
mode-dependent relaxation times.

Figure~\ref{fig:finfet_inverse_relaxation_time_2modes}(a)
presents the evolution of the loss function
and learned relaxation-time scaling factors
for the identical-scaling case.
The loss decreases substantially during
the initial optimization epochs, accompanied
by a gradual increase in both scaling factors
from their initial values toward the prescribed
reference value of unity.
The two parameters exhibit similar optimization
trajectories and progressively approach their
reference values, demonstrating successful
simultaneous recovery of the two relaxation-time
scaling factors.

The accuracy of the reconstructed thermal
fields is further evaluated by comparing
the predicted temperature and heat-flux
distributions with the corresponding target
solutions.
Figure~\ref{fig:finfet_inverse_relaxation_time_2modes}(b)
shows the relative $L_2$ errors in temperature
and heat flux, evaluated along the centerline
and over the computational domain.
The small reconstruction errors indicate
close agreement between the predicted and
target thermal fields, demonstrating that
the learned relaxation-time parameters
reproduce the prescribed thermal response
with high accuracy.

The second case considers unequal reference
scaling factors,
$\alpha_{\tau,1}=0.8$ and
$\alpha_{\tau,2}=1.2$.
As shown in
Fig.~\ref{fig:finfet_inverse_relaxation_time_2modes}(c),
the loss decreases progressively as both
parameters evolve toward their respective
reference values.
Unlike the identical-scaling case, the
two parameters exhibit different convergence
trajectories. In particular, the first
scaling factor initially increases beyond
its prescribed reference value before
gradually decreasing toward
$\alpha_{\tau,1}=0.8$, whereas the second
scaling factor continues increasing toward
$\alpha_{\tau,2}=1.2$.

This nonmonotonic parameter evolution
indicates coupled optimization dynamics
associated with the simultaneous identification
of multiple relaxation times.
Although the individual parameter trajectories
are not necessarily monotonic, the optimization
process progressively reduces the discrepancy
between the predicted and target temperature
fields. This behavior illustrates that
successful multi-parameter inverse learning
does not require each unknown parameter to
approach its reference value monotonically.

Figure~\ref{fig:finfet_inverse_relaxation_time_2modes}(d)
presents the corresponding relative $L_2$
errors in the reconstructed temperature
and heat-flux fields for the unequal-scaling
case. The small reconstruction errors
demonstrate that the learned relaxation-time
parameters accurately reproduce the target
thermal response, even when the two phonon
modes possess different reference scaling
factors.

Overall, these results demonstrate the
capability of JAX-Nano to simultaneously
recover multiple mode-dependent relaxation-time
scaling factors using reverse-mode automatic
differentiation and gradient-based optimization.
The successful recovery of both identical
and unequal scaling factors establishes
the effectiveness of the proposed framework
for identifying material transport parameters.
In the following subsection, the inverse-learning
capability is further examined for a spatially
distributed heat-generation problem involving
a larger number of unknown parameters.

\subsubsection{Spatially Distributed Heat Sources in a
$3\times3$ FinFET Array}

\begin{figure}
\begin{center}
\includegraphics[width=1.\columnwidth]{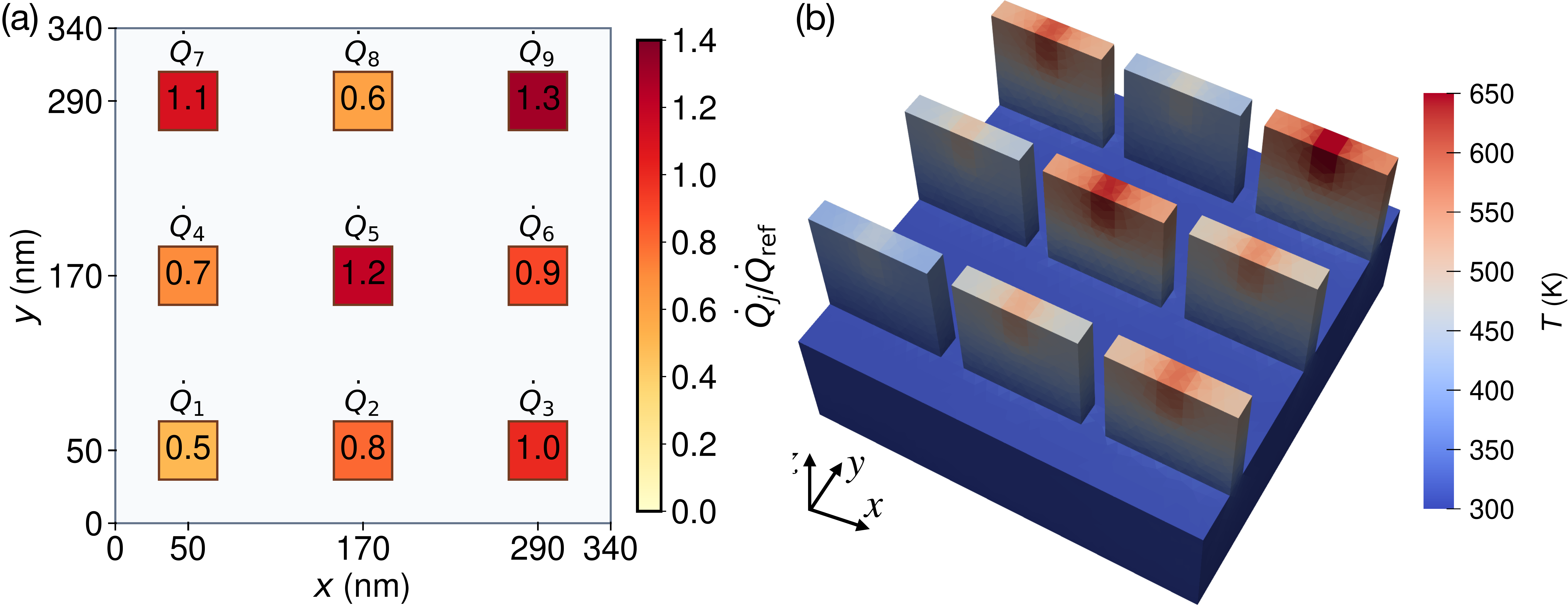}
\end{center}
\caption{
Spatially distributed volumetric heat generation in a
$3\times3$ FinFET array.
(a) Spatial arrangement of the nine heat-generation regions
with independently prescribed volumetric heat-generation
rates, normalized by $\dot{Q}_{\mathrm{ref}}=10^{19}$
W\,m$^{-3}$.
The numerical values indicate the prescribed normalized
heat-generation rates $\dot{Q}_j/\dot{Q}_{\mathrm{ref}}$.
(b) Three-dimensional steady-state temperature distribution
obtained using the prescribed nonuniform heat-generation
rates, serving as the reference solution for inverse
parameter identification.
}
\label{fig:finfet_array_inverse_configuration}
\end{figure}

\begin{figure}
\begin{center}
\includegraphics[width=1.\columnwidth]{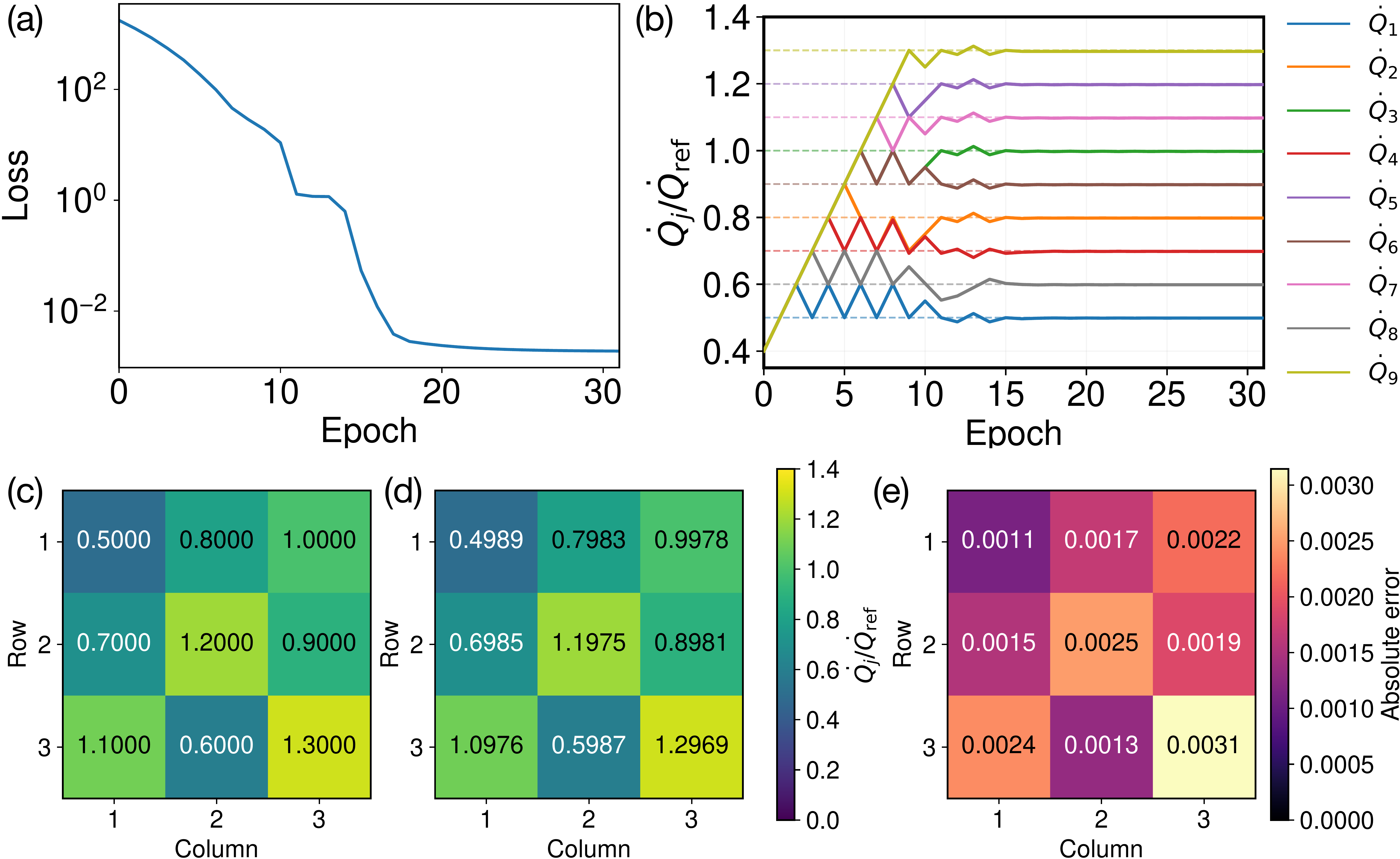}
\end{center}
\caption{
Gradient-based inverse identification of nine independently
distributed volumetric heat-generation rates in the
$3\times3$ FinFET array.
(a) Evolution of the temperature-based loss function
during optimization.
(b) Optimization histories of the nine normalized
heat-generation rates $\dot{Q}_j/\dot{Q}_{\mathrm{ref}}$, with horizontal
dashed lines indicating their prescribed reference values.
(c) Prescribed normalized heat-generation rates.
(d) Recovered normalized heat-generation rates after
optimization.
(e) Absolute errors between the prescribed and recovered
normalized heat-generation rates.
}
\label{fig:finfet_array_inverse_optimization}
\end{figure}

\begin{figure}
\begin{center}
\includegraphics[width=1.\columnwidth]{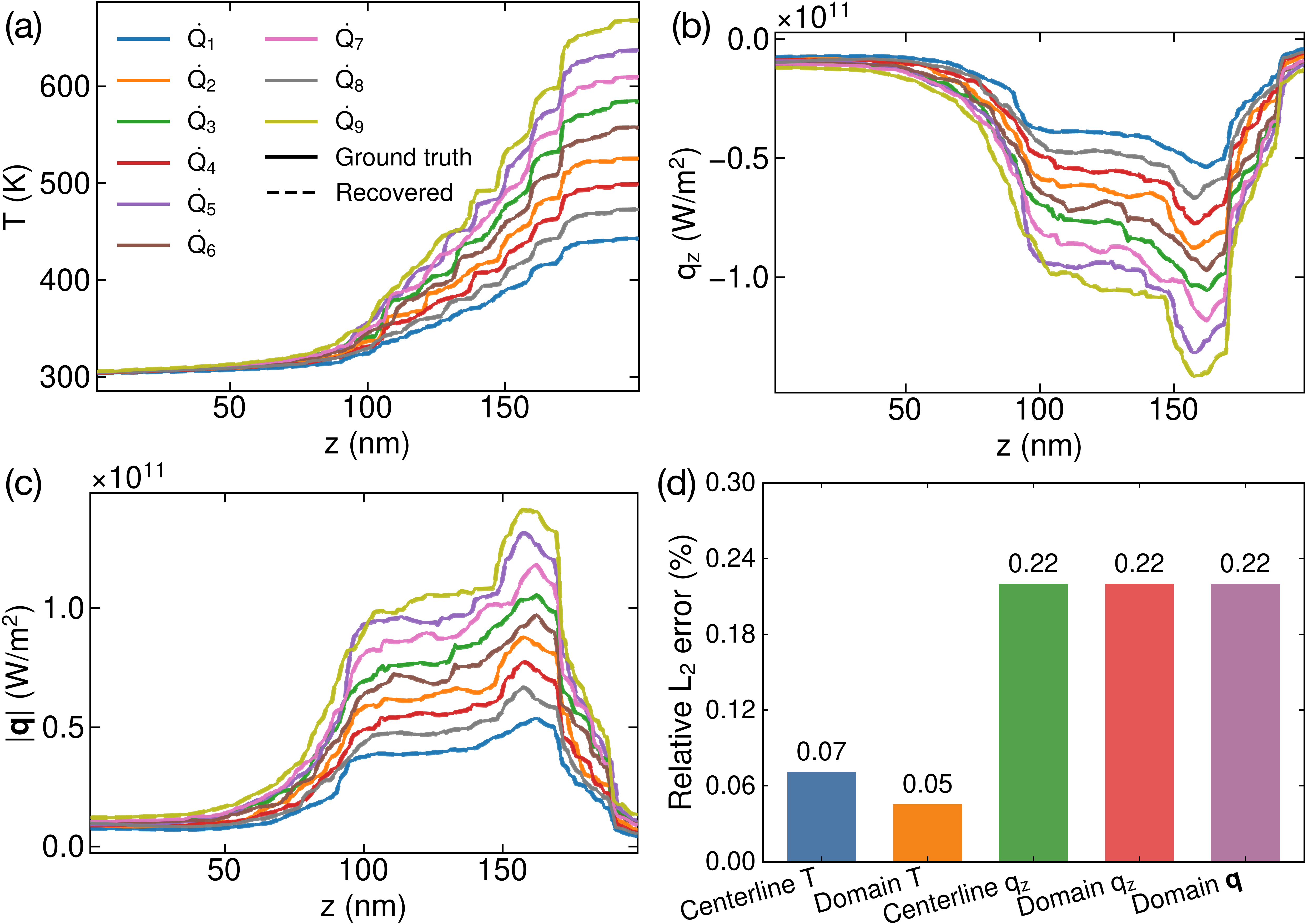}
\end{center}
\caption{
Comparison of reference and reconstructed thermal fields
using the recovered volumetric heat-generation rates in
the $3\times3$ FinFET array.
(a) Temperature profiles along the nine vertical centerlines
passing through the individual heat-generation regions.
(b) Corresponding vertical heat-flux component $q_z$.
(c) Heat-flux magnitude $|\mathbf{q}|$.
Solid and dashed lines represent the reference and
reconstructed solutions, respectively, with matching
colors identifying the nine heat-generation regions.
(d) Relative $L_2$ errors for the combined centerline
temperature profiles, full-domain temperature field,
combined centerline vertical heat-flux profiles,
full-domain vertical heat flux, and full-domain
heat-flux vector.
}
\label{fig:finfet_array_inverse_reconstruction}
\end{figure}

To further assess the capability of JAX-Nano for
multi-parameter inverse identification, we consider
the simultaneous recovery of nine spatially distributed
volumetric heat-generation rates in a $3\times3$
FinFET array. The computational geometry and boundary
conditions follow those employed in the preceding
forward-validation study, while a coarser tetrahedral mesh
consisting of 66,441 cells is adopted, with one
phonon band and 64 angular directions. In contrast
to the uniform heat-generation configuration
considered previously, the nine source regions are
assigned different volumetric heat-generation rates,
as illustrated in
Fig.~\ref{fig:finfet_array_inverse_configuration}(a).

The unknown parameter vector is defined as
\begin{equation}
\boldsymbol{\theta}
=
\left[
\theta_1,\theta_2,\ldots,\theta_9
\right]^{\mathsf T},
\qquad
\theta_i=\frac{\dot{Q}_j}{\dot{Q}_{\mathrm{ref}}},
\end{equation}
where $\dot{Q}_j$ denotes the volumetric heat-generation
rate within the $j$th source region and
$\dot{Q}_{\mathrm{ref}}=10^{19}\,\mathrm{W\,m^{-3}}$
is the reference heat-generation rate.
The prescribed normalized parameter values are
\begin{equation}
\boldsymbol{\theta}_{\mathrm{ref}}
=
\left[
0.5,\,
0.8,\,
1.0,\,
0.7,\,
1.2,\,
0.9,\,
1.1,\,
0.6,\,
1.3
\right]^{\mathsf T}.
\end{equation}
The corresponding steady-state temperature
distribution, obtained by solving the forward
phonon BTE with the prescribed heat-generation
rates, is shown in
Fig.~\ref{fig:finfet_array_inverse_configuration}(b).
The nonuniform heat-generation rates produce
distinct temperature distributions across the
individual fins, providing the reference thermal
response for inverse parameter identification.

The nine normalized heat-generation rates are
identified simultaneously through gradient-based
optimization of the temperature-based loss function.
The gradients are computed using forward-mode
automatic differentiation, with individual gradient
components evaluated through separate
Jacobian--vector products (JVPs).
The optimization employs projected gradient descent
with gradient clipping and backtracking line search,
using an initial learning rate of $\eta=0.1$.
All nine parameters are initialized to $0.4$.
Each forward BTE evaluation employs 200 iterations
to obtain a converged steady-state solution.

Figure~\ref{fig:finfet_array_inverse_optimization}
presents the optimization histories and
parameter-identification results.
As shown in
Fig.~\ref{fig:finfet_array_inverse_optimization}(a),
the loss function decreases rapidly during the
initial optimization epochs, followed by a more
gradual reduction toward a nearly stationary value.
The corresponding evolution of the nine normalized
heat-generation rates is shown in
Fig.~\ref{fig:finfet_array_inverse_optimization}(b).
Despite differences in their optimization
trajectories, all nine parameters approach their
prescribed reference values within approximately
15 epochs and remain nearly constant thereafter.

Figures~\ref{fig:finfet_array_inverse_optimization}(c)
and (d) compare the prescribed and recovered
normalized heat-generation rates. The recovered
parameter vector is
\begin{equation}
%\frac{\boldsymbol{\theta}_{\mathrm{rec}}}
%{Q_{\mathrm{ref}}}
\theta_{\mathrm{rec}}
=
\left[
0.4989,\,
0.7983,\,
0.9978,\,
0.6985,\,
1.1975,\,
0.8981,\,
1.0976,\,
0.5987,\,
1.2969
\right]^{\mathsf T}.
\end{equation}
The recovered values closely reproduce the
prescribed spatial distribution of heat generation.
The corresponding absolute identification errors,
shown in
Fig.~\ref{fig:finfet_array_inverse_optimization}(e),
range from $0.0011$ to $0.0031$ in normalized
heat-generation rate. The largest absolute error
occurs for $\dot{Q}_9$, while all nine normalized
parameters are recovered with absolute errors
below $0.0032$. These results demonstrate that
the differentiable solver can simultaneously
identify multiple spatially distributed source
strengths from their combined thermal response.

To further evaluate the accuracy of the identified
parameters, the reconstructed temperature and
heat-flux fields are compared with the reference
solutions. Figure~\ref{fig:finfet_array_inverse_reconstruction}
presents the temperature, vertical heat-flux
component, and heat-flux magnitude along nine
vertical centerlines passing through the
individual heat-generation regions. The
centerline locations correspond to the source
positions shown in
Fig.~\ref{fig:finfet_array_inverse_configuration}(a).
As shown in
Fig.~\ref{fig:finfet_array_inverse_reconstruction}(a),
the reconstructed temperature profiles closely
follow the reference profiles over the entire
sampled vertical range. The differences in
temperature among the nine centerlines reflect
the nonuniform heat-generation rates, with
larger source strengths generally producing
higher local temperatures.

Figures~\ref{fig:finfet_array_inverse_reconstruction}(b)
and (c) compare the corresponding vertical
heat-flux component $q_z$ and heat-flux magnitude
$|\mathbf{q}|$, respectively. The reconstructed
solutions reproduce the spatial variations in
both quantities, including the pronounced
heat-flux variations near the heat-generation
regions. The agreement in heat flux is
particularly relevant because the inverse
objective is based on temperature observations,
whereas the heat-flux fields provide an additional
assessment of the recovered thermal transport
response.

The reconstruction accuracy is quantified using
the relative $L_2$ errors summarized in
Fig.~\ref{fig:finfet_array_inverse_reconstruction}(d).
The combined temperature centerlines exhibit
a relative $L_2$ error of approximately $0.07\%$,
while the full-domain temperature error is
approximately $0.05\%$. For the vertical
heat-flux component, the relative errors along
the combined centerlines and over the full
computational domain are both approximately
$0.22\%$. The full-domain heat-flux vector
also exhibits a relative $L_2$ error of
approximately $0.22\%$.

These results demonstrate that JAX-Nano can
simultaneously recover nine independently
distributed volumetric heat-generation rates
with small parameter-identification errors.
Furthermore, the recovered parameters reproduce
both the reference temperature field and the
associated heat-flux distributions with relative
$L_2$ errors below $0.25\%$. This example
illustrates the applicability of the differentiable
phonon BTE framework to multi-parameter inverse
problems involving spatially heterogeneous
thermal sources in three-dimensional nanoscale
devices.

\section{Computational Performance}

\subsection{GPU Performance and Scalability}
\label{subsec:gpu_performance}

\begin{figure}
\begin{center}
\includegraphics[width=1.\columnwidth]{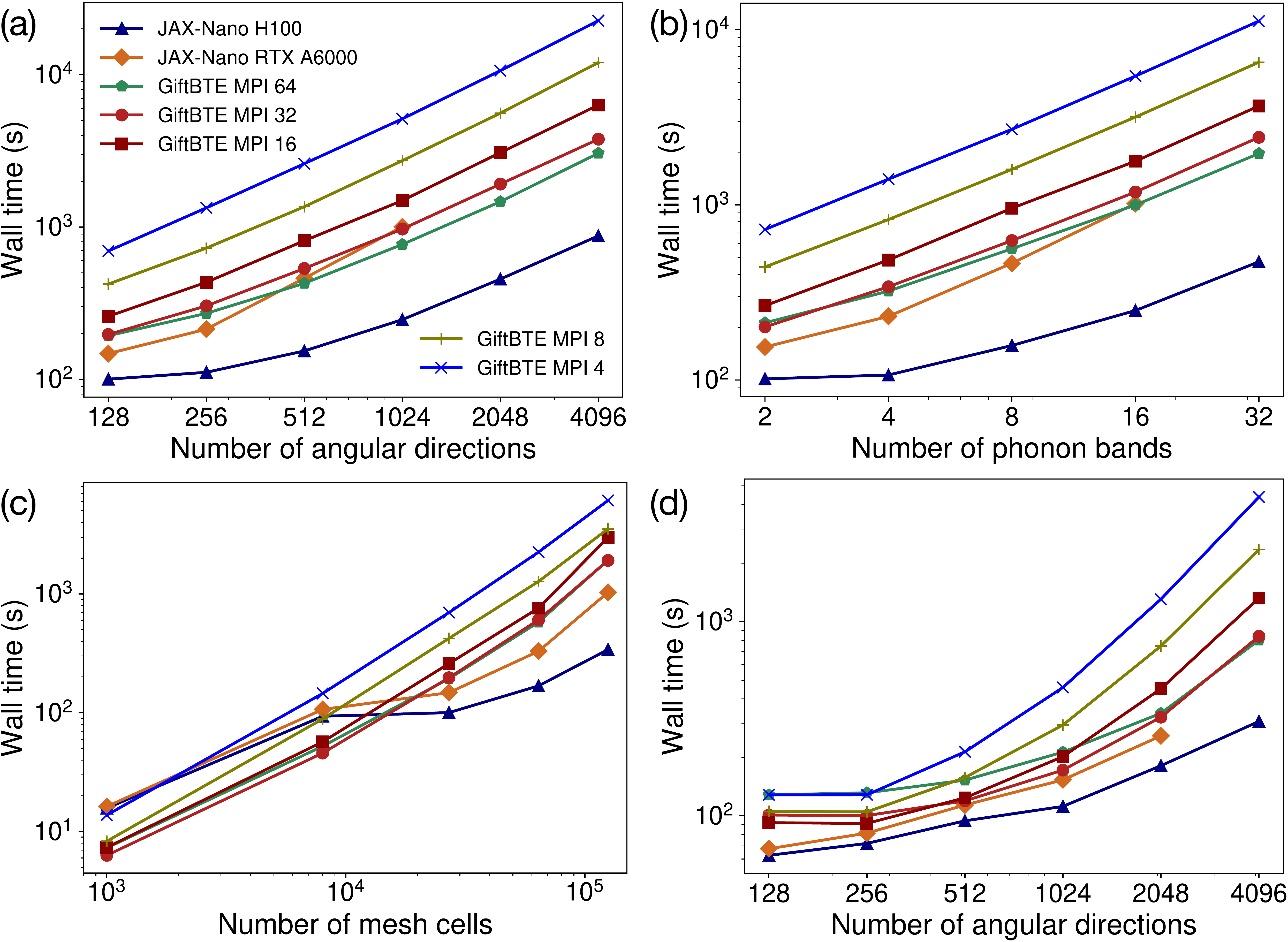}
\end{center}
\caption{
Computational performance and scalability of JAX-Nano
and GiftBTE for steady-state and transient phonon transport
simulations in a three-dimensional cubic silicon domain. Wall-clock time as a function of
(a) the number of angular directions for one phonon band
and 27,000 computational cells,
(b) the number of phonon bands for 64 angular directions
and 27,000 computational cells, and
(c) the number of computational mesh cells for 128 angular
directions and one phonon band.
Panels (a)--(c) correspond to steady-state simulations.
(d) Wall-clock time as a function of the number of angular
directions for transient simulations using explicit time
integration in both JAX-Nano and GiftBTE, with one phonon
band and 27,000 computational cells.
JAX-Nano is executed on NVIDIA H100 and RTX A6000 GPUs,
while GiftBTE is evaluated using 4, 8, 16, 32, and
64 MPI processes.
}
\label{fig:gpu_speedup}
\end{figure}

The computational performance and scalability of JAX-Nano
are evaluated using phonon transport simulations in a
three-dimensional cubic silicon domain. Thermalizing boundary
conditions are imposed on the top and bottom surfaces,
maintained at 800~K and 300~K, respectively, while
diffuse-reflection boundary conditions are applied to
the four lateral walls.

The benchmarks are conducted on a workstation equipped
with two Intel Xeon Gold 6438Y+ processors, providing
64 physical CPU cores and 128 logical CPUs. The GPU
hardware consists of one NVIDIA H100 NVL with 94~GB
of memory and three NVIDIA RTX A6000 GPUs, each with
48~GB of memory. JAX-Nano is evaluated on individual
H100 NVL and RTX A6000 GPUs, whereas the CPU-based
GiftBTE solver is benchmarked using 4, 8, 16, 32,
and 64 MPI processes.

The benchmarks investigate the effects of angular
resolution, spectral discretization, and spatial
mesh resolution on computational cost. The corresponding
wall-clock times are presented in Fig.~\ref{fig:gpu_speedup}.
Panels (a)--(c) consider steady-state phonon transport,
where both JAX-Nano and GiftBTE employ the BiCGSTAB
iterative solver with a fixed number of 200 iterations
for consistent performance comparisons. Panel (d)
examines transient simulations using explicit time
integration in both solvers, with an identical time
step of $\Delta t=2\times10^{-13}$~s and 200 time steps.

Figure~\ref{fig:gpu_speedup}(a) compares the wall-clock
times of JAX-Nano and GiftBTE as the number of angular
directions increases from 128 to 4096, using a fixed
mesh of 27,000 computational cells and one phonon band.
For JAX-Nano executed on the NVIDIA H100 NVL GPU,
the computational time increases from 100.28~s to
875.59~s, whereas GiftBTE using 64 MPI processes
requires 193.42~s and 3052.92~s, respectively.
Consequently, the speedup of JAX-Nano relative to
GiftBTE increases from 1.93 to 3.49 as the angular
resolution increases. At 1024 and 2048 angular
directions, the corresponding speedup factors are
3.12 and 3.22, respectively, indicating that the
computational advantage of the H100 becomes more
pronounced for larger angular workloads.

The NVIDIA RTX A6000 also demonstrates competitive
performance at moderate angular resolutions.
At 512 angular directions, JAX-Nano on the RTX A6000
requires 461.06~s, compared with 534.67~s for
GiftBTE using 32 MPI processes, corresponding to
a speedup of 1.16. At 1024 angular directions,
the execution times become comparable, with
1003.37~s for the RTX A6000 and 970.94~s for
GiftBTE using 32 MPI processes, representing a
difference of approximately 3.3\%.
These results demonstrate that the computational
benefits of GPU acceleration depend on both the
angular resolution and the GPU hardware, with
the H100 providing a consistently greater
performance advantage over the investigated
CPU configurations.

The measured speedup also depends on the number
of MPI processes employed by GiftBTE. Relative
to GiftBTE using four MPI processes, the H100
implementation achieves speedup factors ranging
from 6.95 to 25.87 as the number of angular
directions increases from 128 to 4096.
When compared with GiftBTE using 64 MPI
processes, the corresponding speedup factors
range from 1.93 to 3.49. The latter provides
a more demanding CPU baseline, utilizing all
64 physical CPU cores available on the
benchmark workstation.

The available GPU memory also imposes a practical
constraint on the maximum angular resolution.
While the NVIDIA H100 NVL GPU, equipped with
94~GB of memory, successfully completes simulations
with up to 4096 angular directions, the RTX A6000,
with 48~GB of memory, is limited to 1024 angular
directions in the present benchmark due to
GPU memory constraints. This highlights the
importance of device memory capacity for
high-dimensional phonon BTE simulations,
particularly as the number of angular
degrees of freedom increases.

The influence of spectral discretization is examined in
Fig.~\ref{fig:gpu_speedup}(b), where the number of phonon
bands increases from 2 to 32 while maintaining 64 angular
directions and 27,000 computational cells. Increasing the
number of bands introduces additional spectral transport
equations and consequently increases the computational
workload. For JAX-Nano executed on the NVIDIA H100 NVL GPU,
the wall-clock time increases from 101.40~s for two bands
to 472.84~s for 32 bands, whereas GiftBTE using 64 MPI
processes requires 211.95~s and 1963.65~s, respectively.
Consequently, the H100 achieves speedup factors increasing
from 2.09 to 4.15 relative to GiftBTE using 64 MPI
processes. When compared with GiftBTE using four MPI
processes, the corresponding speedup increases from
7.13 to 23.69. These results demonstrate that the
computational advantage of GPU acceleration becomes
more pronounced as the number of phonon bands increases.

The RTX A6000 also exhibits competitive performance
for multiband simulations. Its wall-clock time increases
from 154.22~s for two bands to 1016.57~s for 16 bands.
At 16 bands, the RTX A6000 exhibits performance
comparable to GiftBTE using 64 MPI processes, which
requires 997.87~s, corresponding to a difference of
approximately 1.9\%. However, the 32-band configuration
exceeds the available device memory of the RTX A6000,
whereas the H100 NVL successfully completes the
simulation. These findings highlight both the
computational advantages of GPU acceleration for
multiband phonon transport and the importance of
device memory capacity when increasing the spectral
degrees of freedom.

Figure~\ref{fig:gpu_speedup}(c) examines the influence
of spatial resolution on computational performance,
with 128 angular directions and one phonon band.
The number of computational cells increases from
1000 to 125,000, resulting in a corresponding
increase in the number of transport unknowns.
For relatively small meshes, the CPU-based GiftBTE
solver exhibits shorter execution times than
JAX-Nano. At 1000 cells, GiftBTE using 64 MPI
processes requires only 7.30~s, compared with
15.78~s and 16.27~s for JAX-Nano on the H100 NVL
and RTX A6000 GPUs, respectively. 
This behavior indicates that GPU acceleration is less
beneficial for small computational workloads, where
the overhead associated with JAX just-in-time (JIT)
compilation, GPU initialization, and warm-up can
account for a substantial fraction of the total
execution time.

A clear performance crossover is observed between
8000 and 27,000 computational cells. At 8000 cells,
GiftBTE using 64 MPI processes completes the simulation
in 52.32~s, compared with 93.50~s and 106.72~s
for JAX-Nano on the H100 NVL and RTX A6000 GPUs,
respectively. At 27,000 cells, however, the H100
and RTX A6000 require 100.28~s and 147.64~s,
respectively, compared with 193.42~s for GiftBTE.
The corresponding GPU speedup factors are 1.93
and 1.31, respectively. This crossover indicates
that GPU acceleration becomes advantageous beyond
a sufficiently large spatial computational workload,
where the increased degree of parallelism can offset the overheads
associated with GPU execution.

The computational advantage of JAX-Nano becomes
more pronounced as the mesh resolution increases
further. At 64,000 cells, the H100 completes the
simulation in 168.44~s, compared with 578.42~s
for GiftBTE using 64 MPI processes, yielding a
speedup of 3.43. For the largest mesh containing
125,000 cells, the H100 requires 340.20~s,
whereas GiftBTE using 64 MPI processes requires
1928.16~s, corresponding to a speedup of 5.67.
Relative to GiftBTE using four MPI processes,
the H100 achieves a speedup of 17.96 for the
largest mesh. The RTX A6000 also demonstrates
a computational advantage at this resolution,
requiring 1031.51~s and achieving a speedup of
1.87 relative to GiftBTE using 64 MPI processes.
These results demonstrate that GPU acceleration
becomes increasingly effective with growing
spatial computational workloads, highlighting
the suitability of JAX-Nano for large-scale
three-dimensional phonon BTE simulations.

The computational performance of explicit transient
simulations is examined in Fig.~\ref{fig:gpu_speedup}(d),
where the number of angular directions increases
from 128 to 4096 for a fixed mesh of 27,000 cells
and one phonon band. Both JAX-Nano and GiftBTE
employ identical explicit time-integration settings,
ensuring a consistent physical simulation duration
and a direct comparison of computational performance.
As the angular resolution increases, the computational
cost rises because the directional transport equations
must be evaluated at each time step.

At 128 angular directions, JAX-Nano on the
H100 NVL requires 62.66~s, compared with
128.25~s for GiftBTE using 64 MPI processes,
corresponding to a speedup of 2.05. At 1024
directions, the execution times are 111.94~s
and 213.05~s, respectively, yielding a speedup
of 1.90. For the largest configuration with
4096 directions, the H100 completes the
simulation in 306.69~s, whereas GiftBTE using
64 MPI processes requires 804.13~s, achieving
a speedup of 2.62. Relative to GiftBTE using
four MPI processes, the H100 speedup reaches
14.29 at this angular resolution.

The RTX A6000 also demonstrates a computational
advantage at moderate angular resolutions.
At 128 directions, its execution time is
67.75~s, compared with 128.25~s for GiftBTE
using 64 MPI processes, corresponding to a
speedup of 1.89. At 2048 directions, the
RTX A6000 requires 258.18~s, compared with
337.60~s for GiftBTE using 64 MPI processes,
yielding a speedup of 1.31. 
The 4096-direction configuration exceeds the
available GPU memory of the RTX A6000 and
therefore cannot be completed on this device.

The GiftBTE results further indicate that increasing
the number of MPI processes does not necessarily
improve computational performance at low angular
resolutions. For example, in the explicit transient
benchmark with 128 angular directions, GiftBTE
requires 92.25~s using 16 MPI processes, compared
with 128.25~s using 64 MPI processes. This behavior
is consistent with increased interprocess
communication and synchronization overhead
relative to the computational work assigned to
each process. As the angular resolution increases,
the larger computational workload can improve
the utilization of additional CPU processes.
At 4096 directions, GiftBTE using 64 MPI processes
achieves the shortest execution time among the
tested CPU configurations. These observations
highlight the importance of selecting an
appropriate MPI process count according to
the angular resolution.

It should be noted that the present transient
benchmark considers only 200 time steps, and
the reported wall-clock times include JAX
just-in-time (JIT) compilation, GPU initialization,
and warm-up overhead. These one-time costs can
represent a substantial fraction of the total
execution time for relatively short simulations.
For longer transient simulations involving more
time steps, these fixed overheads are expected
to become less significant relative to the
cumulative time-integration cost. 
Consequently, the relative contribution of these
one-time overheads decreases for longer transient
simulations, potentially enhancing the computational
advantage of JAX-Nano. The performance of JAX-Nano
and GiftBTE for longer transient simulations
approaching steady state is further examined
in the following subsection.

Overall, the benchmarks demonstrate the
effectiveness of GPU acceleration for both
steady-state and explicit transient phonon
transport, particularly at higher spatial,
angular, and spectral resolutions. The results
also highlight the influence of computational
workload and GPU memory capacity on achievable
performance. The computational efficiency of
explicit and implicit transient formulations
is examined in the following subsection.

\subsection{Explicit versus Implicit Transient Performance}

\begin{figure}
\begin{center}
\includegraphics[width=1.\columnwidth]{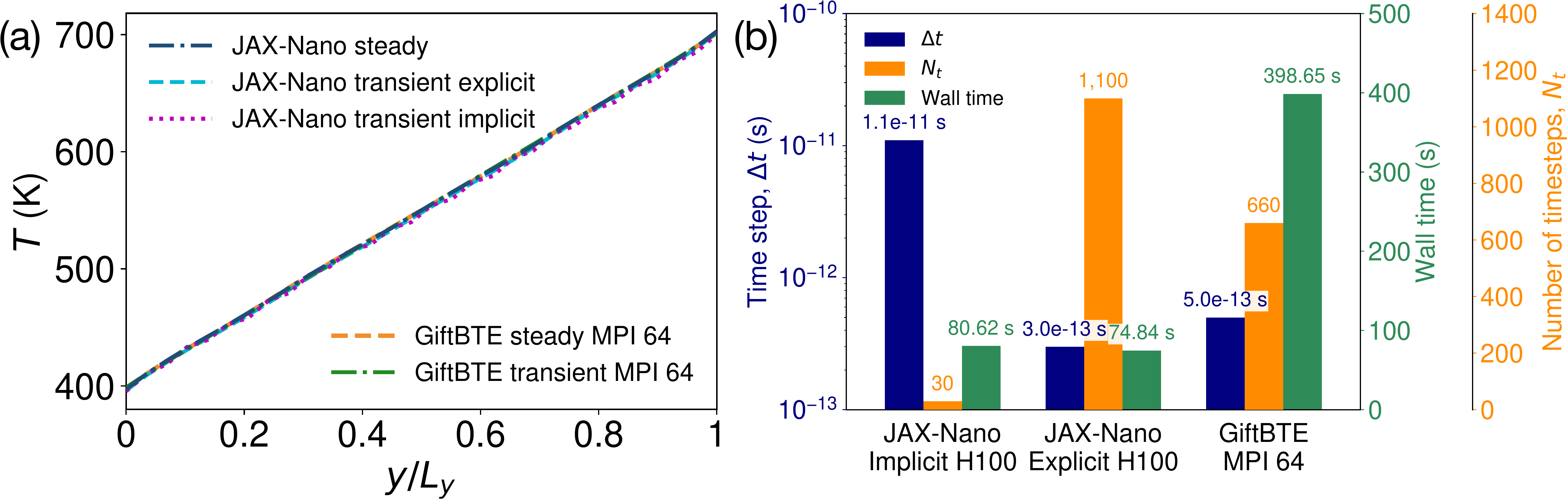}
\end{center}
\caption{
Validation and computational performance of
JAX-Nano and GiftBTE for a three-dimensional
cubic silicon domain with 27,000 computational cells,
128 angular directions, and one phonon band.
(a) Comparison of centerline temperature
distributions obtained using the steady-state,
transient explicit, and transient implicit
formulations of JAX-Nano and the corresponding
steady-state and transient solutions of GiftBTE.
(b) Comparison of time-step size $\Delta t$,
number of physical time steps $N_t$, and total
wall-clock time for JAX-Nano implicit and explicit
integration on an NVIDIA H100 NVL GPU and GiftBTE
explicit integration using 64 MPI processes.
All transient simulations reach the same final
physical time of $3.3\times10^{-10}$~s, at which
steady-state conditions are attained.
}
\label{fig:explicit_implicit_giftbte}
\end{figure}

The computational efficiency and numerical consistency
of the steady-state and transient formulations are
evaluated using a three-dimensional cubic silicon domain
containing 27,000 computational cells, 128 angular
directions, and one phonon band. The steady-state
and transient explicit and implicit formulations
of JAX-Nano are executed on a single NVIDIA H100
NVL GPU, while the corresponding steady-state and
transient simulations of GiftBTE are performed
using 64 MPI processes on the CPU platform.
The top and bottom surfaces are maintained at
800~K and 300~K, respectively, while diffuse
reflection conditions are imposed on the four
lateral surfaces. The steady-state solutions
serve as references for evaluating the transient
formulations. All transient simulations are
advanced to a common final physical time of
$3.3\times10^{-10}$~s, at which the temperature
field has reached steady state.

Figure~\ref{fig:explicit_implicit_giftbte}(a)
compares the centerline temperature distributions
obtained from the steady-state and transient
formulations of JAX-Nano and GiftBTE.
The temperature profiles exhibit close agreement
across the domain, demonstrating that both
explicit and implicit transient formulations
recover the corresponding steady-state solution.
This agreement also confirms the consistency
between the GPU-based JAX-Nano implementation
and the CPU-based GiftBTE solver.

Figure~\ref{fig:explicit_implicit_giftbte}(b)
compares the time-step sizes, numbers of time
steps, and total wall-clock times.
The JAX-Nano implicit formulation employs a
time step of $\Delta t=1.1\times10^{-11}$~s,
requiring only 30 time steps to reach the
final physical time, with a total execution
time of 80.62~s. In contrast, the JAX-Nano
explicit formulation uses
$\Delta t=3.0\times10^{-13}$~s and 1100 time
steps, completing the simulation in 74.84~s.
GiftBTE employs explicit time integration
with $\Delta t=5.0\times10^{-13}$~s and
660 time steps, requiring 398.65~s using
64 MPI processes.

Despite requiring substantially more time
steps, the JAX-Nano explicit formulation
achieves a slightly shorter total execution
time than its implicit counterpart.
This behavior reflects the additional
computational cost associated with solving
the implicit transport equations at each
time step. Relative to GiftBTE using
64 MPI processes, the JAX-Nano explicit
and implicit formulations achieve speedup
factors of approximately 5.33 and 4.95,
respectively.

These results demonstrate that both transient
formulations of JAX-Nano can efficiently
recover steady-state phonon transport
solutions. Although implicit time integration
permits substantially larger time steps,
its computational advantage depends on
the balance between the number of time steps
and the cost of each implicit solve.
For the present configuration, explicit
time integration provides slightly better
overall computational performance, while
both GPU implementations substantially
outperform the CPU-based GiftBTE solver. 

%===============================================================================
% Conclusion
%===============================================================================
\section{Conclusion}
\label{sec:conclusion}

This study presented JAX-Nano, a GPU-accelerated
and differentiable finite-volume framework for
steady-state and transient phonon Boltzmann
transport simulations. The framework integrates
structured and unstructured spatial discretizations,
mode-resolved phonon transport, multiple boundary
conditions, and explicit and implicit time-integration
schemes within a unified JAX implementation.
A matrix-free sparse iterative solution strategy,
combined with just-in-time compilation and
GPU-accelerated numerical operations, enables
three-dimensional phonon transport simulations
while retaining compatibility with automatic
differentiation. This integration provides a
computational foundation for both forward thermal
analysis and gradient-based inverse parameter
identification.

The numerical accuracy of JAX-Nano was systematically
examined using canonical phonon transport benchmarks.
For steady-state cross-plane transport, the predicted
temperature and heat-flux distributions agreed
closely with analytical solutions and DUGKS results
over a broad range of Knudsen numbers, capturing
the transition from diffusive to ballistic transport.
For transient thermal grating problems, JAX-Nano
reproduced the characteristic temperature evolution
in both diffusive and ballistic regimes, including
the oscillatory behavior associated with ballistic
phonon propagation. Spatial refinement studies
demonstrated approximately second-order convergence
of temperature and heat flux. These results establish
the numerical consistency of the framework across
different transport regimes.

The framework was further validated for
three-dimensional phonon transport in cubic
domains and realistic FinFET configurations.
Comparisons with GiftBTE demonstrated close
agreement in temperature and heat-flux predictions
under different boundary conditions, spatial
discretizations, and spectral resolutions.
The simulations included single-fin and
double-fin transistors as well as a
$3\times3$ FinFET array with localized
volumetric heat generation. These examples
demonstrate the applicability of JAX-Nano
to geometrically complex nanoscale devices
and spatially heterogeneous thermal transport
problems.

The differentiable capabilities of JAX-Nano
were demonstrated through a series of
gradient-based inverse parameter identification
problems. Single-parameter experiments examined
the recovery of characteristic domain size,
thermal boundary temperature, volumetric
heat-generation rate, and phonon relaxation
time. Multi-parameter experiments further
investigated the simultaneous identification
of two mode-dependent relaxation-time scaling
factors and nine independently distributed
volumetric heat-generation rates in a
$3\times3$ FinFET array. The relaxation-time
experiments demonstrated successful recovery
of both the original mode-dependent properties
and independently scaled relaxation times.
For the spatially distributed heat-source
problem, the reconstructed temperature and
heat-flux fields exhibited relative $L_2$
errors below $0.25\%$. These results demonstrate
the feasibility of using automatic differentiation
through the phonon BTE solver to identify
multiple physical parameters from thermal
observations.

The computational performance of JAX-Nano
was evaluated on NVIDIA H100 NVL and RTX A6000
GPUs and compared with the MPI-based GiftBTE
solver. Systematic benchmarks examined the
effects of spatial resolution, angular
discretization, and the number of phonon
bands. The results demonstrated increasing
benefits from GPU acceleration as the
computational workload increased. For
steady-state simulations with 125,000
computational cells, JAX-Nano on the H100
achieved a speedup of approximately $5.67$
relative to GiftBTE using 64 MPI processes.
For 4096 angular directions, the corresponding
speedup was approximately $3.49$.
Comparisons of transient integration schemes
further showed that implicit integration
substantially reduced the required number
of physical time steps, although its additional
linear-solver cost resulted in a slightly
longer execution time than explicit integration
for the investigated configuration.
These findings highlight the importance
of computational workload, numerical
formulation, and hardware characteristics
in determining the performance of
GPU-accelerated phonon transport simulations.

Overall, JAX-Nano establishes a unified
framework that combines accurate phonon
BTE simulation, GPU acceleration, and
automatic differentiation for nanoscale
thermal transport analysis. Beyond forward
prediction, its ability to identify material
properties, boundary conditions, geometric
parameters, and spatially distributed
heat sources demonstrates the potential
of differentiable kinetic transport
simulations for thermal characterization
and inverse modeling. Future work will
focus on extending the framework to
more comprehensive phonon dispersion
and scattering models, improving
computational and memory efficiency
for large-scale differentiable simulations,
and enabling gradient-based thermal
design and optimization of complex
nanoscale devices.

%===============================================================================
%	Acknowledgments and bibliography
%===============================================================================
\section*{Acknowledgements}
%The author thanks the anonymous reviewers for their comments and suggestions that helped to promote the quality and clarity of the manuscript.  
This research was supported by the Division of Research and Innovation Partnership Commitments (RIPS) at Northern Illinois University. 

\section*{Data availability}
The data that support the findings of the present work are available from the corresponding author upon reasonable request. 

\section*{References}
\bibliographystyle{unsrtnat}
\bibliography{refs}

\end{document}